\documentclass{aa} 
\usepackage{times,epsfig,amssymb,amsmath}

\title{The cluster gas mass fraction as a cosmological probe: \\ a revised study}
\titlerunning{Cosmological constraints from the Cluster Gas Fraction}

\author{S. Ettori\inst{1,2} \and A. Morandi\inst{3,4} \and P. Tozzi\inst{5,6} 
 \and I. Balestra\inst{7} \and S. Borgani\inst{5,6,8} \and P. Rosati\inst{9}
 \and L. Lovisari\inst{3,10} \and F. Terenziani\inst{3}}
\authorrunning{S. Ettori et al.}

\institute{
 INAF, Osservatorio Astronomico di Bologna, via Ranzani 1, I-40127 Bologna, Italy
 \and INFN, Sezione di Bologna, viale Berti Pichat 6/2, I-40127 Bologna, Italy
 \and Dipartimento di Astronomia, Universit\`a di Bologna, via Ranzani 1, I-40127 Bologna, Italy
 \and Dark Cosmology Centre, Niels Bohr Institute, University of Copenhagen, Juliane Maries Vej
   30, DK-2100 Copenhagen, Denmark
 \and INAF, Osservatorio Astronomico di Trieste, via G.B. Tiepolo 11, I-34131 Trieste, Italy
 \and INFN, Sezione di Trieste, I-34100 Trieste, Italy
 \and MPE, Karl-Schwarzschild-Str. 2, D-85741 Garching, Germany
 \and Dipartimento di Astronomia, Universit\`a di Trieste, via Tiepolo 11, I-34131 Trieste, Italy
 \and ESO, Karl-Schwarzschild-Str. 2, D-85748 Garching, Germany
 \and Institut f\"ur Astro- und Teilchenphysik, Universit\"at Innsbruck, Technikerstrasse 25, 
   A-6020 Innsbruck, Austria
}

\offprints{S. Ettori}

\mail{stefano.ettori@oabo.inaf.it}

\date{Accepted on March 28, 2009}

\begin{document}

\abstract
{We present the analysis of the baryonic content of 52 X-ray luminous galaxy clusters 
observed with Chandra in the redshift range 0.3 -- 1.273.}
{Our study aims at resolving the gas mass fraction in these objects to place constraints
on the cosmological parameters $\Omega_{\rm m}$, $\Omega_{\Lambda}$ and the ratio between the
pressure and density of the dark energy, $w$. }
{We deproject the X-ray surface brightness profiles to recover the gas
  mass profiles and fit a single thermal component to the spectrum
  extracted from a region around the cluster that maximizes the
  signal-to-noise ratios in the observation. The measured values of the gas
  temperature are used to evaluate the temperature profile with a 
  given functional form and to estimate the total gravitating mass in
  combination with the gas density profiles. These measured quantities
  are then used to statistically estimate the gas fraction and the
  fraction of mass in stars. By assuming that galaxy clusters are
  representative of the cosmic baryon budget, the distribution of the
  cluster baryon fraction in the hottest ($T_{\rm gas}> 4$ keV)
  systems as a function of redshift is used to constrain the
  cosmological parameters.  We discuss how our constraints are
  affected by several systematic effects, namely the isothermality, the assumed baryon
  fraction in stars, the depletion parameter and the sample selection.}
{By using only the cluster baryon fraction as a proxy for the
  cosmological parameters, we obtain that $\Omega_{\rm m}$ is very
  well constrained at the value of $0.35$ with a relative statistical
  uncertainty of 11\% ($1 \sigma$ level; $w=-1$) and a further
  systematic error of about $(-6,+7)$\%. On the other hand,
  constraints on $\Omega_{\Lambda}$ (without the prior of flat geometry)
  and $w$ (using the prior of flat geometry)
  are definitely weaker due to the presence of greater statistical and 
  systematic uncertainties (of the order of 40 per cent on $\Omega_{\Lambda}$ 
  and greater than 50 per cent on $w$).  
 If the WMAP 5-year best-fit results are assumed to fix the cosmological parameters, 
we limit the contributions expected from non-thermal pressure support 
and ICM clumpiness to be lower than about 10 per cent, 
also leaving room to accommodate baryons not accounted for either 
in the X-ray emitting plasma or in stars of the order
of 18 per cent of the total cluster baryon budget.
This value is lowered to zero for a no-flat Universe with $\Omega_{\Lambda}>0.7$.
}
{}

\keywords{galaxies: cluster: general -- galaxies: fundamental
parameters -- intergalactic medium -- X-ray: galaxies: clusters -- cosmology:
observations -- dark matter.}

\maketitle

\section{Introduction}

Several tests have been suggested to constrain the geometry
and the relative amounts of the matter and energy constituents of the 
Universe (see Peebles \& Ratra 2003 and references
therein). A method that is robust and complementary to others is
obtained using the gas mass fraction, $f_{\rm gas} =
M_{\rm gas}/M_{\rm tot}$, as inferred from X-ray observations
of clusters of galaxies. In this work, we consider
two independent methods for our purpose: 
(i) we compare the relative amount of baryons with respect to the
total mass observed in galaxy clusters to the cosmic baryon fraction
to provide a direct constraint on $\Omega_{\rm m}$
(this method was originally adopted by White et al. 1993 to show the limitation
of the standard cold dark matter scenario in an Einstein-de Sitter Universe), 
(ii) we limit the parameters that describe the geometry of the
universe by assuming that the gas fraction is constant 
in time, as first suggested by Sasaki (1996).

Starting from these pioneering works, many studies have followed this approach
to constrain the cosmological parameters (White \& Fabian 1995, David
et al. 1995, Ettori \& Fabian 1998, Rines et al. 1999, Roussel et
al. 2000, Allen et al. 2002, Ettori, Tozzi \& Rosati 2003,
Castillo-Morales \& Schindler 2003, Sadat et al. 2005, Allen et
al. 2008).  We present here a revised and updated version of the work
discussed in Ettori, Tozzi \& Rosati (2003, Paper I).  The main
differences with respect to Paper I are the following: (1) the present dataset
contains 60 objects that spans homogeneously the redshift range
between 0.06 and 1.27, whereas in the previous analysis 17 objects
were selected, 9 at $z \la 0.1$ and 8 at $z>0.7$; (2) several
assumptions on, e.g., the star mass fraction and the depletion
parameter are tested against different measurements from observations
and numerical simulations; (3) a more extensive analysis is performed
to assess the systematics that affect our results on the constraints
of the cosmological parameters.

The outline of our work is the following.  In Section~2, we describe
the cosmological framework that allows us to formulate the
cosmological dependence of the cluster gas mass fraction. In
Section~3, the dataset is presented and the physical quantities under
examination are computed.  The method based on using the gas mass fraction as
a cosmological probe is discussed in Section~4. The results on the
cosmological parameters are shown in Section~5, along with a discussion on
their robustness against systematic effects.  The summary of our
findings and the conclusions are drawn in Section~6.  Throughout this
work, if not otherwise stated, we plot and tabulate values, with
errors quoted at the 68.3 per cent ($1 \sigma$) level of confidence,
that are estimated by assuming
$H_0= 70 h_{70}^{-1}$ km s$^{-1}$ Mpc$^{-1}$ and $\Omega_{\rm m}=
1-\Omega_{\Lambda}=0.3$.

\begin{table*}[ht]
\caption{Properties of the $z>0.3$ cluster sample described in Balestra et al. (2007).
Column (1): name; column (2): adopted redshift;
column (3): best-fit gas temperature within $R_{\rm spec}$;
column (4): best-fit parameters of the gas density profile in equation~\ref{eq:ngas}
fitted over the radial range $0-R_{\rm spat}$;
column (5): outer radius $R_{\rm spat}$ where the signal-to-noise ratio is above 2;
column (6): radius $R_{\rm spec}$ of the circular region adopted in the spectral fit;
column (7--8): $R_{500}$ and $f_{\rm gas}$ estimated by assuming
the temperature profile in eq.~\ref{eq:tr};
column (9--10): $R_{500}$ and $f_{\rm gas}$ estimated by assuming a constant temperature
equal to $kT_{\rm gas}$.
A cosmology of $(H_0, \Omega_{\rm m}, \Omega_{\Lambda}) = (70$ km s$^{-1}$ Mpc$^{-1},
0.3, 0.7)$ is adopted here. }
\hspace*{-0.7cm}
\vspace*{-0.2cm}
\begin{tabular}{l@{\hspace{.5em}} c@{\hspace{.5em}} c@{\hspace{.5em}} c@{\hspace{.5em}} c@{\hspace{.3em}} 
 c@{\hspace{.3em}} c@{\hspace{.5em}} c@{\hspace{.5em}} c@{\hspace{.5em}} c@{\hspace{.5em}} }
\hline \\
 Cluster & z & $kT_{\rm gas}$ & $(a_0,a_1,a_2/10^{-3},a_3,a_4,a_6)$ & $R_{\rm spat}$ & $R_{\rm spec}$ & $R_{500}$  &  $f_{\rm gas}$  &  $\hat{R}_{500}$  &  $\hat{f}_{\rm gas}$  \\ 
  &  & keV &  & kpc & kpc & kpc & $(<R_{500})$ & kpc & $(<R_{500})$ \\
\hline \\
MS$1008.1-1224$ & 0.306 & $6.90\pm0.35$ & $(0.928,0.028,3.9,0.39,1.13,0.44)$ & 688 & 490 & $1176\pm116$ & $0.135\pm0.045$ & $1303\pm179$ & $0.112\pm0.044$ \\
MS$2137.3-2353$ & 0.313 & $5.05\pm0.11$ & $(0.028,0.204,93.3,0.37,0.53,0.47)$ & 698 & 362 & $940\pm26$ & $0.152\pm0.016$ & $997\pm35$ & $0.135\pm0.016$ \\
A1995 & 0.320 & $8.61\pm0.35$ & $(0.142,0.290,7.3,0.32,0.30,1.56)$ & 855 & 479 & $1317\pm46$ & $0.091\pm0.011$ & $1429\pm60$ & $0.076\pm0.010$ \\
MACS-J$0308.9+2645$ & 0.324 & $12.73\pm1.02$ & $(0.104,0.230,11.2,0.30,0.30,0.96)$ & 1039 & 577 & $1423\pm68$ & $0.131\pm0.020$ & $1491\pm77$ & $0.120\pm0.017$ \\
ZW$1358.1+6245$ & 0.328 & $6.78\pm0.28$ & $(0.006,0.400,143.5,0.00,0.35,1.40)$ & 869 & 419 & $1157\pm73$ & $0.106\pm0.021$ & $1256\pm100$ & $0.090\pm0.019$ \\
MACS-J$0404.6+1109$ & 0.355 & $7.38\pm0.90$ & $(0.105,0.125,4.6,0.00,0.30,0.19)$ & 1398 & 782 & $887\pm63$ & $0.205\pm0.039$ & $850\pm68$ & $0.215\pm0.038$ \\
RXJ$0027.6+2616$ & 0.367 & $8.76\pm1.69$ & $(0.676,0.163,2.0,0.35,0.89,0.62)$ & 642 & 550 & $1299\pm269$ & $0.067\pm0.054$ & $1421\pm351$ & $0.055\pm0.048$ \\
MACS-J$1720.2+3536$ & 0.391 & $6.46\pm0.33$ & $(0.010,0.533,161.4,0.00,0.40,2.03)$ & 974 & 546 & $1200\pm104$ & $0.103\pm0.034$ & $1345\pm143$ & $0.078\pm0.028$ \\
ZW$0024.0+1652$ & 0.395 & $4.32\pm0.32$ & $(0.375,0.010,13.2,0.17,0.66,0.60)$ & 672 & 341 & $879\pm100$ & $0.098\pm0.035$ & $957\pm144$ & $0.083\pm0.031$ \\
V$1416+4446$ & 0.400 & $3.43\pm0.20$ & $(0.008,0.544,72.5,0.00,0.36,2.19)$ & 677 & 397 & $797\pm100$ & $0.135\pm0.058$ & $894\pm148$ & $0.110\pm0.058$ \\
MACS-J$0159.8-0849$ & 0.405 & $9.43\pm0.61$ & $(0.011,0.302,167.0,0.09,0.40,0.86)$ & 994 & 639 & $1214\pm62$ & $0.139\pm0.025$ & $1286\pm81$ & $0.125\pm0.022$ \\
MACS-J$2228.5+2036$ & 0.412 & $8.25\pm0.59$ & $(0.240,0.079,6.4,0.42,0.48,0.23)$ & 1211 & 753 & $1052\pm61$ & $0.217\pm0.033$ & $1091\pm82$ & $0.205\pm0.032$ \\
MS$0302.7+1658$ & 0.424 & $4.78\pm0.60$ & $(0.015,0.361,50.7,0.00,0.37,2.50)$ & 369 & 328 & $1041\pm92$ & $0.055\pm0.023$ & $1167\pm104$ & $0.041\pm0.016$ \\
MS$1621.5+2640$ & 0.426 & $7.22\pm0.75$ & $(0.248,0.212,3.9,0.12,0.30,0.96)$ & 1003 & 658 & $977\pm79$ & $0.153\pm0.040$ & $1006\pm106$ & $0.146\pm0.039$ \\
MACS-J$0417.5-1154$ & 0.440 & $10.84\pm0.98$ & $(0.015,1.000,137.1,0.00,0.40,1.58)$ & 1259 & 785 & $1340\pm105$ & $0.201\pm0.043$ & $1465\pm141$ & $0.171\pm0.036$ \\
MACS-J$1206.2-0847$ & 0.440 & $11.98\pm0.85$ & $(0.012,0.316,82.9,0.00,0.30,1.11)$ & 1259 & 785 & $1336\pm75$ & $0.161\pm0.028$ & $1416\pm95$ & $0.144\pm0.025$ \\
V$1701+6414$ & 0.453 & $4.36\pm0.28$ & $(0.011,0.842,66.4,0.00,0.39,2.50)$ & 729 & 370 & $826\pm98$ & $0.146\pm0.051$ & $910\pm150$ & $0.123\pm0.050$ \\
CL$1641+4001$ & 0.464 & $4.74\pm0.54$ & $(1.000,0.135,1.7,0.44,0.30,1.34)$ & 341 & 287 & $848\pm139$ & $0.079\pm0.059$ & $902\pm189$ & $0.070\pm0.055$ \\
MACS-J$1621.4+3810$ & 0.465 & $6.62\pm0.74$ & $(0.301,0.010,139.1,0.00,0.30,1.14)$ & 893 & 458 & $980\pm102$ & $0.152\pm0.041$ & $1037\pm146$ & $0.137\pm0.037$ \\
MACS-J$1824.3+4309$ & 0.487 & $9.69\pm2.82$ & $(0.034,0.343,7.6,1.50,0.30,0.00)$ & 765 & 506 & $930\pm189$ & $0.136\pm0.072$ & $873\pm223$ & $0.145\pm0.069$ \\
MACS-J$1311.0-0311$ & 0.492 & $8.17\pm0.86$ & $(0.086,0.029,39.8,0.00,0.44,0.59)$ & 631 & 478 & $1055\pm72$ & $0.106\pm0.026$ & $1112\pm85$ & $0.096\pm0.023$ \\
V$1525+0958$ & 0.516 & $5.54\pm0.40$ & $(0.740,0.092,4.2,0.00,1.20,0.59)$ & 834 & 490 & $901\pm113$ & $0.137\pm0.050$ & $1002\pm168$ & $0.115\pm0.048$ \\
MS$0451.6-0305$ & 0.540 & $9.11\pm0.45$ & $(0.481,0.082,14.4,0.02,0.82,0.71)$ & 1167 & 625 & $1292\pm86$ & $0.129\pm0.025$ & $1446\pm124$ & $0.099\pm0.020$ \\
MS$0015.9+1609$ & 0.541 & $9.59\pm0.43$ & $(1.000,0.256,3.5,0.38,0.54,1.18)$ & 1408 & 626 & $1177\pm54$ & $0.180\pm0.020$ & $1279\pm78$ & $0.155\pm0.018$ \\
MACS-J$1149.5+2223$ & 0.544 & $13.10\pm1.12$ & $(0.769,0.225,5.0,0.00,1.20,0.46)$ & 1147 & 944 & $1536\pm150$ & $0.103\pm0.031$ & $1734\pm201$ & $0.079\pm0.025$ \\
MACS-J$1423.8+2404$ & 0.545 & $7.55\pm0.59$ & $(0.019,0.875,164.6,0.00,0.52,2.50)$ & 805 & 504 & $1176\pm138$ & $0.091\pm0.041$ & $1343\pm187$ & $0.067\pm0.033$ \\
V$1121+2327$ & 0.562 & $5.24\pm0.37$ & $(0.811,0.590,1.4,0.36,1.20,2.50)$ & 647 & 447 & $1053\pm167$ & $0.071\pm0.043$ & $1217\pm225$ & $0.050\pm0.031$ \\
SC$1120-1202$ & 0.562 & $6.35\pm1.15$ & $(0.039,0.037,12.9,0.00,0.30,0.47)$ & 297 & 318 & $795\pm132$ & $0.083\pm0.057$ & $795\pm163$ & $0.082\pm0.058$ \\
RXJ$0848.7+4456$ & 0.570 & $3.42\pm0.35$ & $(0.150,0.119,2.8,0.44,0.30,0.90)$ & 380 & 196 & $648\pm91$ & $0.074\pm0.039$ & $679\pm127$ & $0.068\pm0.039$ \\
MACS-J$2129.4-0741$ & 0.570 & $9.17\pm0.90$ & $(0.105,0.830,14.4,0.32,0.48,2.50)$ & 1199 & 639 & $1339\pm81$ & $0.089\pm0.014$ & $1526\pm92$ & $0.066\pm0.009$ \\
MS$2053.7-0449$ & 0.583 & $5.77\pm0.51$ & $(1.000,0.191,1.0,0.71,0.30,1.21)$ & 568 & 357 & $903\pm132$ & $0.075\pm0.036$ & $975\pm183$ & $0.065\pm0.033$ \\
MACS-J$0647.7+7015$ & 0.584 & $12.79\pm1.59$ & $(1.000,0.311,2.7,0.57,0.30,1.92)$ & 832 & 584 & $1440\pm149$ & $0.077\pm0.027$ & $1586\pm192$ & $0.061\pm0.021$ \\
RXJ$0956.0+4107$ & 0.587 & $7.59\pm2.00$ & $(1.000,0.138,3.5,0.00,1.18,0.91)$ & 450 & 424 & $1046\pm332$ & $0.088\pm0.130$ & $1152\pm457$ & $0.072\pm0.122$ \\
CL$0542.8-4100$ & 0.634 & $8.36\pm1.09$ & $(1.000,0.168,2.2,0.41,0.30,1.03)$ & 714 & 539 & $946\pm116$ & $0.126\pm0.045$ & $984\pm164$ & $0.119\pm0.043$ \\
RCS-J$1419.2+5326$ & 0.640 & $4.35\pm0.74$ & $(0.053,0.043,29.0,0.00,0.30,0.72)$ & 276 & 303 & $685\pm140$ & $0.157\pm0.114$ & $709\pm192$ & $0.148\pm0.107$ \\
MACS-J$0744.9+3927$ & 0.686 & $9.58\pm0.67$ & $(0.010,0.217,159.7,0.00,0.36,0.59)$ & 1079 & 695 & $963\pm43$ & $0.192\pm0.026$ & $995\pm53$ & $0.183\pm0.024$ \\
RXJ$1221.4+4918$ & 0.700 & $8.37\pm0.82$ & $(1.000,0.308,2.1,0.28,0.30,1.63)$ & 901 & 562 & $990\pm81$ & $0.114\pm0.026$ & $1064\pm103$ & $0.101\pm0.022$ \\
RXJ$1113.1-2615$ & 0.730 & $5.92\pm0.76$ & $(0.327,0.035,12.9,0.00,1.15,0.79)$ & 290 & 286 & $934\pm177$ & $0.039\pm0.058$ & $1046\pm238$ & $0.030\pm0.049$ \\
RXJ$2302.8+0844$ & 0.734 & $8.39\pm1.35$ & $(0.116,0.103,7.0,0.00,0.30,0.78)$ & 930 & 393 & $856\pm84$ & $0.090\pm0.028$ & $876\pm95$ & $0.087\pm0.026$ \\
MS$1137.5+6624$ & 0.782 & $6.87\pm0.52$ & $(0.067,0.427,20.4,0.00,0.47,2.18)$ & 590 & 367 & $893\pm123$ & $0.089\pm0.040$ & $968\pm172$ & $0.075\pm0.037$ \\
RXJ$1317.4+2911$ & 0.805 & $4.51\pm1.17$ & $(0.110,1.000,3.6,0.45,0.60,1.20)$ & 202 & 184 & $693\pm112$ & $0.042\pm0.026$ & $751\pm130$ & $0.037\pm0.021$ \\
RXJ$1350.0+6007$ & 0.810 & $4.44\pm0.67$ & $(1.000,0.538,0.6,0.78,0.30,2.50)$ & 526 & 483 & $615\pm146$ & $0.151\pm0.102$ & $660\pm219$ & $0.140\pm0.121$ \\
RXJ$1716.4+6708$ & 0.813 & $7.04\pm0.81$ & $(1.000,0.254,1.5,0.67,0.30,1.79)$ & 565 & 408 & $844\pm124$ & $0.110\pm0.055$ & $904\pm166$ & $0.097\pm0.051$ \\
RXJ$0152.7-1357$S & 0.828 & $9.43\pm2.35$ & $(0.367,0.533,3.6,0.35,1.20,1.20)$ & 420 & 400 & $1162\pm248$ & $0.030\pm0.037$ & $1326\pm309$ & $0.022\pm0.026$ \\
MS$1054.4-0321$ & 0.832 & $7.91\pm0.48$ & $(1.000,0.323,5.6,0.00,0.30,2.39)$ & 848 & 599 & $962\pm92$ & $0.147\pm0.038$ & $1067\pm128$ & $0.120\pm0.031$ \\
RXJ$0152.7-1357$N & 0.835 & $6.74\pm1.03$ & $(0.073,0.642,7.8,0.00,0.30,2.50)$ & 582 & 442 & $866\pm179$ & $0.117\pm0.094$ & $975\pm265$ & $0.095\pm0.094$ \\
1WGA-J$1226.9+3332$ & 0.890 & $12.14\pm1.38$ & $(0.077,0.257,29.1,0.00,0.46,0.87)$ & 677 & 497 & $1126\pm166$ & $0.085\pm0.040$ & $1222\pm226$ & $0.071\pm0.035$ \\
CL$1415.1+3612$ & 1.030 & $6.93\pm0.74$ & $(0.040,0.316,18.2,1.02,0.30,2.50)$ & 425 & 318 & $857\pm93$ & $0.059\pm0.026$ & $960\pm121$ & $0.045\pm0.020$ \\
RDCS-J$0910+5422$ & 1.106 & $6.43\pm1.41$ & $(0.147,1.000,3.4,0.58,0.36,1.20)$ & 230 & 201 & $576\pm144$ & $0.117\pm0.092$ & $570\pm208$ & $0.116\pm0.096$ \\
RDCS-J$1252-2927$ & 1.235 & $7.57\pm1.18$ & $(1.000,0.089,4.0,0.32,0.30,1.20)$ & 329 & 287 & $636\pm123$ & $0.079\pm0.061$ & $662\pm172$ & $0.073\pm0.061$ \\
RDCS-J$0849+4452$ & 1.261 & $5.21\pm1.31$ & $(0.046,1.000,10.3,0.00,0.36,1.20)$ & 222 & 197 & $504\pm152$ & $0.103\pm0.100$ & $512\pm216$ & $0.099\pm0.103$ \\
RDCS-J$0848+4453$ & 1.273 & $3.75\pm1.86$ & $(0.069,1.000,6.8,0.00,0.60,1.20)$ & 112 & 165 & $432\pm151$ & $0.052\pm0.295$ & $456\pm192$ & $0.047\pm0.436$ \\
\hline \\
\end{tabular}

\label{tab:highz}
\end{table*}

\section{The cosmological framework}

We refer to $\Omega_{\rm b}$ as the {\it baryon matter} density, to
$\Omega_{\rm m}$ as the {\it total matter} density (i.e.  $\Omega_{\rm
  m} = \Omega_{\rm b} + \Omega_{\rm c}$, where $\Omega_{\rm c}$ is the
{\it cold dark matter} component), to $\Omega_{\Lambda}$ as the {\it
  dark energy} density, that we consider both in its static and
homogeneous form as a {\it cosmological constant} and with an equation
of state varying with cosmic time as ``quintessence'' (e.g. Turner \&
White 1997, Caldwell et al. 1998).  All these densities are expressed
in units of the critical density, $\rho_{\rm c} = 3 H_0^2/ (8 \pi G)$,
where $H_0$ is Hubble's constant and $G$ is the gravitational
constant.  In our computation, we neglect (i) the energy associated with
the cosmic radiation, $\Omega_{\rm r} \approx 4.16 \times 10^{-5}
(T_{\rm CMB}/2.726K)^4$, and (ii) any possible contributions from
light neutrinos, $\Omega_{\nu} h_{70}^2 = \sum {\rm m}_{\nu} / 45.5
{\rm eV}$, that is expected to be less than $0.01$ for a total mass in
neutrinos, $\sum {\rm m}_{\nu}$, lower than 0.62 eV (see, e.g.,
Hannestad \& Raffelt 2006).  Therefore, we adopt the Einstein equation
in the form $\Omega_{\rm m} + \Omega_{\Lambda} +\Omega_{\rm k} = 1,$
where $\Omega_{\rm k}$ accounts for the curvature of space.  From,
e.g., Carroll, Press \& Turner (1992, cf. eq.~25), we can then write
the angular diameter distance as
\begin{eqnarray}
d_{\rm ang} = & \frac{d_{\rm lum}}{(1+z)^2} = 
\frac{c}{H_0 (1+z)} \frac{S(\omega)}{|\Omega_{\rm k}|^{1/2}},   \nonumber  \\
\omega = & |\Omega_{\rm k}|^{1/2} \int^z_0 \frac{d \zeta}{E(\zeta)},
\label{eq:dang}
\end{eqnarray}
where $d_{\rm lum}$ is the luminosity distance, 
$S(\omega)$ is sinh$(\omega)$, $\omega$, $\sin(\omega)$ for
$\Omega_{\rm k}$ greater than, equal to and less than 0, respectively. 
In addition, we define
\begin{eqnarray}
E(z) & = & \left[\Omega_{\rm m} (1+z)^3 +\Omega_{\rm k} (1+z)^2 +
\Omega_{\Lambda} \lambda(z) \right]^{1/2} \nonumber \\
\lambda(z) & = & \exp\left(3 \int_0^z \frac{1+w(z)}{1+z} dz\right)\,.
\label{eq:ez}
\end{eqnarray}
The above equations (i) do not include the evolution with redshift
of the radiation component, $\Omega_{\rm r} (1+z)^4$, that 
at $z\approx1$ is about $1.4 \times 10^{-3}$ and therefore
negligible in the overall budget, 
(ii) consider the dependence upon the ratio $w$
between the pressure and the energy density in
the equation of state of the dark energy component 
(Caldwell, Dave \& Steinhardt 1998, Wang \& Steinhardt 1998).
Hereafter we consider both a pressure-to-density ratio
$w$ constant in time that implies $\lambda(z) = (1+z)^{3+3 w}$
and a simple parameterization of its evolution with redshift 
(e.g. Rapetti et al. 2005, Firmani et al. 2005)
\begin{equation}
w(z) = w_0 + w_1 \frac{z}{1+z}.
\label{eqn:wz}
\end{equation}
In particular, the case of a cosmological constant requires $w=-1$.

\section{The dataset}

We consider the sample of 52 galaxy clusters at $z>0.3$ presented 
in Balestra et al. (2007; see Table~\ref{tab:highz}).
The preparation, reduction and analysis of this dataset is described below.

As local measurements, we consider the sample of 
8 objects at a redshift between $0.06$ and $0.23$
with a gas temperature higher than 4 keV described in 
Vikhlinin et al. (2006; see Table~\ref{tab:local}).
We use the best-fit results of the gas temperature and density 
profiles obtained by adopting the same functional form that 
is applied in the present analysis to reproduce 
the ICM properties of the high$-z$ sample.
The gas and total mass profiles are then recovered for both
the local and the $z>0.3$ sample with the same method described 
at the end of this section. 
 
\begin{figure}
\epsfig{figure=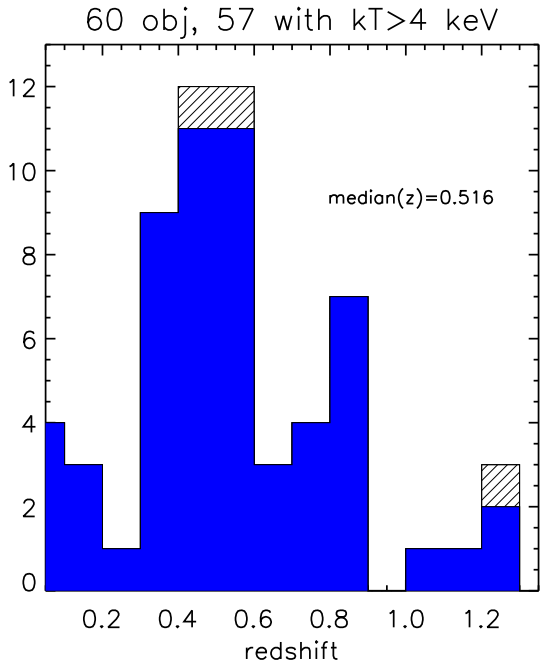,width=0.5\textwidth}
\caption{Redshift distribution of the clusters in the
whole sample (dashed region) and of the selected hottest objects
(shaded regions). 
} \label{fig:sample} \end{figure}

The histogram of the redshift distribution of the two 
samples is shown in Fig.~\ref{fig:sample}.

The only criterion adopted for the selection of these objects is 
the availability of good exposures with {\em Chandra} of 
hot, massive, relaxed galaxy clusters over a significant redshift range.
No selection effect is expected to occur in the application of the 
gas mass fraction method once the clusters are selected 
to ensure (i) the use of the hydrostatic equilibrium equation
to recover the total mass (round, relaxed objects are required
for that), and (ii) a negligible contribution from non-gravitational energy 
in the region of interest to allow the use of cross--calibration 
with numerical simulations, such as the estimate of the depletion parameter 
(the selection of hot, massive systems dominated
energetically by gravitational collapse satisfies this condition).

We have considered only {\em Chandra} exposures of the clusters
listed in Table~\ref{tab:highz}.
Data reduction is performed using the CIAO 3.2 software package 
with a recent version of the Calibration Database (CALDB 3.0.0)
including the correction for the degraded effective area of ACIS--I 
chips due to material accumulated on the ACIS optical blocking filter 
at the epoch of the observation. We also apply the time-dependent gain
correction\footnote{http://asc.harvard.edu/ciao/threads/acistimegain/},
which is necessary to adjust the ``effective gains", which 
drifts with time due to an increasing charge transfer inefficiency (CTI).
The detailed procedure of data reduction is described in Balestra et al. (2007).

A single thermal component (XSPEC model {\tt mekal}; Arnaud 1996) 
absorbed by the Galactic column density is fitted to the spectrum extracted
from a circular region with radius $R_{\rm spec}$ (see Table~\ref{tab:highz})
around the cluster chosen to maximize the signal-to-noise
ratio once the contaminating point sources are masked.
The only free parameters in the spectral fit are the temperature, 
the normalization and the metallicity with respect to the
abundance table from Grevesse \& Sauval (1998) here assumed as the reference.
These temperature measurements are on average
3 per cent higher than the values obtained by assuming the standard
metallicity table of Anders \& Grevesse (1989).

\begin{table}
\caption{Properties of the local cluster sample from Vikhlinin et al. (2006).
Column (1): name; column (2): adopted redshift; column (3): best-fit spectral temperature;
column (4): $R_{500}$; column (5): gas mass fraction within $R_{500}$.
A cosmology of $(H_0, \Omega_{\rm m}, \Omega_{\Lambda}) = (70$ km s$^{-1}$ Mpc$^{-1},
0.3, 0.7)$ is adopted here. }
\begin{tabular}{l@{\hspace{.8em}} c c@{\hspace{.8em}} c@{\hspace{.7em}} c@{\hspace{.7em}} 
 c@{\hspace{.7em}} c@{\hspace{.7em}} c@{\hspace{.7em}}}
\hline \\
 Cluster & z & $kT_{\rm gas}$ &  $R_{500}$ & $f_{\rm gas}$ \\ 
  &  &  keV & kpc & $(<R_{500})$ \\
\hline \\
A133 & 0.057 & $4.15\pm0.07$ & $1027\pm38$ & $0.087\pm0.006$ \\
A1795 & 0.062 & $6.10\pm0.06$ & $1320\pm47$ & $0.108\pm0.006$ \\
A2029 & 0.078 & $8.46\pm0.09$ & $1419\pm32$ & $0.128\pm0.007$ \\
A478 & 0.088 & $7.95\pm0.14$ & $1398\pm62$ & $0.125\pm0.011$ \\
A1413 & 0.143 & $7.38\pm0.12$ & $1377\pm49$ & $0.112\pm0.007$ \\
A907 & 0.160 & $5.96\pm0.08$ & $1149\pm32$ & $0.129\pm0.006$ \\
A383 & 0.188 & $4.80\pm0.12$ & $983\pm34$ & $0.129\pm0.007$ \\
A2390 & 0.230 & $8.90\pm0.17$ & $1489\pm50$ & $0.147\pm0.009$ \\
\hline \\
\end{tabular}

\label{tab:local}
\end{table}

\begin{figure}
\epsfig{figure=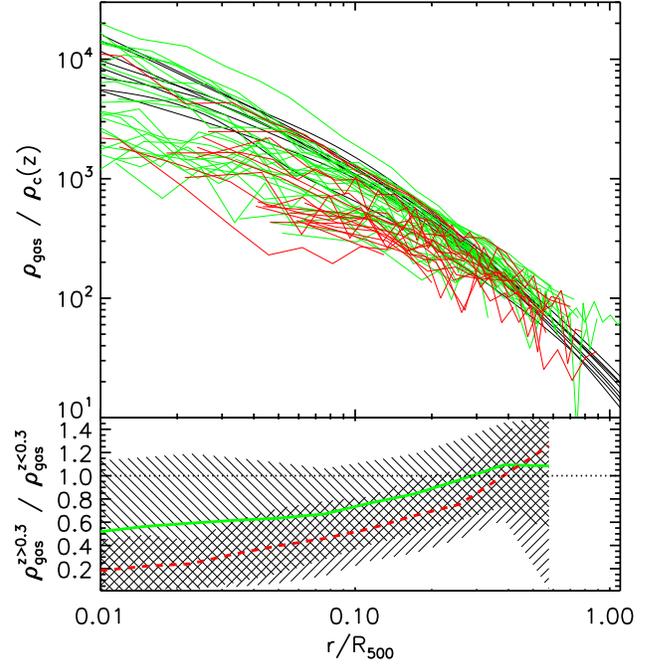,width=0.5\textwidth}
\caption{Radial gas density profiles in units of the critical density at the
cluster redshift. 
Green, black and red lines indicate objects at redshift $<0.3$, $0.3-0.6$ and
$>0.6$, respectively.
The bottom panel shows the ratio (and relative dispersion) of the mean
density profiles observed in the redshift range $0.3-0.6$ (green line) and
$z>0.6$ (red line) with respect to the local profile.
} \label{fig:ngas} \end{figure}

We assume that galaxy clusters are spherically symmetric
gravitationally bound systems.  We deproject the X-ray surface
brightness to obtain the electron density profile, $n_{\rm e}(r)$, by
correcting the emissivity by the contribution from the outer shells
moving inwards.  We consider the profiles from the deprojection of the
background-subtracted surface brightness $S$, with error $\epsilon_S$,
up to the radial limit, $R_{\rm spat}$, beyond which $(S
/ \epsilon_S) < 2$.  The gas density profiles, normalized to the
critical density at the cluster's redshift, are shown in
Fig.~\ref{fig:ngas}.  We note that the average profile measured in
local ($z<0.3$) clusters is definitely more peaked than the ones
observed at higher redshift, with a central gas density that is a
factor of 2 higher than in systems at $0.3<z<0.6$ and almost a factor 
of 5 higher than in objects at $z>0.6$.  
This follows recent evidence that cooling
cores, at relativly higher central density, preferentially form later
in time (e.g. Santos et al. 2008, Ettori \& Brighenti 2008).

To estimate the gas and total mass profiles, we fit the electron
density profiles with a functional form adapted from Vikhlinin et
al. (2005):
\begin{eqnarray}
n_{\rm e}(r) & = & a_2 x_0^{-a_3} \times \left(1 + x_0^2 \right)^{-1.5 a_4 +a_3/2} 
 \times \left(1 + x_1^{a_5} \right)^{-a_6/a_5}  \\
x_0 & = & r/a_0  \nonumber \\
x_1 & = & r/a_1.  \nonumber 
\label{eq:ngas}
\end{eqnarray}
The number of free parameters is reduced according to the number of
datapoints $N_{\rm dat}$ available, in the following order of
priority: $a_5=3$ (always), $a_6=1.2$ (when $6 \le N_{\rm dat} <8$,
otherwise free to vary between 0 and 2.5), $a_4=0.6$ (when $N_{\rm
  dat} = 5$), $a_3=0$ (when $N_{\rm dat} \le 4$).  The best-fit
parameters that reproduce the electron density profiles up to the
outer radial limit $R_{\rm spat}$ are quoted in
Table~\ref{tab:highz}. All the electron density profiles and the
best-fit lines are shown in Fig.~\ref{fig:profiles}.

\begin{figure}
\epsfig{figure=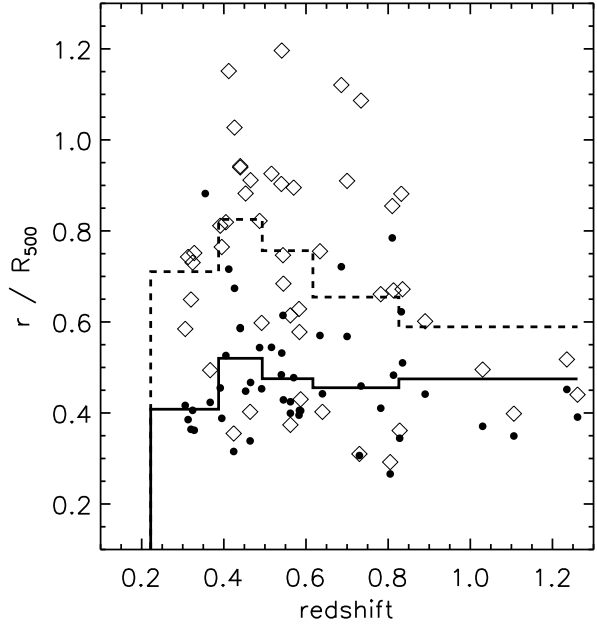,width=0.5\textwidth}
\caption{Ratios between the maximum radial extension in the gas density profile, 
$R_{\rm spat}$, and
$R_{500}$ ({\it diamonds}) and the extraction radius of the circular region used
to map the object in the spectral analysis, $R_{\rm spec}$, and $R_{500}$ ({\it dots}).
The dashed and solid lines show the average values of these ratios, $R_{\rm spat}/R_{500}$
and $R_{\rm spec} / R_{500}$ respectively, in bins of $\sim$10 objects.
} \label{fig:r500} \end{figure}

The total and gas masses are then evaluated at $R_{\Delta}=R_{500}$
that describes the sphere within which the cluster overdensity with
respect to the critical density is $\Delta = 500$.  This value of
overdensity is maintained fixed in all the cosmologies studied and at
any redshift.  Under the assumption of isothermality, we calculate therefore
\begin{eqnarray}
M_{\rm tot}(<r) & = & - \frac{kT_{\rm gas} \, r}{\mu m_{\rm u} G} \frac{d \log n_{\rm e}}{d \log
r}, \nonumber  \\
R_{500} & = & \left( \frac{3 M_{\rm tot}(<R_{500})}{4 \pi 500 \rho_{\rm c,z}} \right)^{1/3}, \nonumber
\\
M_{\rm gas}(<r) & = & \int_0^{R_{500}} 1.155 m_{\rm u} n_{\rm e}(r) \,  4 \pi r^2 dr,
\label{eq:masses}
\end{eqnarray}
where $1.155$ is the value associated to a cosmic mix of hydrogen and
helium with 0.3 times solar abundance in the remaining elements with
a relative contribution that follows Grevesse \& Sauval (1998),
$\mu=0.600$ is the corresponding mean molecular weight, $m_{\rm u} =
1.66 \times 10^{-24}$ g is the atomic mass unit and $\rho_{\rm c, z}$
is the critical density at redshift $z$ and is equal to $3 H_z^2/ (8
\pi G)$ with $H_z = H_0 E(z)$ (see eq.~\ref{eq:ez}).  The gas mass
fraction is then $f_{\rm gas}(R_{\Delta}) = M_{\rm gas}(<R_{\Delta}) /
M_{\rm tot}(<R_{\Delta})$.  The distribution of $R_{\rm spat}/R_{500}$
and of $R_{\rm spec}/R_{500}$ as a function of redshift is shown in
Fig.~\ref{fig:r500}.

All the errors are determined at $1 \sigma$ confidence level from the
16th and 84th percentile of the distribution of the values obtained by
repeating the calculations 1000 times after a new surface brightness
profile and gas temperature are considered by selecting
normally-distributed random values according to the original
measurements and their relative errors.  A new total mass profile is
then estimated and new $R_{500}$ and gas mass measurements are
obtained by following equations~\ref{eq:masses}.

\begin{figure*}
\hbox{
 \epsfig{figure=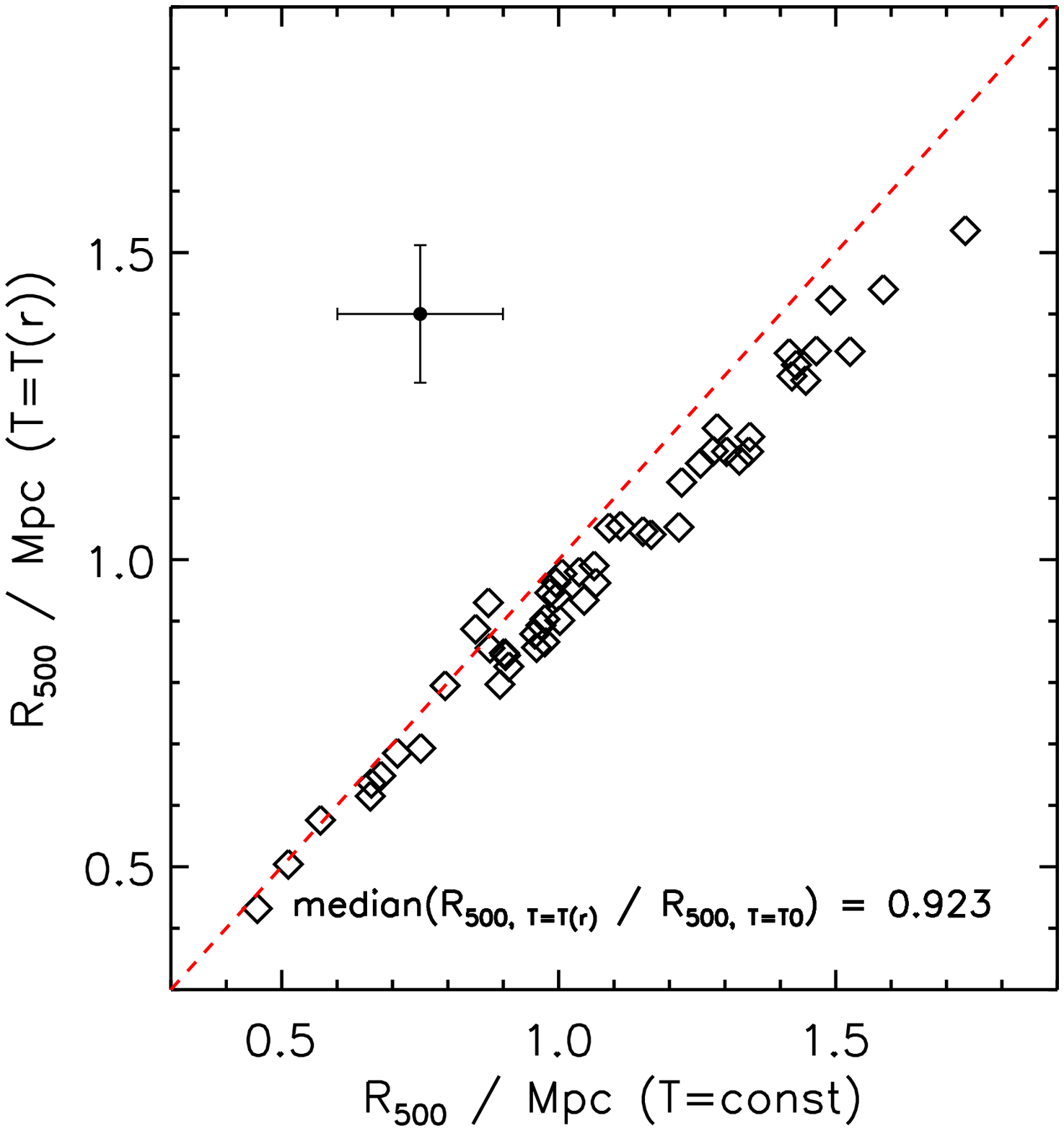,width=0.5\textwidth}
 \epsfig{figure=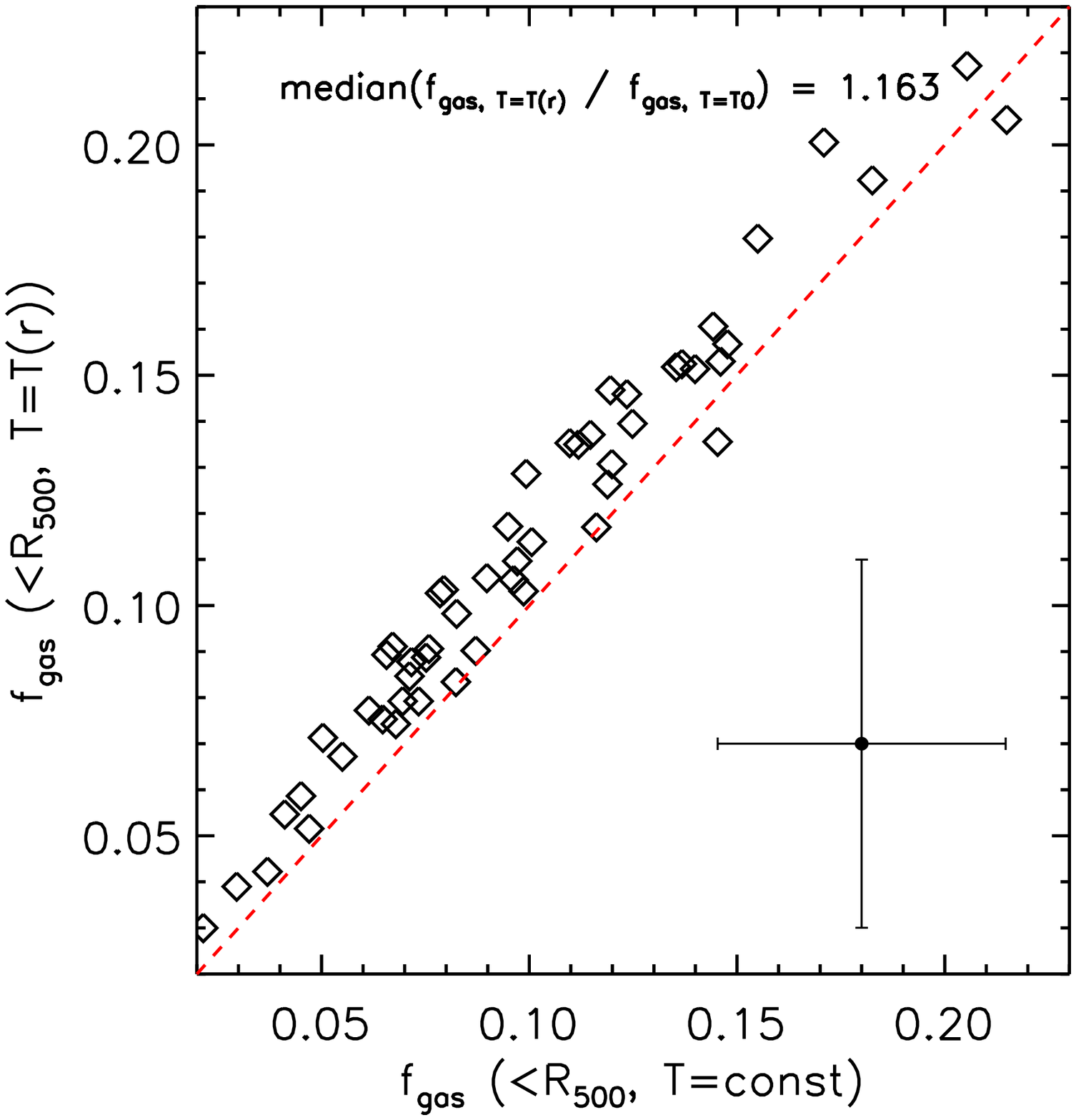,width=0.5\textwidth}
}\caption{
Changes in estimates of $R_{500}$ and $f_{\rm gas}$ once the temperature profile
in eq.~\ref{eq:tr} is assumed.
Median relative error bars are shown.
} \label{fig:fgas_tr} \end{figure*}

Our estimates of the gas fraction rely upon the determination of the
total gravitating mass obtained under the assumption of an isothermal
gas.  To check the effect of the presence of a temperature profile on
the measured total mass and, hence, gas fraction, we assume the
expression for the temperature profile that Vikhlinin et al. (2006)
showed to reproduce well the temperature gradients in nearby relaxed
systems:
\begin{equation}
T(r) = 1.23 \frac{ (x/0.045)^{1.9}+0.45 }{(x/0.045)^{1.9}+1}
  \frac{T_{\rm spec}}{\left[ (x/0.6)^2+1 \right]^{0.45}}, 
\label{eq:tr}
\end{equation}
where $x = r /R_{500}$ and the factor $1.23$ comes from equations~8 and 9 in
Vikhlinin et al. (2006).  Starting from the value of $R_{500}$
measured under the assumption of isothermality, we calculate (i) the
temperature gradient, (ii) the total mass using the
hydrostatic equilibrium equation ($M_{\rm tot} \propto T(r) \times r
\times d \log (n_{\rm e}\,T)/d \log r$) and (iii) a new estimate of
$R_{500}$. The calculations are repeated until $R_{500}$, that is
evaluated from the total mass profiles and adopted in
equation~\ref{eq:tr}, converges. This happens after 2--6 iterations.
At the newly estimated $R_{500}$, we measure the gas and total
masses.  We find that introducing a temperature profile decreases
$R_{500}$ by 7 per cent, on average, and increases $f_{\rm gas}(<R_{500})$ 
by about 16 per cent with respect to the estimates
obtained under the assumption of isothermality.  
The changes induced in the determination of $R_{500}$ and of $f_{\rm gas}$
are shown in Fig.~\ref{fig:fgas_tr}.

As a reference, for $(h_{70}, \Omega_{\rm m}=1-\Omega_{\Lambda}) = (1, 0.3)$,
in the overall sample of 52 objects at $z>0.3$, we measure $f_{\rm gas, 500} = 
(0.106, 0.113, 0.044)$ (median / mean / dispersion), when a gas temperature 
profile is assumed, and $(0.096, 0.099, 0.043)$, when a constant $kT$ is adopted.

In the following analysis, we make use of the estimates obtained with a
temperature profile to limit the cosmological parameters for the case of 
reference, and leave to the discussion on the systematics the case
with a constant temperature.

\subsection{Comparison with previous work}

We compare our results with recent determinations of the gas mass
fraction obtained from a similar X-ray dataset.

Allen et al. (2008) presented an extensive cosmological study which uses
the gas mass fraction in 42 hot ($kT > 4$ keV) relaxed clusters.
When we consider the 16 objects in common with the present work, 
the mean (dispersion) of the ratios of the estimated $f_{\rm gas}$ 
at $R_{2500}$ between their measurements and our estimates obtained 
by assuming a temperature profile is $1.14 (0.24)$.
Also within the scatter observed in this small sample used for comparison
is the ratio between the spectroscopic measurements of the gas
temperature ($T_{\rm Allen} / T_{\rm this\, work} = 1.06$, r.m.s $=0.17$).

We recall here that our gas temperatures are measured in spectra 
obtained to maximize the signal-to-noise ratio,
including the contribution from the core emission. Any cool
component in the core is expected to bias downwards the emission-weighted
cluster temperature. We have checked the statistical effect of this
bias by considering the sample of 90 clusters with temperatures $> 4$
keV for which Maughan et al. (2007) obtained from {\it Chandra}
observations a measure of the temperature both in the range $(0-1)
R_{500}$ and in the range $(0.15-1) R_{500}$.  We obtain a mean
$T_{0.15-1} / T_{0-1}$ of 1.01, with a dispersion of 0.14, suggesting
that, even though on average a cool core does not significantly affect
the cluster temperature, the scatter in the distribution of
the measured temperatures can be high.  

Overall, these results suggest that some systematics
(e.g., the background subtraction, the energy range adopted
for spectral fitting; see discussion in Appendix of Balestra et al. 2007),
of the order of 10-20 per cent and comparable to the mean statistical 
error of 10 per cent, affect the estimates of the global temperatures.
These uncertainties propagate also to the measurements of the 
gas mass fraction.
To assess the role of these errors on the robustness of our
constraints, we investigate in Sect.~\ref{sect:syst} how our
results depend upon the sample selected according to the measured
relative error on the gas mass fraction, an error that is propagated
from the uncertainties on $T_{\rm gas}$.

\section{The cluster gas fraction as cosmological probe: the method}

We start with a grid of values for the cosmological parameters we want
to investigate, specifically $\Omega_{\rm m}$ and $\Omega_{\Lambda}$ or $w$. 
We compute then
\begin{eqnarray}
\chi^2 & = & \sum_{i=1}^{N_{\rm dat}}
\frac{(f_{{\rm bar}, i} / b_i - \Omega_{\rm b}/\Omega_{\rm m})^2}
{\epsilon_{{\rm bar}, i}^2/b_i^2 +(f_{{\rm bar}, i} \epsilon_{b_i} / b_i^2)^2 +
\epsilon_{\Omega_{\rm b}}^2/\Omega_{\rm m}^2}.
\label{eq:chi2}
\end{eqnarray}
In the above equation, we use the following definitions:
\begin{enumerate}
\item $f_{\rm bar} = f_{\rm gas}+f_{\rm cold}$, where the gas mass
  fraction $f_{\rm gas}$ is directly measured from our X-ray
  observation and depends upon the cosmological parameters through the
  angular diameter distance, $d_{\rm ang}$, defined in
  equation~\ref{eq:dang}, being $f_{\rm gas} = M_{\rm gas} / M_{\rm
    tot} \propto n_{\rm gas} R^3 / R \propto d_{\rm ang}^{5/2} /
  d_{\rm ang} \propto d_{\rm ang}(\Omega_{\rm m}, \Omega_{\Lambda}, w)^{3/2}$ 
  (see Fig.~\ref{fig:dang} for
  the relative dependence upon different cosmologies), while the mass
  fraction in cold baryons, $f_{\rm cold} = M_{\rm cold} / M_{\rm
    tot}$, is estimated statistically as described in the next
  subsection;

\begin{figure}
\epsfig{figure=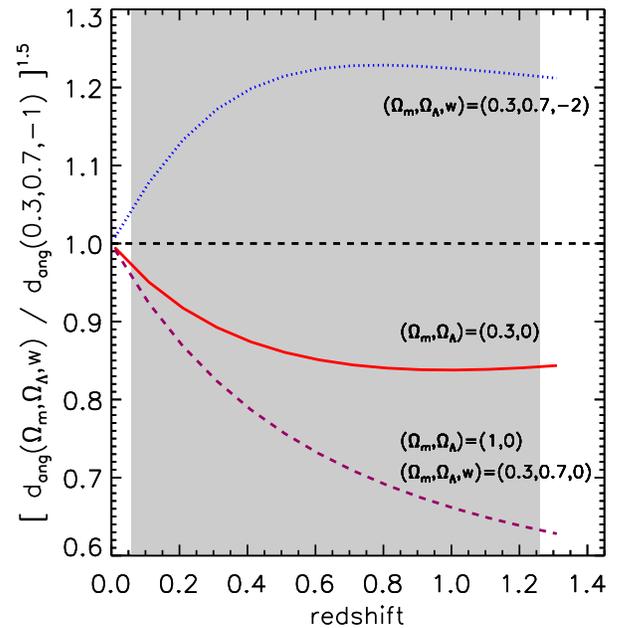,width=0.5\textwidth}
\caption{Sensitivity of the cluster baryon fraction method to the variation
of the cosmological parameters. The shaded region indicates the redshift
range considered in the present study.
} \label{fig:dang} \end{figure}

\item the error on $f_{\rm bar}$, $\epsilon_{\rm bar}$, is the sum in quadrature
of the uncertainties on $f_{\rm gas}$, $f_{\rm star}$ and on the assumed 
value of Hubble's constant $H_0$ (see below) propagated through the 
following dependence: $f_{\rm gas} \propto H_0^{-1.5}$ and 
$f_{\rm cold} \propto H_0^{-1}$;

\item the depletion parameter $b$ (with error $\epsilon_b$)
represents the fraction of cosmic baryons that fall in the cluster dark matter halo 
and is estimated from hydrodynamical simulations as discussed in section~4.2;

\item we assume $\Omega_{\rm b} h_{70}^2 = 0.0462 \pm 0.0012$ (error at $1 \sigma$ level)
from the best-fit results of the joint analysis in Komatsu et al. (2008) of 
(i) the power spectrum of the temperature anisotropy measured from the 
Wilkinson Microwave Anisotropy Probe five year data release 
(WMAP 5-year, Dunkley et al. 2008), (ii) a combined set of magnitudes
of Type Ia supernovae (Riess et al. 2007, Astier et al. 2006,
Wood-Vasey et al. 2007), (iii) the Baryon Acoustic Oscillations measured
in a survey of galaxies at $z=0.2$ and $z=0.35$ (Percival et al. 2007).

We note that the recent compilation of the best-fit results from the
Primordial Nucleosythesis calculations on $\Omega_{\rm b}$ 
provides comparable constraints (see, e.g., Steigman 2006, page 34:
$\Omega_{\rm b} h_{70}^2 = 0.0454 \pm 0.0045$). 
We consider the WMAP 5-year limits in the following analysis.

We adopt a present-day Hubble's constant of $H_0=72 \pm 8$ km s$^{-1}$ (error at $1 \sigma$ level)
measured in the Hubble Space Telescope Key Project by Freedman et al. (2001).
This value, assumed also as a Gaussian prior in the joint cosmological analysis
from Komatsu et al. (2008), is in agreement with their final result
of $H_0=70.1 \pm 1.3$ km s$^{-1}$, but is marginally in conflict with the value 
of $H_0 = 62.3 \pm 1.3 {\rm (random)} \pm 5.0 {\rm (systematic)}$
km s$^{-1}$ Mpc$^{-1}$ recently determined from Type Ia supernovae calibrated with
Cepheid variables in the nearby galaxies that hosted them (Sandage et al. 2006).
We make use of these estimates of $H_0$ in section~5, where we discuss
the robustness of our cosmological constraints.
\end{enumerate}

For a set of cosmological parameters $\{\Omega_{\rm m}, \Omega_{\Lambda},
w\}$, the gas mass, the total mass and the critical density $\rho_{c, z}$
are evaluated by considering all their cosmological dependencies.
A new gas fraction for each cluster in the selected sample is then
estimated at the fixed overdensity of $\Delta=500$ following
equations~\ref{eq:masses}.

\subsection{The stellar mass fraction}
\label{sect:fcold}

\begin{figure}
\epsfig{figure=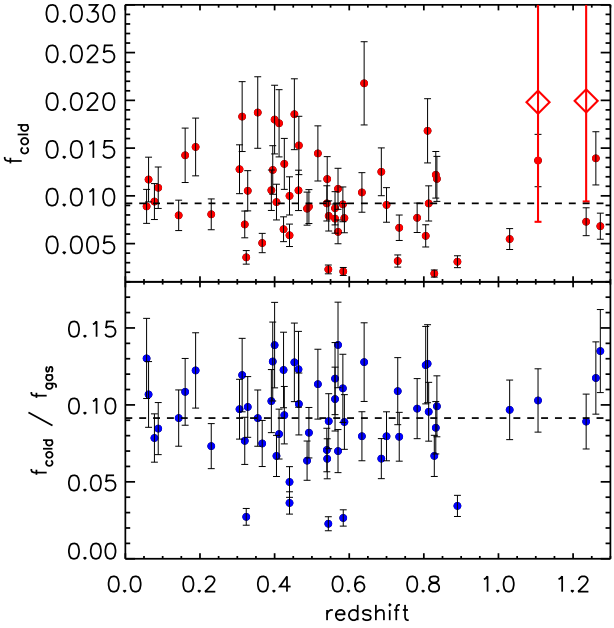,width=0.5\textwidth}
\caption{Stellar mass fractions (cf. equation~\ref{eq:fstar}) and 
ratios between their values and the estimated gas mass fraction at $R_{500}$ 
as function of redshift. 
Dashed lines indicate the median value: 
$f_{\rm cold}=0.009$ and $f_{\rm cold}/f_{\rm gas}=0.091$.
The two diamonds indicate the values for RDCS-J0910 and RDCS-J1252 with
the stellar masses estimated from the near-infrared luminosity function
in Strazzullo et al. (2006).
A cosmology of $(H_0, \Omega_{\rm m}, \Omega_{\Lambda}) = (70$ km s$^{-1}$ Mpc$^{-1},
0.3, 0.7)$ is adopted here.
} \label{fig:fgal} \end{figure}

The stellar mass fraction, $f_{\rm star} = M_{\rm star}/M_{\rm tot}$, 
cannot be computed individually for each single object for which most 
of the information of, e.g., the luminosity function of many clusters 
at medium and high redshift is lacking. Therefore, a statistical approach is 
generally used by estimating $f_{\rm star}$ as a function of other 
observed quantities for which it has been possible to confirm a robust 
correlation on a small and local sample of galaxy clusters.

From the measurements of the optical luminosity and {\it Rosat} X-ray
surface brightness in Coma, White et al. (1993) found $f_{\rm star} =
0.18 (\pm 0.05) f_{\rm gas}$. 
Arnaud et al. (1992) quote $f_{\rm star} = 0.17 (\pm 0.03)
f_{\rm gas}$ when their values are converted using a mass-to-light
ratio for stars in the V band of $6 (M/L)_{\odot}$ (see also Voevodkin \&
Vikhlinin 2004).  
Similar results have been obtained from Fukugita, Hogan \& Peebles (1998) 
by estimating the global budget of cosmic baryons.
In the following analysis, we will refer to these
very similar recipes as $f_{\rm star, W} = 0.18 (\pm 0.05) f_{\rm gas}$.

Lin, Mohr \& Stanford (2003) use the near-infrared light as a tracer of
the total stellar mass within cluster galaxies.  They find a good
correlation between the stellar mass fraction obtained from the K-band
luminosity function measured using the Two Micron All Sky Survey
(2MASS) data and X-ray properties of 27 nearby systems.  Considering
that their results apply to all the sample that spans a gas
temperature between 2 and 9 keV and that they rely on total mass estimates
obtained from scaling relations taken from the literature, we 
use their measurements of $M_{\rm star}$ and their assumed $T_{\rm
  gas}$ to recover an empirical relation valid for the clusters with
$T_{\rm gas} > 4$ keV that we will consider in our analysis.  We obtain
that, for these 10 very massive clusters, the ratio $M_{\rm star, 500}
/ T_{\rm gas}^{1.5}$ is almost independent of the temperature itself
and is equal to $5.09 (\pm 1.73) \times 10^{12} M_{\odot}$ (5 keV)$^{-1.5}$
(weighted mean and dispersion after propagation of the
errors on both $M_{\rm star, 500}$, for which we assume the relative
error on the estimates of the luminosity, and $T_{\rm gas}$).
The estimated stellar masses are then added to the measured gas masses 
to evaluate the total baryon mass. We will refer to $f_{\rm star, L}$ 
to indicate this method calculating the stellar contribution to the 
total baryon budget.

An additional component of the cold baryon budget is the intracluster
light (ICL) at very low surface brightness, making the total cold
baryonic content of galaxy clusters equal to $f_{\rm cold} = f_{\rm
  star} + f_{\rm ICL}$.  Numerical simulations (e.g. Willman et
al. 2004, Murante et al. 2004, 2007) suggest that the ICL mostly
originates from the merging processes taking place during the assembly
of massive cluster galaxies. 
Consistently with observational results for the Virgo cluster
(Feldmeier et al. 2004), simulations predict $f_{\rm ICL} \approx
0.2 f_{\rm star}$ in lower mass systems, with an increasing fraction
up to $f_{\rm ICL} \approx f_{\rm star}$ in $10^{15} M_{\odot}$
clusters (e.g. Lin \& Mohr 2004).

Gonzalez, Zaritsky \& Zabludoff (2007) have presented a census of the
baryons in local systems with total masses in the range $6 \times
10^{13} M_{\odot}-10^{15} M_{\odot}$.  Including the ICL luminosity
well mapped within 300 kpc, but probably underestimated in the
outskirts if there is any significant contribution at larger radii,
they found a well defined correlation between the stellar mass
fraction and the total mass at $\Delta=500$ in the form of $f_{\rm
  cold} = (0.009 \pm 0.002) (M_{500}/10^{15} M_{\odot})^{-(0.64 \pm
  0.13)}$.

Recently, Lagana et al. (2008) discussed in detail the baryonic
content of five massive galaxy clusters, including an ICL
contribution.  They conclude that the stellar-to-gas mass ratio within
$R_{500}$ anti-correlates slightly with the gas temperature and can be
expressed through the relation
\begin{equation}
f_{\rm cold} = f_{\rm star} + f_{\rm ICL} =
\left( 0.18 - 0.012 T_{\rm gas} \right) f_{\rm gas}, 
\label{eq:fstar}
\end{equation}
where $T_{\rm gas}$ is measured in keV.  For clusters with $T_{\rm
  gas}>4$ keV, that is one of our selection criteria, $f_{\rm
  cold}/f_{\rm gas} < 0.12$.  These values lie on the lower end of the
stellar mass distribution presented in the above-mentioned work,
whereas $f_{\rm star, W}$ represents an upper limit.  However,
considering that the analysis has been performed self-consistently for
the baryonic component for a sample of massive galaxy clusters
that well fit the properties of the objects studied in our work, we
adopt this correlation between $f_{\rm cold}$ and $f_{\rm gas}$ and
propagate the estimated 20\% uncertainty on the $f_{\rm cold}$ value.

As a reference, for $(h_{70}, \Omega_{\rm m}=1-\Omega_{\Lambda}) = (1, 0.3)$, 
we obtain for the selected sample of massive objects 
$f_{\rm cold, 500} = (0.091 / 0.090 / 0.027) f_{\rm gas, 500}$
(median / mean / dispersion; see Fig.~\ref{fig:fgal}) 
which represents about 10\% of the total
cluster baryon budget (see discussion in Section~\ref{sect:wmap5}). 

We study further the effect of (i) a different recipe to evaluate
$f_{\rm star}$ and (ii) the possibility of a larger, up to 20 per
cent, contribution of the ICL, when we discuss the robustness of our
results to some systematic errors in Sect.\ref{sect:syst}.

\subsection{The depletion parameter}

\begin{figure}
\epsfig{figure=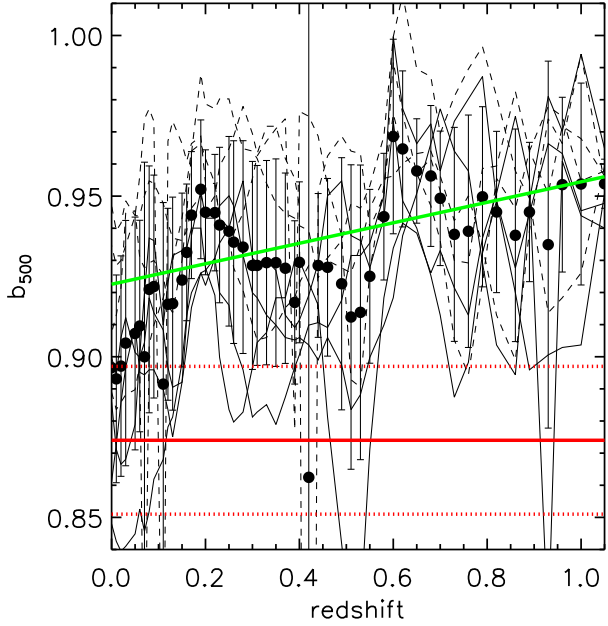,width=0.5\textwidth}
\caption{Depletion parameter as a function of redshift as measured in
  cosmological hydrodynamical simulations in Ettori et al. (2006). 
 The solid and dashed lines show results for non radiative simulations 
and for radiative simulations, which include star formation and galactic winds, 
respectively. These estimates are averaged over simulations of four 
massive clusters with $M_{200}>10^{15} M_{\odot}/h$ at $z=0$. 
The dots and errorbars indicate the average $b_{500}$ and the scatter around it.
The red solid line indicates the best-fit constant value $b_{500}$ 
estimated at $z=0$ in non-radiative simulations (dotted lines
show the dispersion around the mean).
The green solid line represents $b_{500, z}$ of equation~(\ref{eq:bz})
adopted in our analysis. 
} \label{fig:ypar} \end{figure}

The depletion parameter $b = f_{\rm bar} / (\Omega_{\rm b}/\Omega_{\rm m})$ 
indicates the amount of cosmic baryons that are thermalized
within the cluster potential.
When measured over well-representative regions of a galaxy cluster,  
this value tends to be lower than unity because not all the accreting
shock-heated baryons relax within the dark matter halo. 
However, the depletion parameter can presently only be inferred from numerical simulations,
where both the input cosmic baryons are known and the amount of them accreting into the
cluster dark matter halo can be traced.
For instance,  Eke, Navarro \& Frenk (1998) find in
their smoothed particle hydrodynamics non--radiative simulations 
that the gas fraction within $R_{\rm vir}$ is, on average, 
87 per cent of the cosmic value.  
On the other side, recent work with grid-based numerical codes estimates
that a larger amount of baryons are captured in the cluster potential
(e.g. Kravtsov, Nagai \& Vikhlinin 2005).
In a set of SPH and Eulerian simulations of a single cluster 
presented in the Santa Barbara Comparison Project, Frenk et al. (1999) 
measure at the virial radius $b = 0.92 \pm 0.07$.  
In Ettori et al. (2006), we analyze the $b$ parameter in a set of four
massive ($M_{\rm 200} > 10^{15} h^{-1} M_{\odot}$ at $z=0$) galaxy clusters
simulated by using the Tree+SPH code {\tt GADGET-2}. 
We consider here the distribution of the values of $b(<R_{500})$
measured in two sets of hydrodynamical simulations, one with
gravitational heating only, the other including radiative processes,
such as cooling and star formations with feedback provided from weak winds 
(Fig.~\ref{fig:ypar}).
We measure $b(<R_{500}) = 0.874 \pm 0.023$ at $z=0$ and $b(<R_{500}) = 0.947 \pm
0.037$ at $z=1$ in the first set, and $b(<R_{500}) = 0.920 \pm 0.026$ at $z=0$
and $b(<R_{500}) = 0.961 \pm 0.028$ at $z=1$, when cooling and feedback are
considered. If we average over the redshift range 0--1,
we obtain $b (<R_{500}) = 0.940 \pm 0.023$,
or, by fitting a polynomial at 1st order in redshift,
$b (<R_{500}) = 0.923 (\pm 0.006) + 0.032 (\pm 0.010) z$.
Considering that the four objects under consideration reached 
$M_{\rm 200} > 10^{15} h^{-1} M_{\odot}$ at $z=0$
after they accreted mass by a factor 2--6 from $z=1$, 
we adopt two values for the depletion parameter, encompassing the results
of our simulated dataset (see Fig.~\ref{fig:ypar}):
\begin{equation}
b_{500} = 0.874 \pm 0.023,
\end{equation}
that is related to the most massive systems, subjected only to the
gravitational heating, in the studied simulation 
and, thus, particularly appropriate for the high-temperature objects 
that are analyzed in the present work over radial regions, beyond their cores, 
where radiative processes (like cooling, star formation, stellar feedback) 
are not expected to influence significantly the ICM physiscs;
\begin{equation}
b_{500, z} = 0.923 (\pm 0.006) + 0.032 (\pm 0.010) z,
\label{eq:bz}
\end{equation}
that is the quantity measured when all the simulated dataset are
considered over the redshift range of interest.

\section{Results on the cosmological parameters}

We measure the gas mass fraction at $R_{500}$ in 52 $z>0.3$ objects with a
median relative error $\epsilon_{\rm gas}$ of 36 per cent ($1 \sigma$).
In the local sample, the relative error is about 6 per cent.

In the following analysis, we focus on the most massive objects
present in our sample to work under the ideal hypothesis that
the physics of the X-ray emitting plasma is only determined by the
gravitational collapse.
To define a sample of massive systems whose baryon content is
expected to be representative of the cosmic budget, we select the 49
objects at $z>0.3$ with $T_{\rm gas} > 4$ keV (see Fig.~\ref{fig:sample}
and Table~\ref{tab:highz}).
We also consider the 8 galaxy clusters with  $T_{\rm gas} > 4$ keV
in the local sample of Vikhlinin et al. (2006; see Table~\ref{tab:local})

\begin{figure*} \begin{center}
\hbox{
 \epsfig{figure=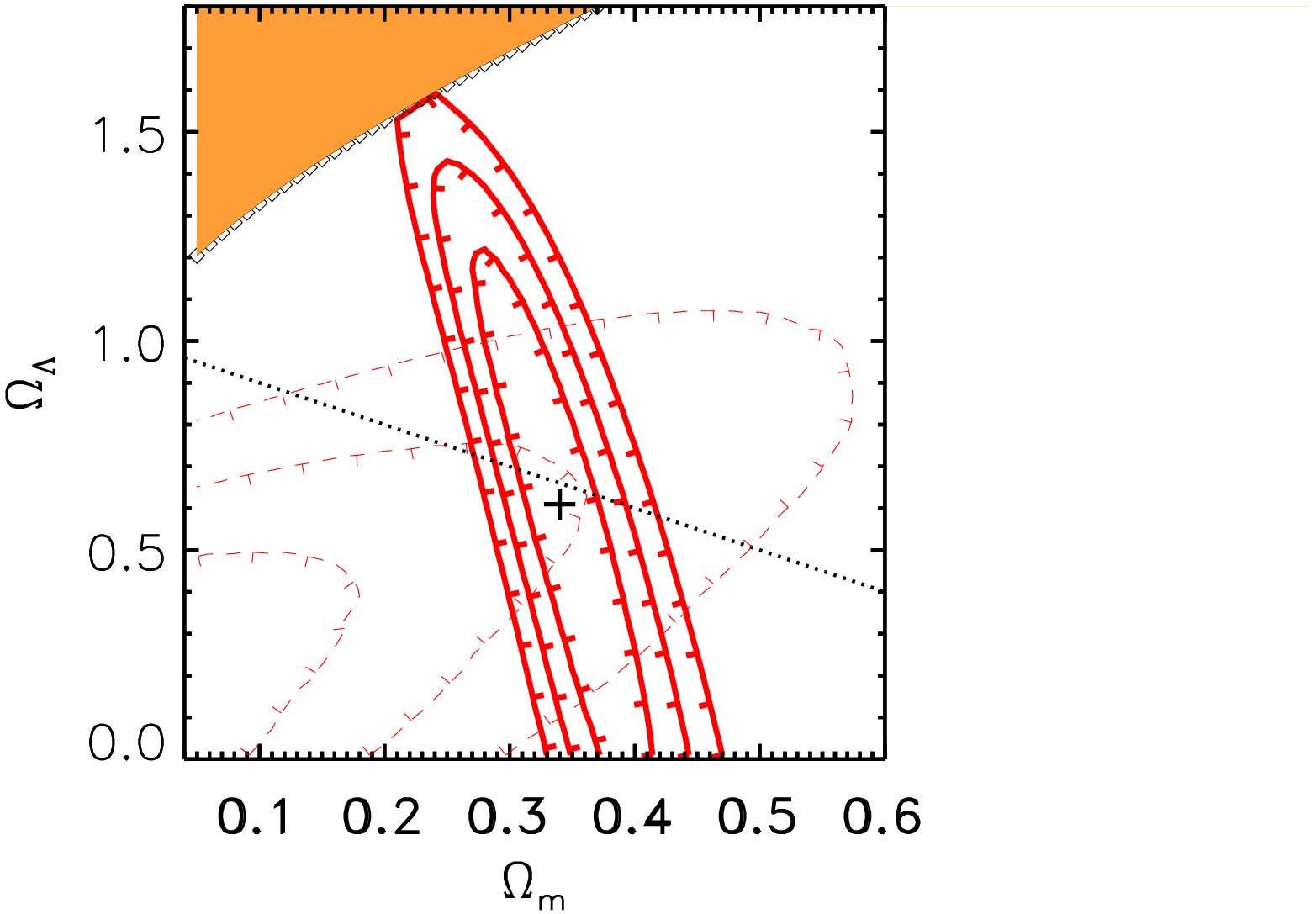,width=0.33\textwidth}
 \epsfig{figure=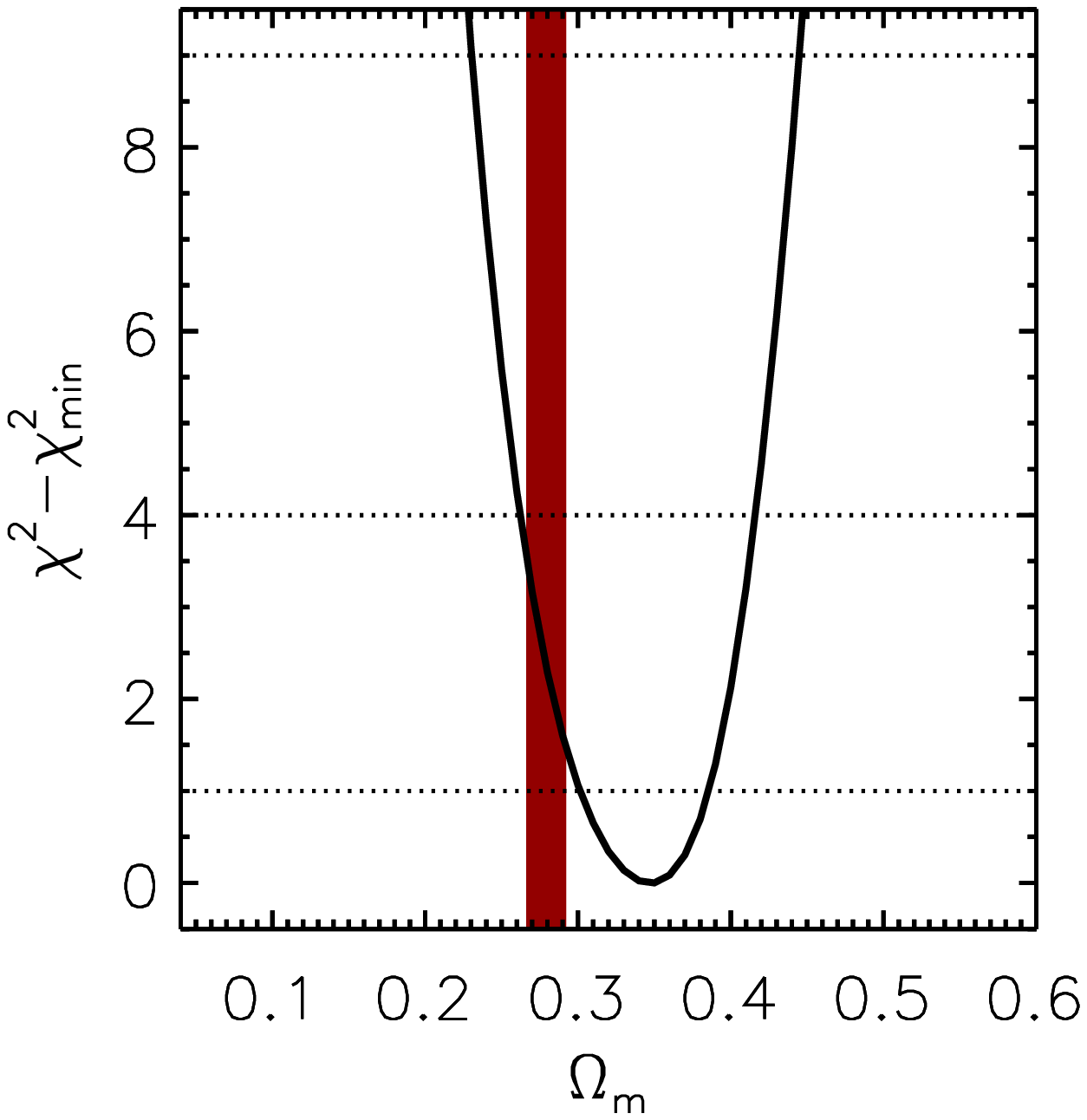,width=0.33\textwidth}
 \epsfig{figure=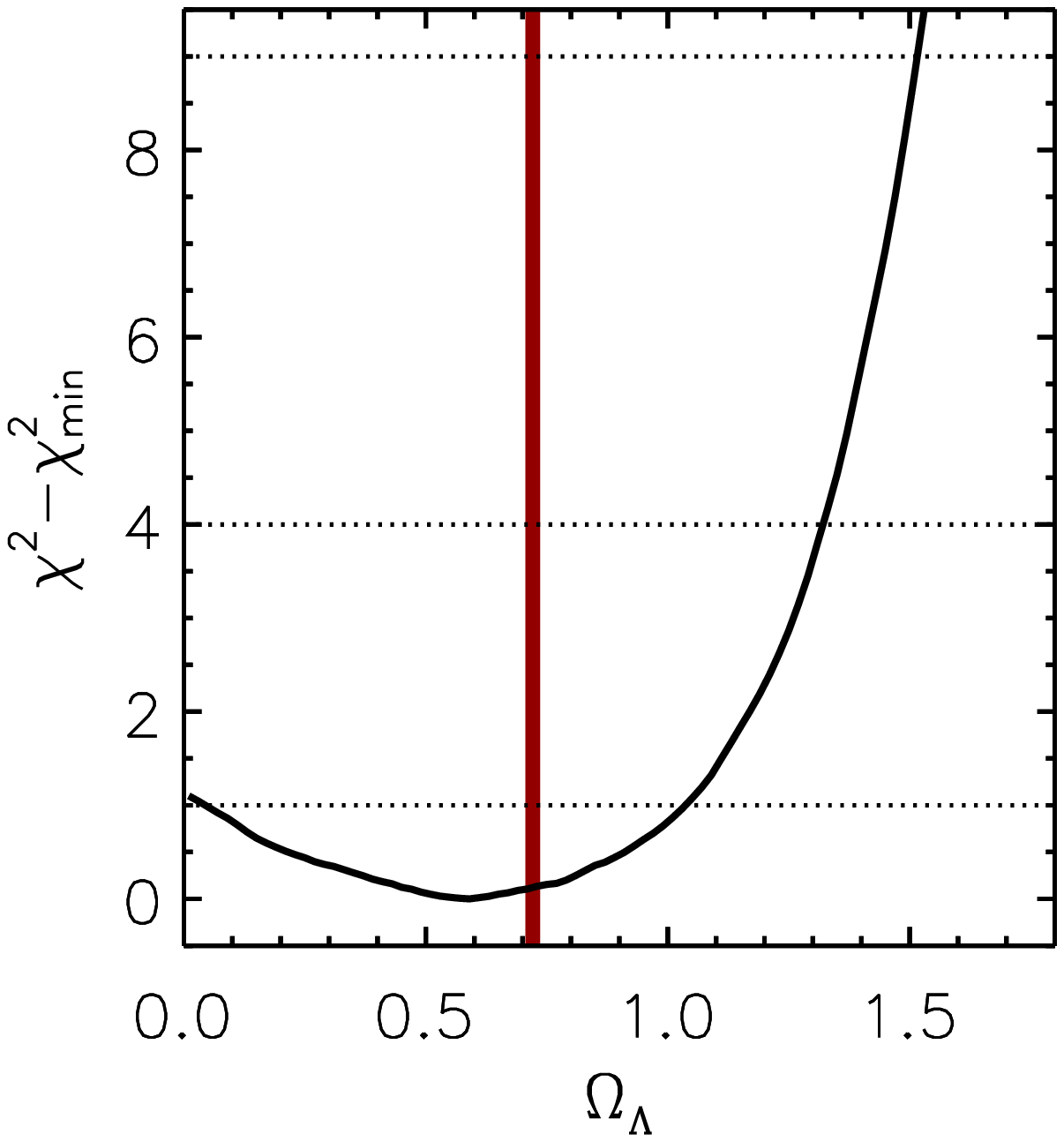,width=0.33\textwidth}
} 
\caption{Constraints in the $\Omega_{\rm m}-\Omega_{\Lambda}$ plane
  ($w=-1$) provided from the sample of 57 objects.  
  ({\it Left}) Likelihood contours at 1, 2
  and 3 $\sigma$ ($\Delta \chi^2 = 2.30, 6.17, 11.8$, respectively,
  for 2 degrees of freedom) for the constraints obtained through
  equation~\ref{eq:chi2} (solid lines) and equation~\ref{eqn:fgas_z}
  by using the gas fraction evolution only (dashed lines).  ({\it
    Middle}) Marginalized probability distribution for $\Omega_{\rm
    m}$.  ({\it Right}) Marginalized probability distribution for
  $\Omega_{\Lambda}$.  The shaded regions show the constraints from
  the joint cosmological analysis of CMB+SnIa+BAO data in Komatsu et
  al. (2008).  } \label{fig:oml} \end{center} \end{figure*}

\begin{figure*} \begin{center}
\hbox{
 \epsfig{figure=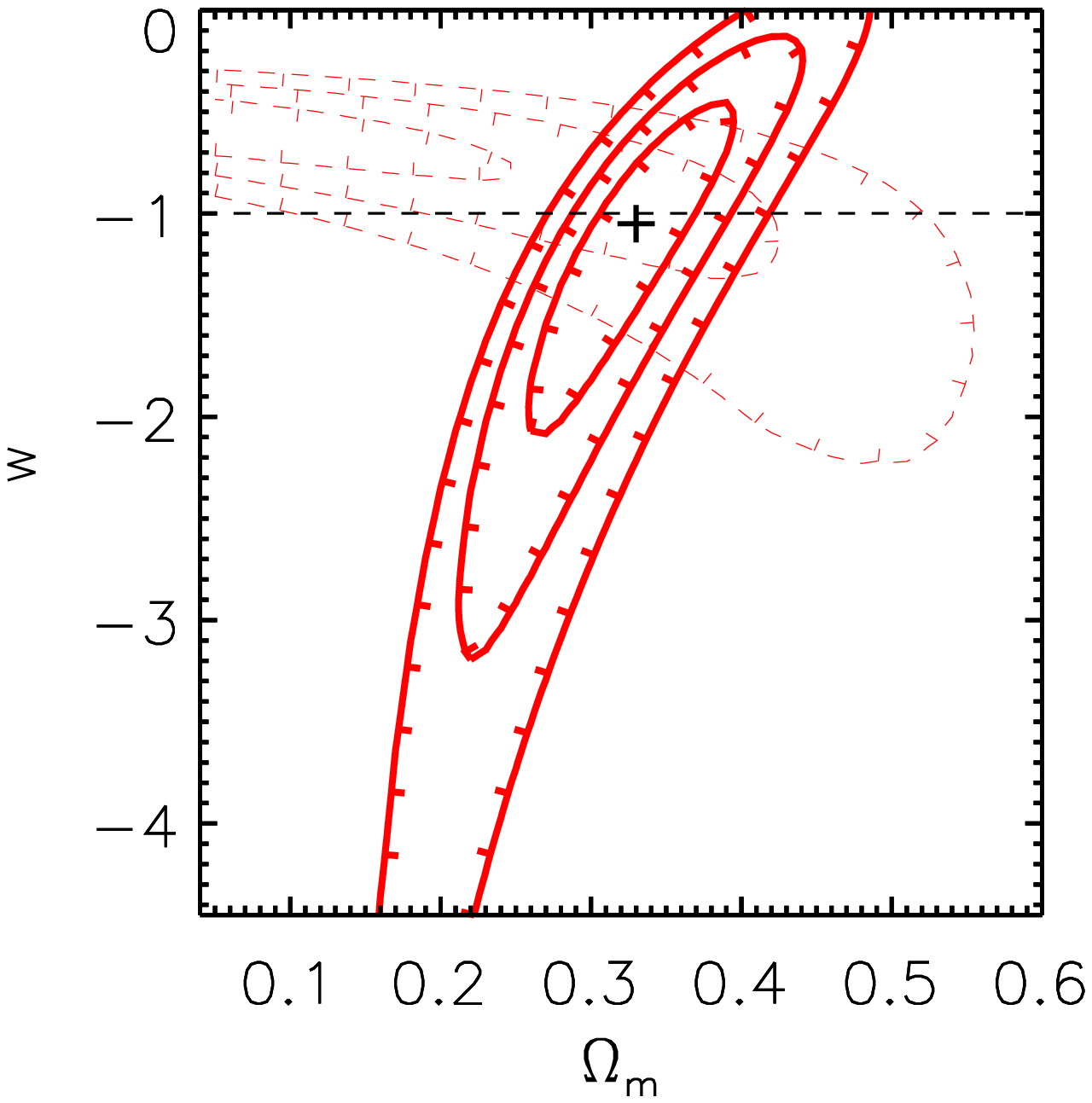,width=0.33\textwidth}
 \epsfig{figure=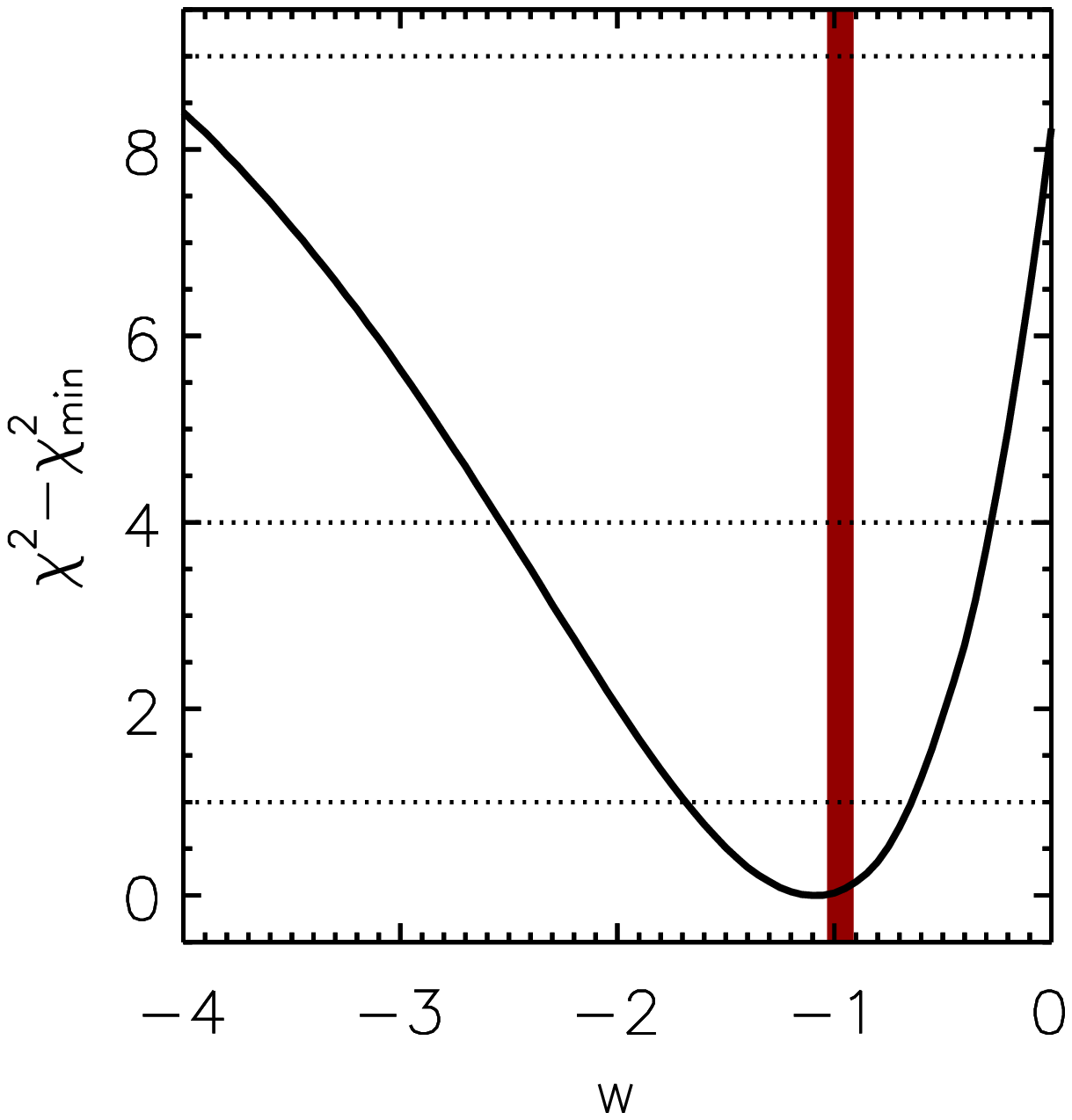,width=0.33\textwidth}
 \epsfig{figure=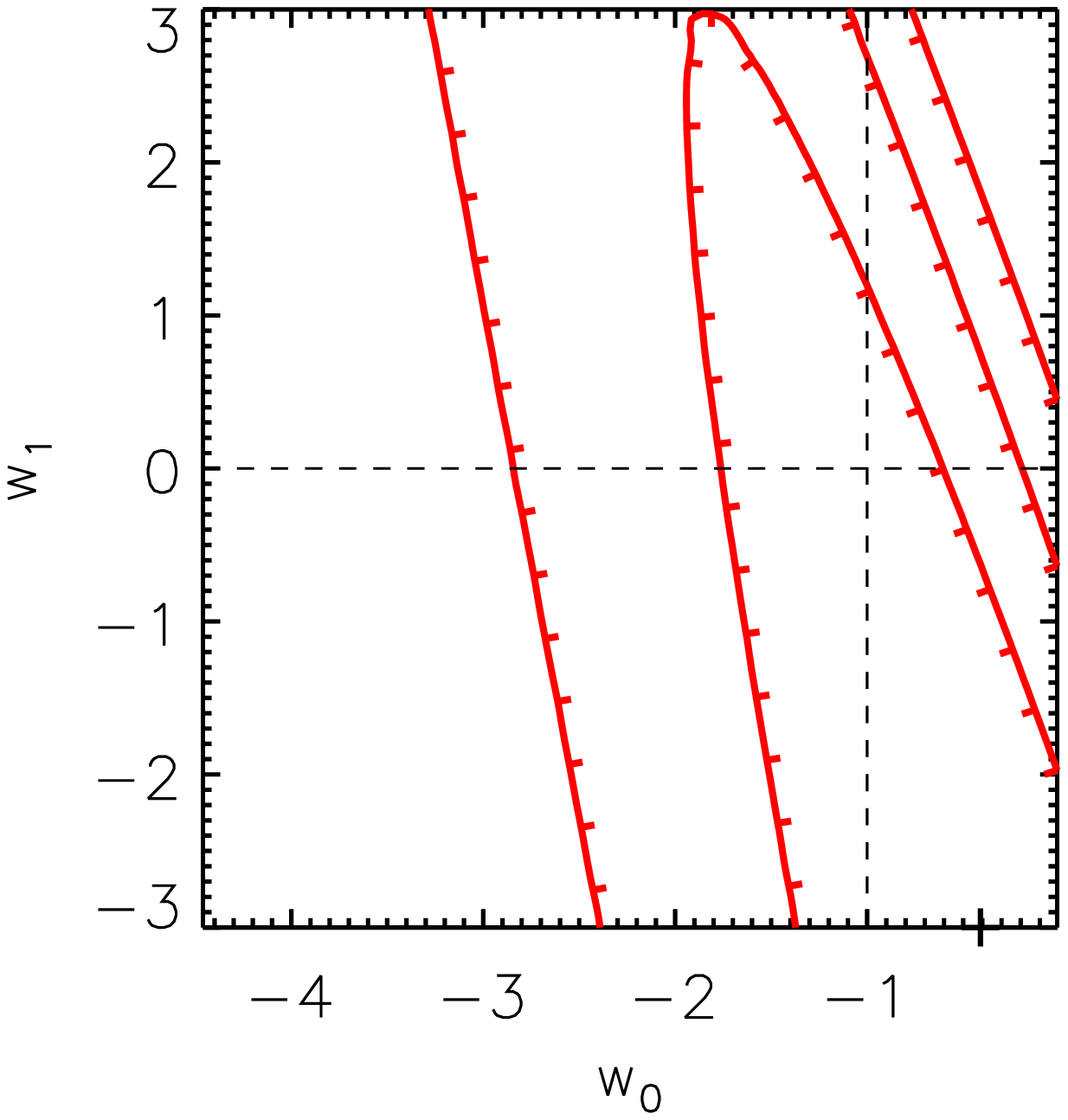,width=0.33\textwidth}
}
\caption{({\it Left}) Constraints in the $\Omega_{\rm m}-w$ plane
($\Omega_{\rm m}+\Omega_{\Lambda}=1$) plane provided from the sample of 57 objects.
Likelihood contours at 1, 2 and 3 $\sigma$
($\Delta \chi^2 = 2.30, 6.17, 11.8$, respectively, for 2 degrees of freedom)
for the constraints obtained through equation~\ref{eq:chi2} (solid lines) and
equation~\ref{eqn:fgas_z} by using the gas fraction evolution only (dashed lines).
({\it Middle}) Marginalized probability distribution on $w$ overplotted
on the best-fit constraints obtained from the joint cosmological analysis
of CMB+SnIa+BAO data in Komatsu et al. (2008).
({\it Right}) Constraints in the $w_0-w_1$ plane from the definition
adopted in equation~\ref{eqn:wz}.
} \label{fig:omw} \end{center} \end{figure*}

\subsection{Constraints on $\Omega_{\rm m}$ and $\Omega_{\Lambda}$}
\label{sect:oml}

We assume hereafter that $\Omega_{\rm b} h_{70}^2 = 0.0462 \pm 0.0012$, 
$H_0=72 \pm 8$ km s$^{-1}$ Mpc$^{-1}$, 
$b_{500} = 0.874 (\pm 0.023)$ and 
$f_{\rm cold} = \left( 0.18 - 0.012 T_{\rm gas} \right) f_{\rm gas}$, 
with an associated 20 per cent relative error (at $1 \sigma$ level). 
From the estimated mean gas mass fraction within $R_{500}$ of 0.11
(see Sect.~3),
the resulting matter density parameter is 
expected to measure (see eq.~\ref{eq:chi2})
\begin{equation}
\Omega_{\rm m} = \frac{b \, \Omega_{\rm b}}{f_{\rm gas}+f_{\rm cold}} 
\approx 0.32.
\end{equation}

\begin{figure} \begin{center}
 \epsfig{figure=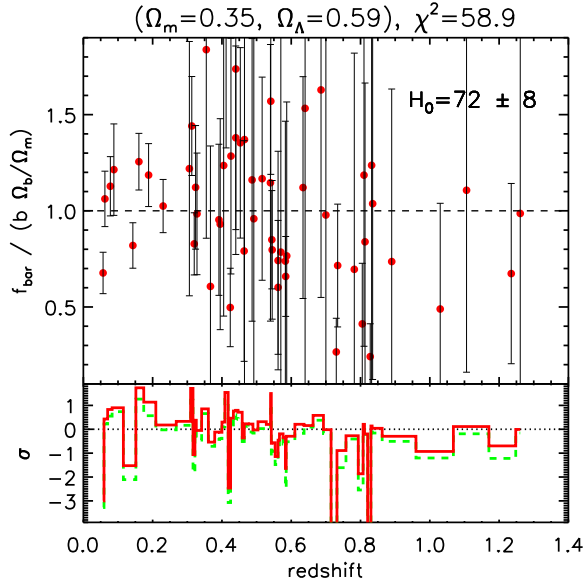,width=0.5\textwidth}
 \caption{Distribution of the measured baryon fractions in the sample
   of 57 objects (49 of which are at $z>0.3$) with $T_{\rm gas}>4$ keV.
The error bars at $1 \sigma$ include all the components listed in Section~4.
The best-fit results of the cosmological model
used as the reference are assumed.
In the lower panel, the residuals with respect to the assumed value of
$f_{\rm bar}/b \times \Omega_{\rm m}/\Omega_{\rm b} = 1$ are shown
({\it solid line}).  The {\it dashed line} refers to the distribution
obtained by fixing all the cosmological parameters to the best-fit
results and assuming $\Omega_{\Lambda}=0$.
} \label{fig:chi2} \end{center}
\end{figure}

To make extensive use of the information present in the gas mass
fraction distribution, we minimize eq.~\ref{eq:chi2} for a 
grid of values of $\{\Omega_{\rm m}, \Omega_{\Lambda}\}$ that
satisfy the condition $\Omega_{\rm m}+\Omega_{\Lambda}+\Omega_k=1$
($w=-1$). We obtain the best-fit results presented in 
Fig.~\ref{fig:oml} and quoted here at the $1 \sigma$ level of 
confidence on one single parameter ($\chi^2-\chi^2_{\rm min} \le 1$) 
after marginalization over the others:
\begin{eqnarray}
\Omega_{\rm m} = & 0.35^{+0.03}_{-0.04}  \nonumber \\
\Omega_{\Lambda} = & 0.59^{+0.46}_{-0.56}.
\end{eqnarray}
The total $\chi^2$ is $58.9$ for $57-2$ degrees of freedom
(see Fig.~\ref{fig:chi2}). 

These results change slightly once the recent determination of 
Hubble's constant, $H_0 = 62.3 \pm 1.3 {\rm (random)} \pm 5.0 {\rm
(systematic)}$ km s$^{-1}$ Mpc$^{-1}$ by Sandage et al. (2006), is
adopted. With a $\chi^2=66.5$, we obtain
\begin{eqnarray}
\Omega_{\rm m} = & 0.37^{+0.04}_{-0.04}  \nonumber \\
\Omega_{\Lambda} = & 0.63^{+0.46}_{-0.48}.
\end{eqnarray}

More stringent results are obtained when tighter constraints on 
Hubble's constant provided with a joint-fit with the cosmic microwave
temperature anisotropies are considered. For example, by using the
best-fit values from WMAP 5-year data analysis by Komatsu et
al. (2008; $H_0=70.1 \pm 1.3$ km s$^{-1}$ Mpc$^{-1}$, $\Omega_{\rm b}
= 0.0462 \pm 0.0015$; labelled ``WMAP5'' in Table~\ref{tab:cosmo}), we
obtain
\begin{eqnarray}
\Omega_{\rm m} = & 0.32^{+0.03}_{-0.02}  \nonumber \\
\Omega_{\Lambda} = & 1.01^{+0.20}_{-0.28},
\end{eqnarray}
but with a $\chi^2=148.9$ mainly due to the propagation of the relative 
error on the Hubble's constant value that is smaller than in our reference
case (2 per cent vs. 11 per cent). 

\subsection{Constraints on the equation of state of the dark energy}
\label{sect:ww}

We investigate the constraints that the cluster baryon fraction can place
on the equation of state of the dark energy, $w = P/\rho$, where $P$ and $\rho$ 
are the pressure and density of the dark energy component, respectively.
We consider both the case $w=$constant and $w=w(w_0, w_1)$ as in
equation~\ref{eqn:wz}.
We present our reference results in Fig.~\ref{fig:omw}.

Under the assumption of a flat Universe ($\Omega_{\rm m}+\Omega_{\Lambda}=1$), 
the best-fit values for our reference model are 
($\chi^2=58.9$; $1 \sigma$ level of confidence on one single parameter):
\begin{eqnarray}
\Omega_{\rm m} = & 0.32^{+0.04}_{-0.05}  \nonumber \\
w = & -1.1^{+0.60}_{-0.45}.
\end{eqnarray}

No significant limits are obtained for an evolving $w$ (see eq.~\ref{eqn:wz}), 
with $w_0 = -1.0^{+0.7}_{-0.7}$ and no constraints at 1 $\sigma$ level on $w_1$
in the range $(-3, 3)$.

Similarly to the case for $\Omega_{\Lambda}$, changes in the 
assumed prior distributions induce appreciable differences
in the central values of the best-fit parameters. 
As shown in Fig.~\ref{fig:oml_syst}, these differences are however
always within $2 \sigma$ for $w$, even in the most extreme cases,
e.g. not considering the local ($z<0.3$) objects or including in the
analysis only the hottest ($kT > 8$ keV) clusters.  For the
measurements of $\Omega_{\rm m}$, the behaviour is the same as
discussed in the previous section.

\subsection{Do the data require $\Omega_{\Lambda}>0$ ?}

Considering our best-fit results presented in Section~\ref{sect:oml},
we obtain that $\Omega_{\Lambda}>0$ with a significance in the range 
$1-3.6 \sigma$ (see, e.g., Table~\ref{tab:cosmo}).
In particular, for our reference case (Fig.~\ref{fig:oml}), our dataset
requires a $\Delta \chi^2$ of about 1 when $\Omega_{\Lambda}$ is
fixed to zero (see Fig.~\ref{fig:chi2}).
Therefore, the baryonic mass fraction in galaxy clusters alone can only
provide marginal evidence for a no null
contribution from a dark energy component.
This contribution is mainly due to the assumption that
the gas mass fraction does not evolve with redshift in the ``correct'' cosmology.
We can write it as
\begin{equation}
\chi^2 = \sum_{i=1}^{N_{\rm dat}} \frac{(f_{{\rm gas}, i} - \overline{f_{\rm gas}})^2}
{\epsilon_{{\rm gas}, i}^2 +\epsilon_{\overline{f}}^2}
\label{eqn:fgas_z}
\end{equation}
where $f_{{\rm gas}, i}$ and $\epsilon_{{\rm gas}, i}$ are the single
gas mass fraction measurements with the relative errors,
$\overline{f_{\rm gas}}$ and $\epsilon_{\overline{f}}$ are the mean
and the error on the mean of the $f_{{\rm gas}, i}$ values. In Fig.~\ref{fig:oml}, 
the likelihood contours are overplotted on the best-fit results. 
From these plots, it is possible to appreciate the
different role played by the two assumptions made to define
the ``correct'' cosmology: while the condition that {\it the gas
  fraction is constant with look-back time} marginally influences 
the value of $\Omega_{\rm m}$ but, at the same time, significantly tilts 
the contours favouring an $\Omega_{\Lambda}>0$ for $\Omega_{\rm m}
\sim 0.35$, the condition that {\it the cluster baryonic fraction is
representative of the cosmic value} has the greatest statistical weight, 
collapsing the probability around
the best-fit value of $\Omega_{\rm m} \approx 0.35$.

\begin{table*}[ht]
\caption{Best-fit results and $1 \sigma$ ($\Delta \chi^2 =1$) errors
of the cosmological parameters constrained in our analysis under
different assumptions. The quoted constraints are obtained
for each parameter after marginalization.
In Fig.~\ref{fig:oml_syst}, the same constraints are plotted
and the number of clusters in the sample and the total $\chi^2$ are indicated.
The single samples are discussed in Sect.~\ref{sect:oml}, \ref{sect:ww}
and \ref{sect:syst}.
}
\begin{center}
\begin{tabular}{l c@{\hspace{.8em}} c@{\hspace{.8em}} c@{\hspace{.8em}} c@{\hspace{.8em}} c@{\hspace{.8em}} l }
\hline \\
 & $\Omega_{\rm m}$ & $\Omega_{\Lambda}$ & & $\Omega_{\rm m}$ & $-w$ & comments \\ 
\hline \\
 & \\
 & \multicolumn{2}{c}{$w=-1$} & & \multicolumn{2}{c}{$\Omega_{\rm k}=0$} & \\
 & \\
REF & $0.35^{+0.03}_{-0.04}$ & $0.59^{+0.44}_{-0.56}$ & & $0.32^{+0.04}_{-0.05}$ & $1.10^{+0.60}_{-0.45}$ & reference case (Sect.~5.1 and 5.2) \\
$T=T_0$ & $0.40^{+0.03}_{-0.02}$ & $0.01^{+0.46}_{-0.01}$ & & $0.29^{+0.08}_{-0.05}$ & $1.90^{+2.50}_{-1.10}$ & isothermal ICM (Sect.~3) \\
$b=b_z$ & $0.36^{+0.05}_{-0.04}$ & $0.65^{+0.48}_{-0.50}$ & & $0.34^{+0.04}_{-0.05}$ & $1.20^{+0.65}_{-0.45}$ & depletion parameter dependent on $z$ (Sect.~4.2) \\
$f_{\rm ICL},0.2$ & $0.34^{+0.03}_{-0.04}$ & $0.59^{+0.44}_{-0.56}$ & & $0.31^{+0.04}_{-0.04}$ & $1.05^{+0.60}_{-0.40}$ & adding 20\% to $f_{\rm cold}$ (Sect.~4.1) \\
$f_{\rm star, W}$ & $0.32^{+0.03}_{-0.03}$ & $0.49^{+0.42}_{-0.48}$ & & $0.30^{+0.03}_{-0.04}$ & $0.95^{+0.50}_{-0.35}$ & $f_{\rm star}$ from White et al. (1993; Sect.~4.1) \\
$f_{\rm star, L}$ & $0.34^{+0.02}_{-0.02}$ & $0.01^{+0.34}_{-0.01}$ & & $0.35^{+0.04}_{-0.04}$ & $0.35^{+0.55}_{-0.35}$ & $f_{\rm star}$ from Lin et al. (2003; Sect.~4.1) \\
$z>0.3$ & $0.36^{+0.04}_{-0.08}$ & $0.19^{+1.12}_{-0.18}$ & & $0.26^{+0.10}_{-0.09}$ & $1.35^{+1.25}_{-0.75}$ & only high$-z$ objects \\
$T>8$ keV & $0.34^{+0.03}_{-0.05}$ & $0.07^{+0.74}_{-0.06}$ & & $0.31^{+0.08}_{-0.09}$ & $0.75^{+0.85}_{-0.60}$ & only hot objects \\
$\epsilon_f<0.5$ & $0.34^{+0.02}_{-0.03}$ & $0.01^{+0.64}_{-0.01}$ & & $0.33^{+0.04}_{-0.04}$ & $0.70^{+0.45}_{-0.45}$ & only measurements with lower uncertainty \\
cen$_0$ & $0.29^{+0.08}_{-0.05}$ & $1.35^{+0.20}_{-0.52}$ & & $0.24^{+0.08}_{-0.05}$ & $2.00^{+2.45}_{-1.15}$ & more relaxed objects (Sect.~5.4) \\
cen$_1$ & $0.32^{+0.04}_{-0.04}$ & $0.89^{+0.36}_{-0.64}$ & & $0.31^{+0.05}_{-0.05}$ & $1.10^{+0.70}_{-0.50}$ & extended ``cen$_0$" sample (Sect.~5.4) \\
$H_0=62$ & $0.37^{+0.04}_{-0.04}$ & $0.63^{+0.46}_{-0.48}$ & & $0.35^{+0.04}_{-0.05}$ & $1.15^{+0.65}_{-0.40}$ & $H_0$ from Sandage te al. (2006) \\
WMAP5 & $0.32^{+0.03}_{-0.02}$ & $1.01^{+0.20}_{-0.28}$ & & $0.33^{+0.02}_{-0.02}$ & $1.25^{+0.35}_{-0.25}$ & $H_0$ and $\Omega_{\rm b}$ from Komatsu et al. (2008) \\
 & \\
\hline \\
\end{tabular}

\end{center}
\label{tab:cosmo}
\end{table*}

\subsection{Robustness to systematic errors}
\label{sect:syst}

\begin{figure*}
  \hbox{ \epsfig{figure=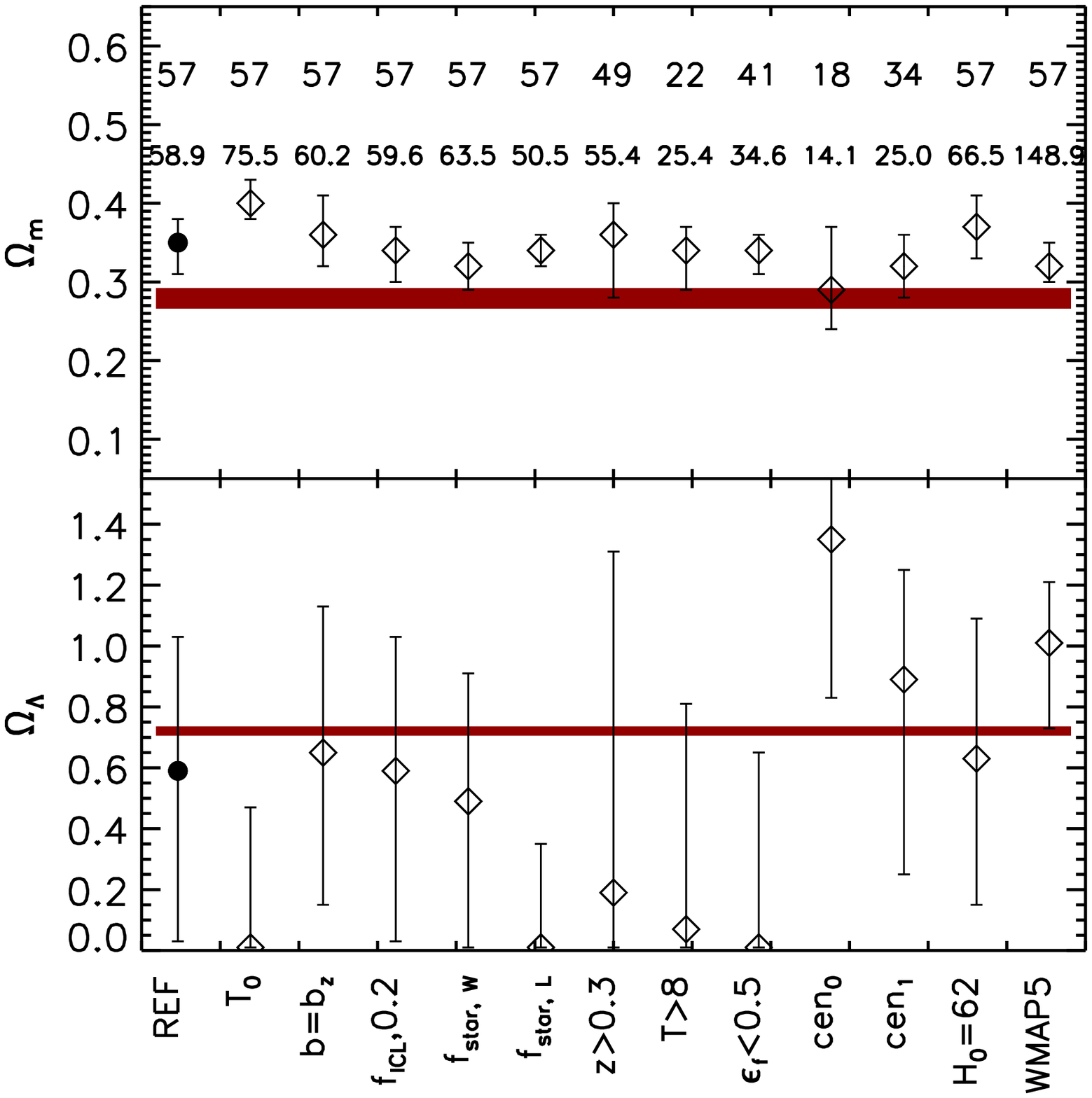,width=0.5\textwidth}
    \epsfig{figure=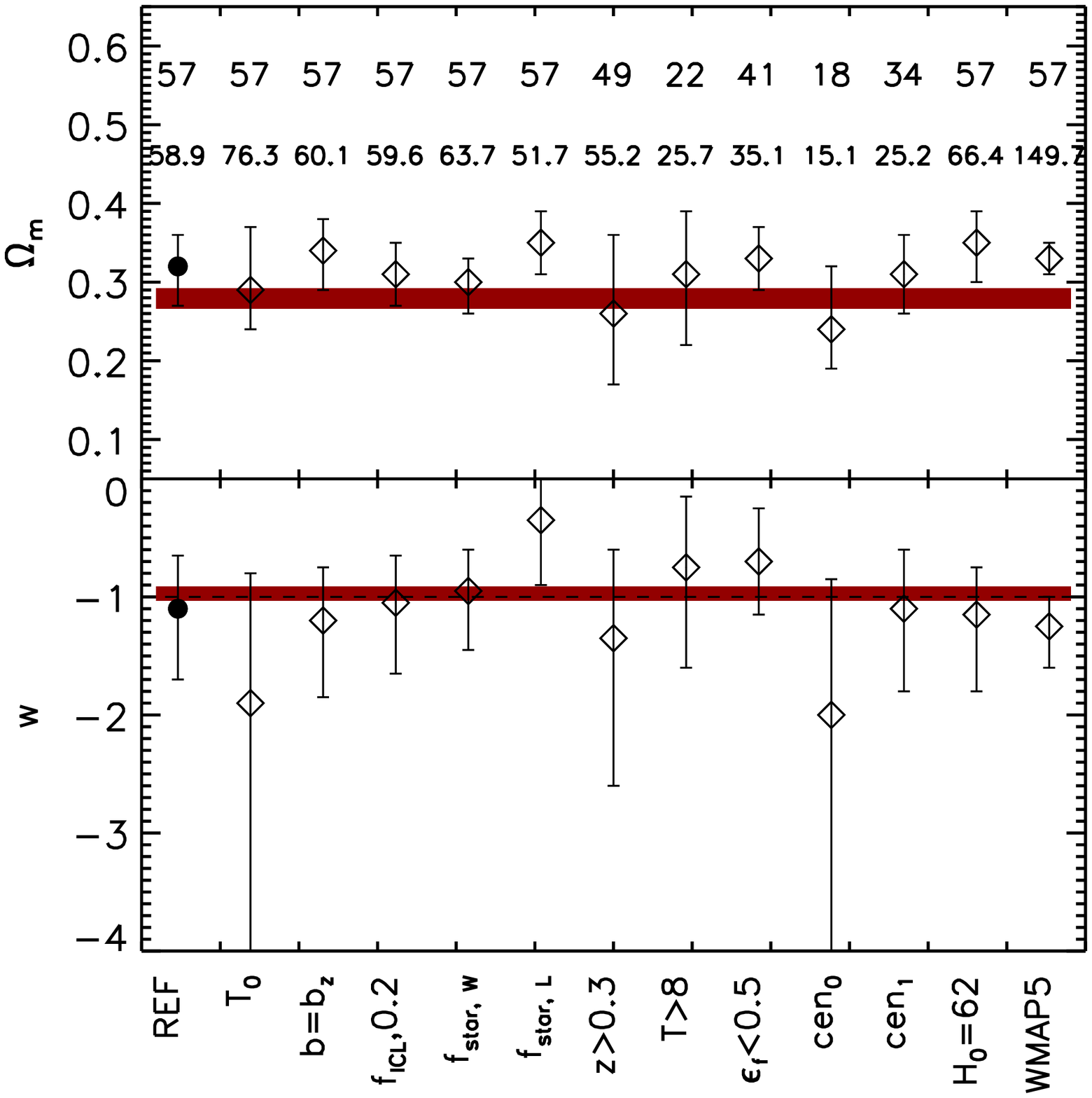,width=0.5\textwidth} }
\caption{Best-fit results on $\Omega_{\rm m}$ and $\Omega_{\Lambda}$ ({\it left})
    and $\Omega_{\rm m}$ and $w$ ({\it right}) for the different sources
    of systematics discussed in the text.  The
    shaded regions show the constraints from WMAP 5-year results
    (Komatsu et al. 2008). All the errors are at the $1 \sigma$ level.
    The number of clusters that meets the selection criteria and the
    total $\chi^2$ are indicated in the top panel.
  } \label{fig:oml_syst} \end{figure*}

In this section, we investigate how our cosmological constraints are
affected by potential systematic errors.  
To evaluate this, we repeat the
analysis described in the previous sections and apply it to several
different subsets defined by changing one of the criteria adopted for
the selection of the reference sample of 57 hot galaxy clusters.

First of all, we do not observe any dependence of $f_{\rm gas}$ on
both temperature and metallicity in our reference sample: once
objects with $T_{\rm gas}>4$ keV are selected, the Spearman's rank
correlation is about 0.1 implying an underlying correlation
significant at less than the $1 \sigma$ confidence level.

Moreover, we notice that the estimated gas mass fraction within $R_{2500}$
is about 75 per cent of the value measured at $R_{500}$ (mean and 
standard deviation after 1000 bootstrap resampling $=75\pm2$; 
median $=73$ per cent), confirming its increased value outwards.
%

As shown in the plots of Fig.~\ref{fig:oml_syst} (and summarized in
Table~\ref{tab:cosmo}), the largest uncertainties are related to the
assumed prior on the temperature profile.
By using the values obtained by assuming an isothermal ICM (see 
$\hat{f}_{\rm gas}$ in Table~\ref{tab:highz}), and applying 
the method described in Section~4, we obtain
(total $\chi^2=75.5$; $\Omega_{\rm m}+\Omega_{\Lambda}+\Omega_k=1,
w=-1$)
\begin{eqnarray}
\Omega_{\rm m} = & 0.40^{+0.03}_{-0.02}  \nonumber \\
\Omega_{\Lambda} < & 0.46,
\end{eqnarray}
and ($\chi^2=76.3$; $\Omega_{\rm m}+\Omega_{\Lambda}=1$)
\begin{eqnarray}
\Omega_{\rm m} = & 0.29^{+0.08}_{-0.05}  \nonumber \\
w < & -0.8.
\end{eqnarray}
The total $\chi^2$ shows a significant increase ($\Delta \chi^2=17$),
suggesting a poorer representation of the distribution of $f_{\rm gas}$ 
and pointing out the necessity for a robust and reliable determination of
the temperature profiles in galaxy clusters, possibly out to the
region where the gas mass fraction measurements are more
representative of the cluster baryon content ($r \ga R_{2500} \approx
0.3 R_{200}$).

The assumption on the stellar mass fraction does not significantly affect 
the estimates of the cosmological parameters.  As discussed in
Section~\ref{sect:fcold}, the stellar mass fraction that we adopt here
from Lagana et al. (2008) lies on the lower end of the distribution
available in the literature.  By assuming instead $f_{\rm cold}/f_{\rm
  gas}=0.18$, independently from the cluster mass, as generally done
in similar work (e.g. Ettori et al. 2003, Allen et al. 2008), 
we measure a decrease in $\Omega_{\rm m}$ by about 9 per cent
(see case labelled ``$f_{\rm star, W}$'').

To select our sample of 57 clusters, we have applied only the
selection criterion of $T_{\rm gas} > 4$ keV to consider only the most
massive systems for which the ICM physics should be mostly determined
by the process of gravitational collapse.

Once we add the further criterion that the relative uncertainty
on the gas mass fraction values has to be lower than 0.5 
($\epsilon_{\rm gas} / f_{\rm gas} < 0.5$), 
the sample is reduced to 41 objects and the good fit
obtained ($\chi^2=34.6$) provides marginally looser constraints on the
cosmological parameters, but in better agreement with WMAP 5-year
limits.  On the other hand, when very hot systems ($T_{\rm gas} > 8$
keV) are considered, the sample is reduced to only 22 objects and
estimates of $\Omega_{\Lambda}$ consistent with zero are obtained. 
Similar constraints are obtained when the local ($z<0.3$) 8 clusters 
are excluded due to the insensitivity of the geometrical part of the 
method to the dark energy component at high redshift.

In our selection, we do not consider the morphological aspect of the
cluster.  To study this effect, we have considered the measurements of
the centroid shift for the 43 objects in common with the sample of 90
clusters analysed by Maughan et al. (2007) with $T_{\rm gas} > 4$ keV.
We consider as relaxed objects the 18 clusters with a centroid shift
lower than the median value of $1.18 \times 10^{-2}$ measured over the
entire sample of 115 clusters (see their Table~2).  These 18 objects
are part of the first sample, labelled $cen_0$.  A second sample,
$cen_1$, includes the 16 objects for which an estimate of the centroid
shift is not available, for a total number of 34 clusters.  Through
these two samples, we obtain very similar results, with a relativly
larger (smaller) contributions from the dark energy (matter) component
compared to the reference values.

Overall, we can attribute a systematic relative error of about $\pm5$ per
cent to the most robust cosmological constraint, $\Omega_{\rm m}$.
It is estimated by considering the standard deviations of the
values plotted in Fig.~\ref{fig:oml_syst} and summarized in
Table~\ref{tab:cosmo} around our best-fit reference value of
$\Omega_{\rm m}=0.35$.  Similarly, we associate a systematic
uncertainty of $(-0.25, +0.30)$ to our best-fit value
$\Omega_{\Lambda}=0.59$ (about $45$ per cent)
and a relative systematic uncertainty
of $(-25, +34)$ per cent to our best-fit value of $w=-1.1$.

\subsection{The reverse situation: 
the baryon distribution in X-ray galaxy clusters after WMAP 5-year} 
\label{sect:wmap5}

In the previous sections, we have looked at the best-fit cosmological
parameters that better reproduce the observed distribution of the 
cluster baryonic mass fraction.
We now reverse the the problem, by fixing the cosmological
parameters to the values suggested from the analysis 
of the anistropies in the cosmic microwave background and
investigating the average physical properties of the baryons 
in our X-ray luminous galaxy clusters.

Given the above measurements of the cluster gas and total masses, and
the estimated contribution from the stellar component, we can estimate
the relative distribution of baryons in galaxy clusters assuming that
the cosmology is defined in its components from best-fit results
obtained after the WMAP 5-year release (Dunkley et al. 2008), combined in a joint
analysis with supernovae Type Ia data and Baryon Acoustic Oscillations
constraints (see Komatsu et al. 2008).  In the present work, we follow 
the analysis described by Ettori (2003) and update the results
discussed there.

\begin{figure*}
\epsfig{figure=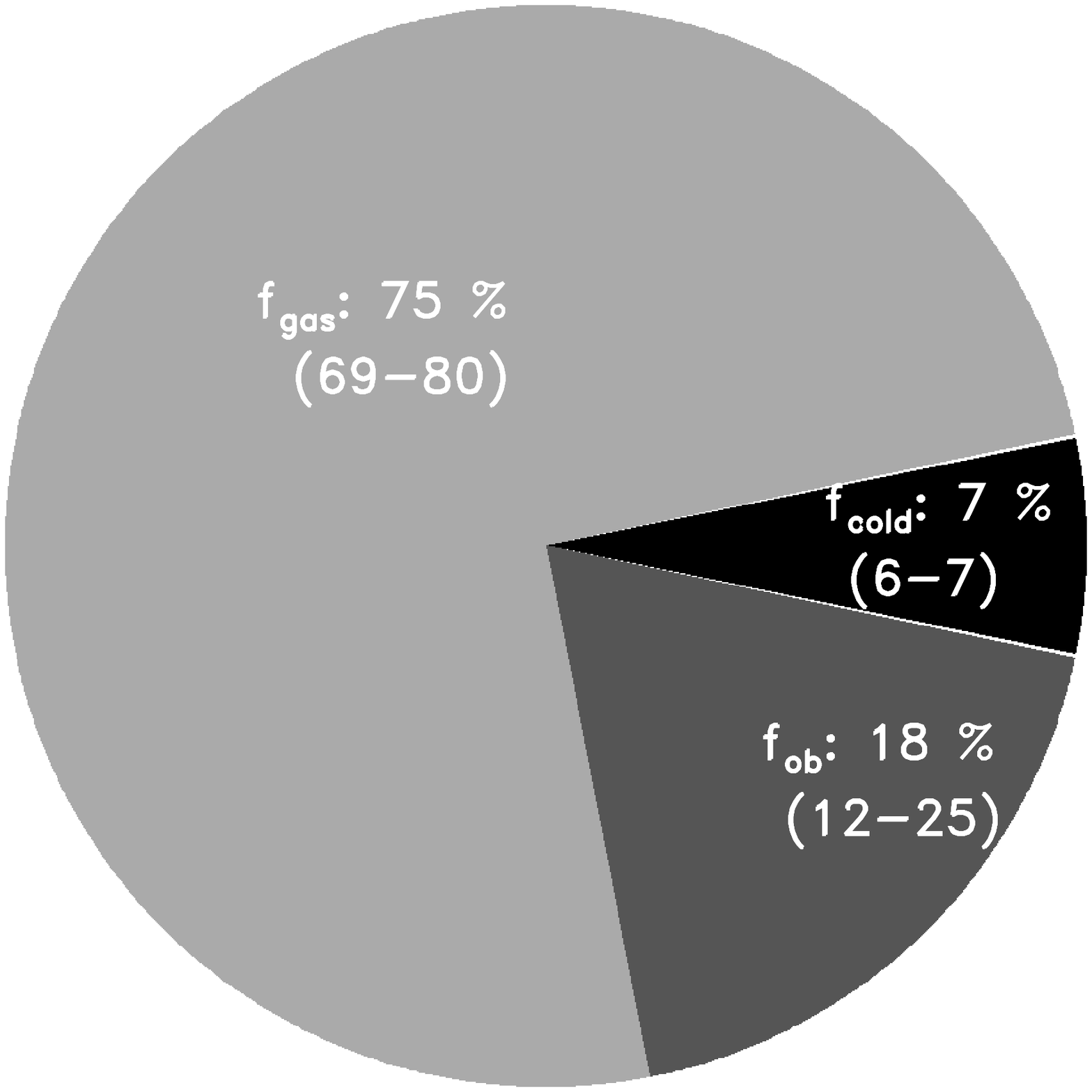,width=0.5\textwidth}
\epsfig{figure=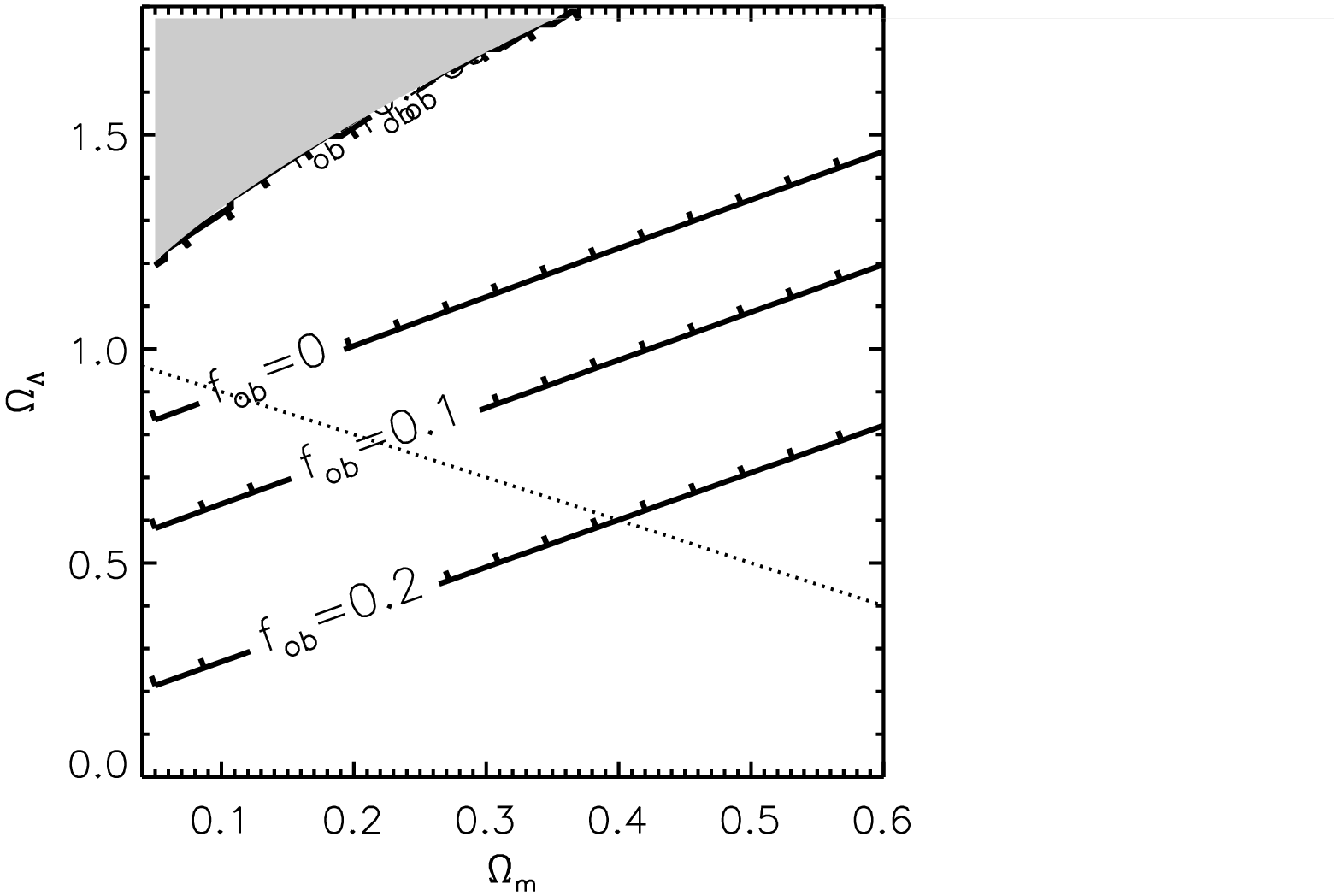,width=0.5\textwidth}
\caption{(Left) Cluster baryonic pie obtained by fixing the cosmological
  parameters to the best-fit results after the WMAP 5-year release (Komatsu et
  al. 2008) and considering the mean values of the hot ($f_{\rm gas}$),
  cold ($f_{\rm cold}$) and other ($f_{\rm ob}$) baryon fractions normalized 
  to the cosmic value, $b (\Omega_{\rm b}/\Omega_{\rm m})_{\rm WMAP}$. 
  Ranges within two standard deviations (95.4 \% level of confidence)
  from the mean (obtained after 1000 bootstrap resamplings) 
  are indicated within the parentheses.
  (Right) Dependence of the average $f_{\rm ob}$ value upon the cosmological
  parameters $(\Omega_{\rm m}, \Omega_{\Lambda})$. The dotted line
  indicates a flat Universe.
} \label{fig:barpie} 
\end{figure*}
 
We estimate the relative contribution of the cluster baryons
to the cosmic value $(\Omega_{\rm b}/\Omega_{\rm m})_{\rm WMAP}
 = 0.165 \pm 0.009$, where the best-fit results after the WMAP 5-year release
are considered ($H_0=70.1 \pm 1.3$ km s$^{-1}$ Mpc$^{-1}$, $\Omega_{\rm m}
= 1 - \Omega_{\Lambda} =0.279 \pm 0.013$, $\Omega_{\rm b} = 0.0462 \pm 0.0015$;
see Table~1 in Komatsu et al. 2008). We correct our baryon fractions
by the depletion parameter with consistent results both 
when we use $b=$constant and $b=b(z)$. 
In the panel on the left of Fig.~\ref{fig:barpie}, we show the mean values
(after 1000 bootstrap resamplings)
of the ratios between the mass fractions in hot / cold / the remaining baryons 
and the cosmic value equals to  $b \times (\Omega_{\rm b}/\Omega_{\rm m})_{\rm WMAP}$.

The requirement for the presence of other baryons apart from the hot,
X-ray emitting component and the cold, stellar phase can be explained
by either (i) an underestimate by a factor of 3 of the mass fraction
in the form of cold baryons, or (ii) an underestimate by about 20 per
cent of $f_{\rm gas}$.

Recently, McCarthy, Bower \& Balogh (2007) obtained similar results by
combining $f_{\rm gas}$ profiles collected from the literature,
and corrected for (i) stellar mass contributions, (ii) violations of the
hydrostatic equilibrium equation and (iii) the depletion parameter,
with the WMAP 3-year results (Spergel et al. 2007), concluding that the
most likely explanation is that $\Omega_{\rm m}$ must lie in the range
$0.28-0.47$, i.e. up to 70 per cent larger than the best-fit from 
the WMAP 5-year analysis.

We emphasize, however, that our conclusion on a no null contribution from
other baryons not accounted for in X-ray emitting plasma and 
cold stars and intracluster light is drawn mainly from fixing
the cosmological parameters to the results of the analysis 
after the WMAP 5-year release that strongly support a flat Universe solution. 
On the other hand, our best-fit results seem to suggest
either a higher $\Omega_{\rm m}$ value or, if $\Omega_{\rm m}$
from the WMAP 5-year constraints is adopted, 
a higher (lower) $\Omega_{\Lambda} (w)$ value, lowering 
the requested amount of $f_{\rm ob}$ (see panel on the
right in Fig.~\ref{fig:barpie}).

Moreover, by assuming that the WMAP 5-year data provide 
the ``correct'' cosmology and,
thus, fixing $(H_0, \Omega_{\rm m}=1-\Omega_{\Lambda})$ to those
values, we can attempt to limit the contribution of other effects that
can further bias the estimates of the gas mass fraction.  Among these,
we consider the level of the ICM clumpiness (e.g. Mathiesen et
al. 1999) and the presence of non-thermal pressure support (e.g. Dolag
\& Schindler 2000).
To weigh their contribution, we look for a minimum $\chi^2$ in the 
parameter space $\{\alpha, C\}$, where $\alpha = P_{\rm Non-thermal} 
/ P_{\rm thermal}$ is the ratio, assumed constant in time
and space, between a pressure of non-thermal origin and the pressure
of the ICM and $C = ( <n_{\rm gas}^2> / <n_{\rm gas}>^2 )^{0.5}$ is
a measure of the clumpiness of the ICM which is observed in brightness
as an average value of the square of the gas density $<n_{\rm gas}^2>$.
These two parameters affect the gas mass fraction estimates through
propagation of corrections on the total and gas masses in a degenerative way:
$M_{\rm tot, new} = (1+\alpha) M_{\rm tot, old}$,
$M_{\rm gas, new} = M_{\rm gas, old} / C$ and, thus,
$f_{\rm gas, new} = f_{\rm gas, old} / [ C (1+\alpha) ]$.
The  net effect is thus to further reduce the gas mass fraction value.
Adding this to the indication that the WMAP 5-year cosmology does not account 
for the observed baryon census will further increase the measured discrepancy.
Indeed, we obtain that, at a $3 \sigma$ level of confidence
with 2 degrees of freedom, our gas mass fraction measurements 
within $R_{500}$ in a WMAP 5-year concordant Universe constrain 
$\alpha < 0.13$ and $C < 1.01$ (the best-fit results
require $\alpha = 0$ and $C = 1$).

\section{Conclusions}

We study the baryonic content in 52 X-ray luminous galaxy clusters 
observed with Chandra in the redshift range 0.3 -- 1.273. 
We include in the sample the measurements of 8 objects 
with $kT_{\rm gas} >4$ keV lying in the redshift range 0.06--0.23
presented in Vikhlinin et al. (2006).
We adopt the same functional form to estimate the gas density
and temperature profiles to recover the gas and total mass radial 
distribution.
We estimate statistically the average contribution
from cold baryons in stars and intracluster light to the total
baryonic budget of each cluster.  By using the baryonic content
determined in this way, we have investigated the robustness of the
cosmological constraints provided from the cluster baryon fraction
alone.

We show that the a-priori knowledge of the cosmic baryon density
$\Omega_{\rm b}$ and the Hubble constant $H_0$ are essential to place
significant limits in the $\Omega_{\rm m}-\Omega_{\Lambda}/w$ plane.
In particular, the determination of their central values are required
to avoid the introduction of systematic differences in the estimate of
the cosmological parameters. This systematic bias cannot be corrected
by simply enlarging the error bars on the assumed central values, but
need a definitive input from other sources (for instance, CMB
measurements, type Ia supernovae data or primordial nucleosynthesis
calculations; see, e.g., Allen et al. 2008).

We find that the gas mass fraction measured in our subsample of 57 hot
($T_{\rm gas}> 4$ keV) galaxy clusters, in combination with a Hubble's
constant value of $H_0=72 \pm 8$ km s$^{-1}$ Mpc$^{-1}$ and
$\Omega_{\rm b} h_{70}^2 = 0.0462 \pm 0.0012$, provides a best-fit
result of $\Omega_{\rm m} = 0.35^{+0.03}_{-0.04}$ and
$\Omega_{\Lambda} = 0.59^{+0.44}_{-0.56}$ ($1 \sigma$ error; $w=-1$)
and $\Omega_{\rm m} = 0.32^{+0.04}_{-0.05}$ and $w =
-1.1^{+0.6}_{-0.5}$, when a flat Universe is considered.

We discuss in detail the systematic effect on our best-fit constraints
originating from (i) the stellar mass fraction component, expected to
be of the order of $0.1-0.2 f_{\rm gas}$, (ii) the redshift dependence
of the depletion parameter $b$, (iii) the isothermality of the ICM.
The latter effect, in particular, increases the total $\chi^2$ by
17 and shifts the best-fit values by about 15 per cent, pointing
out the necessity for a robust and reliable
determination of the temperature profiles in galaxy clusters, possibly
out to the region where the gas mass fraction measurements are more
representative of the cluster baryon content ($r \ga R_{2500} \approx
0.3 R_{200}$).

By fixing the cosmological parameters to the values from the WMAP
5-year data analysis, we limit the contributions expected from
non-thermal pressure support and ICM clumpiness to be lower than about
10 per cent. Otherwise, there is room to accommodate baryons not 
accounted for either in the X-ray emitting plasma or in stars in the order
of 18 per cent of the total cluster baryon budget. However, this value
is lowered to zero once a no-flat Universe is allowed, as suggested
from the cluster gas mass fraction distribution alone.  

Bearing in mind that the $f_{\rm gas}$ method to constrain the
cosmological parameters is not self-consistent and needs a strong
a-priori knowledge of both the Hubble constant value and the cosmic
baryon density, we conclude that it offers well proven and reliable
results on $\Omega_{\rm m}$ (relative statistical error at $1 \sigma$
of $\sim$11\%, with a further systematic effect of $\pm5$\%), 
whereas it presents weakness in terms of the assumptions on the 
astrophysics involved when applied to investigate the dark energy issue.  
On the other hand, this
method provides a complementary and independent constraint of the
parameter space investigated from planned dark energy experiments
(e.g. Rapetti et al. 2008; see also Linder 2007 for a more
pessimistic view) and should be supported for this application.

\section*{Acknowledgements}
We acknowledge the financial contribution from contracts ASI-INAF
I/023/05/0 and I/088/06/0. PT and SB acknowledge the financial support from
the PD51 INFN grant. Cristiano De Boni is thanked for the useful
discussion.

\begin{appendix}

\section{Fit of the deprojected gas density profiles}
We show the electron density profiles, obtained from the deprojection of the 
surface brightness, of the objects in our high$-z$ sample.
We overplot the best-fitted function form in equation~\ref{eq:ngas}
fitted over the radial range $0-R_{\rm spat}$. Both the best-fit parameters
and $R_{\rm spat}$ are quoted in Table~\ref{tab:highz}.

\begin{figure*}
\vspace*{-0.0cm} \hbox{
 \epsfig{figure=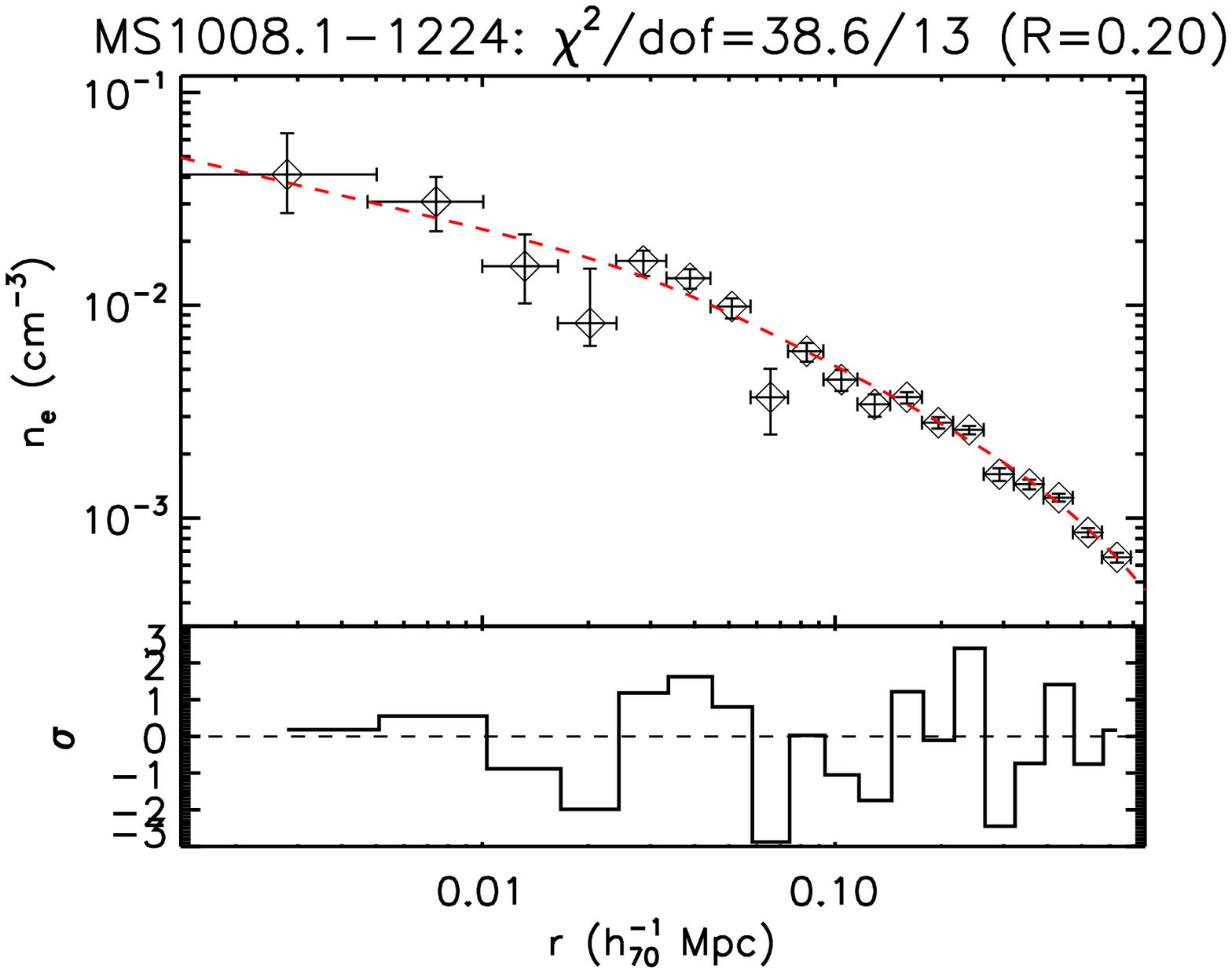,width=0.4\textwidth}
 \epsfig{figure=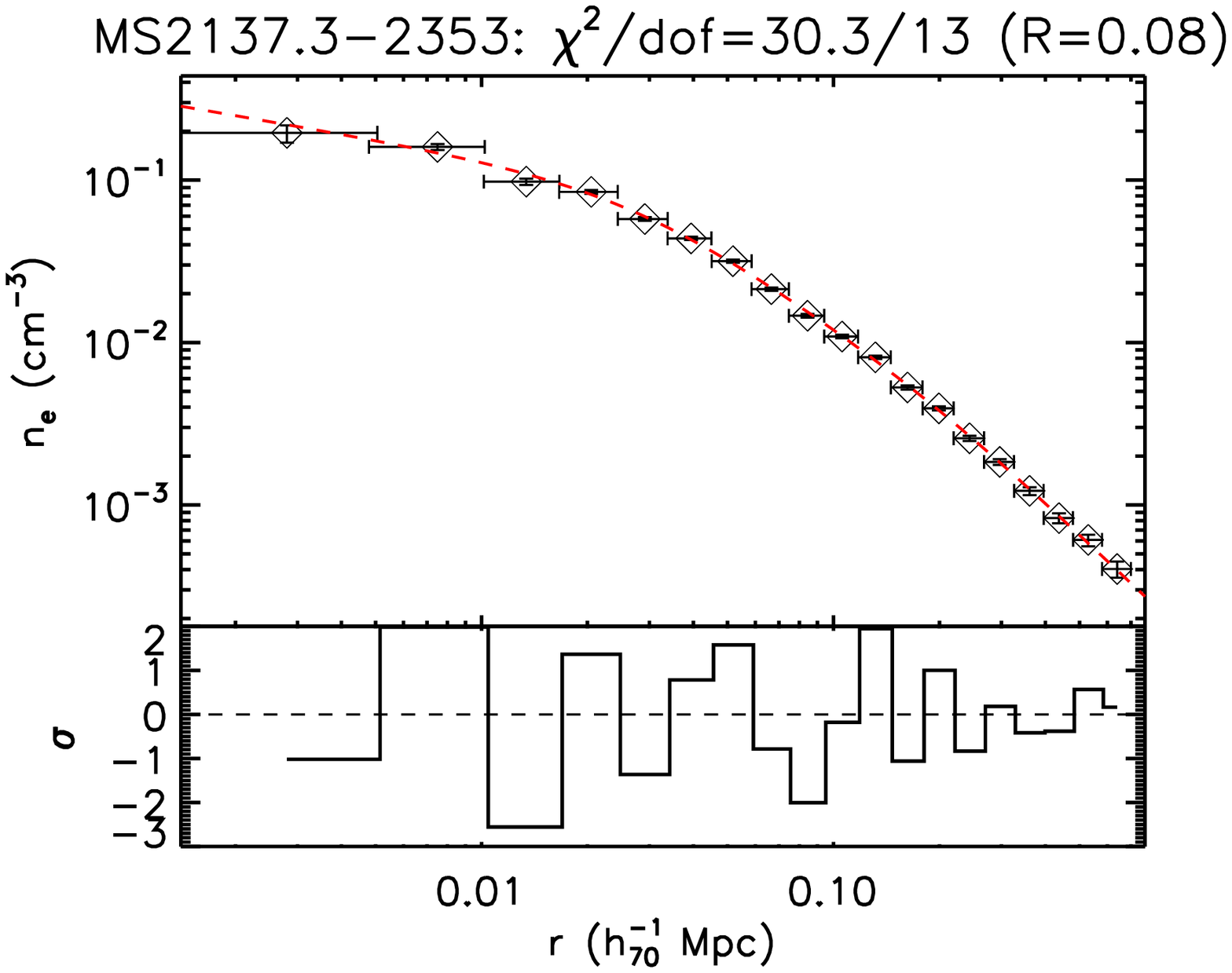,width=0.4\textwidth}
}\vspace*{-0.5cm} \hbox{
 \epsfig{figure=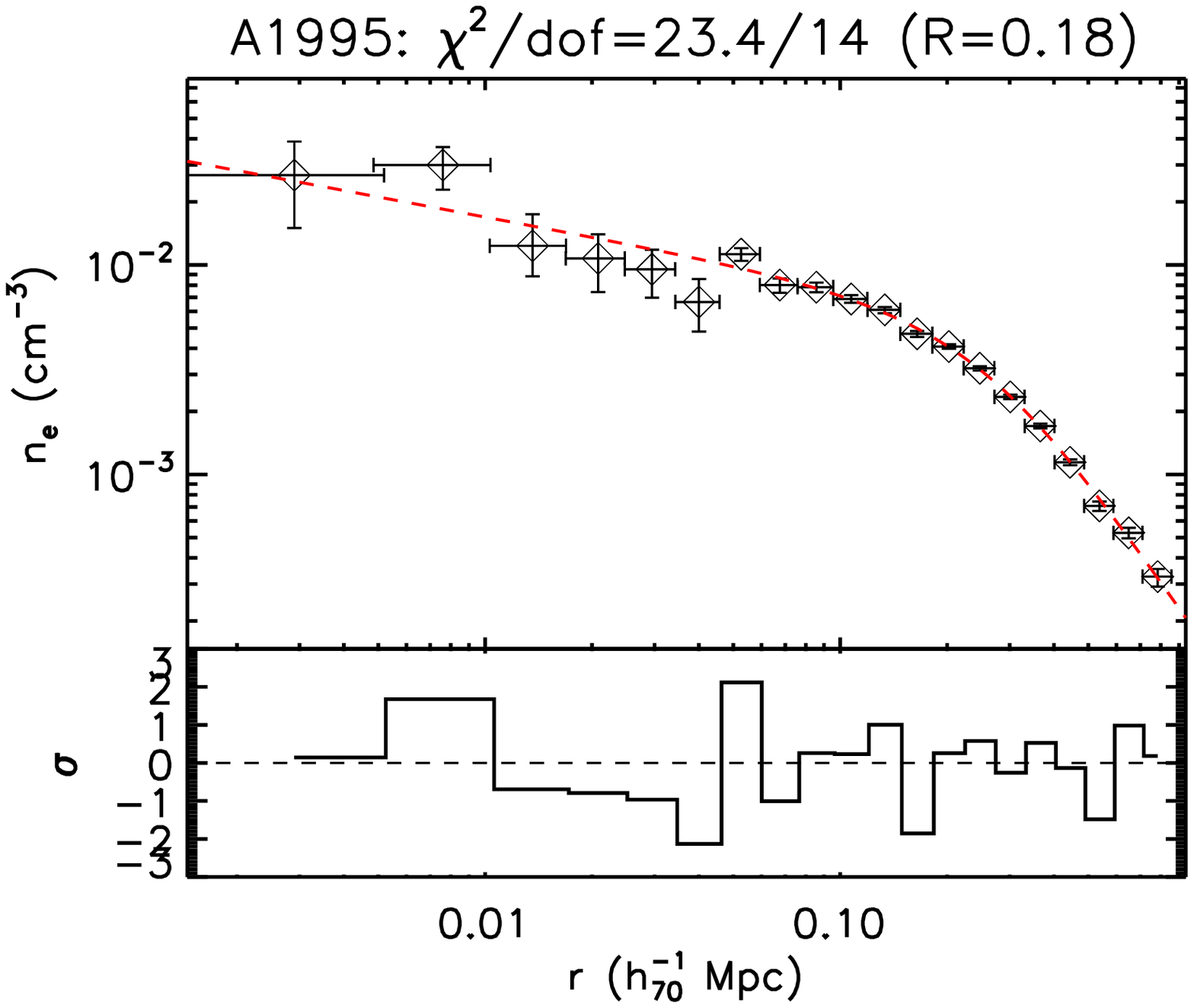,width=0.4\textwidth}
 \epsfig{figure=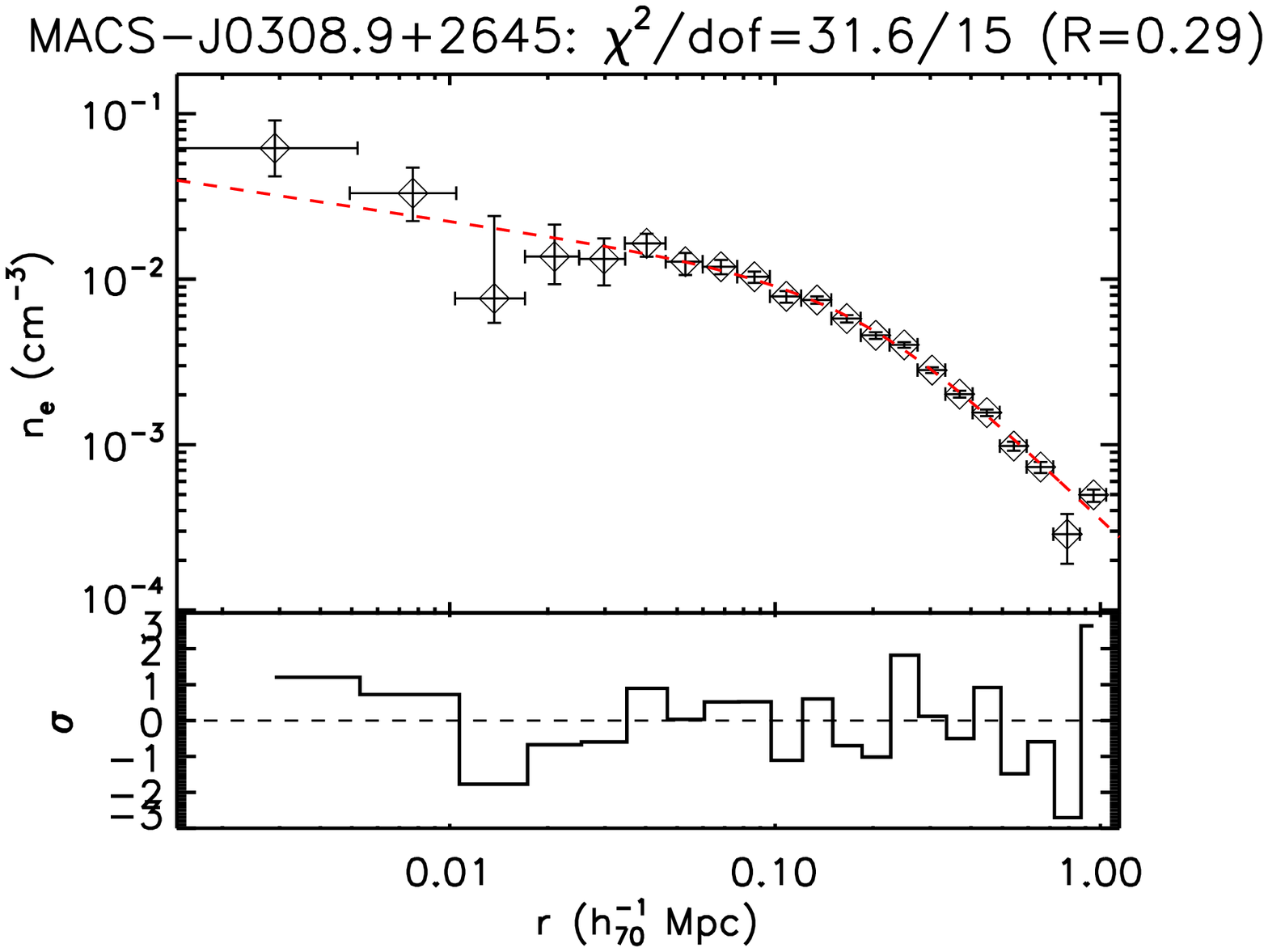,width=0.4\textwidth}
}\vspace*{-0.5cm} \hbox{
 \epsfig{figure=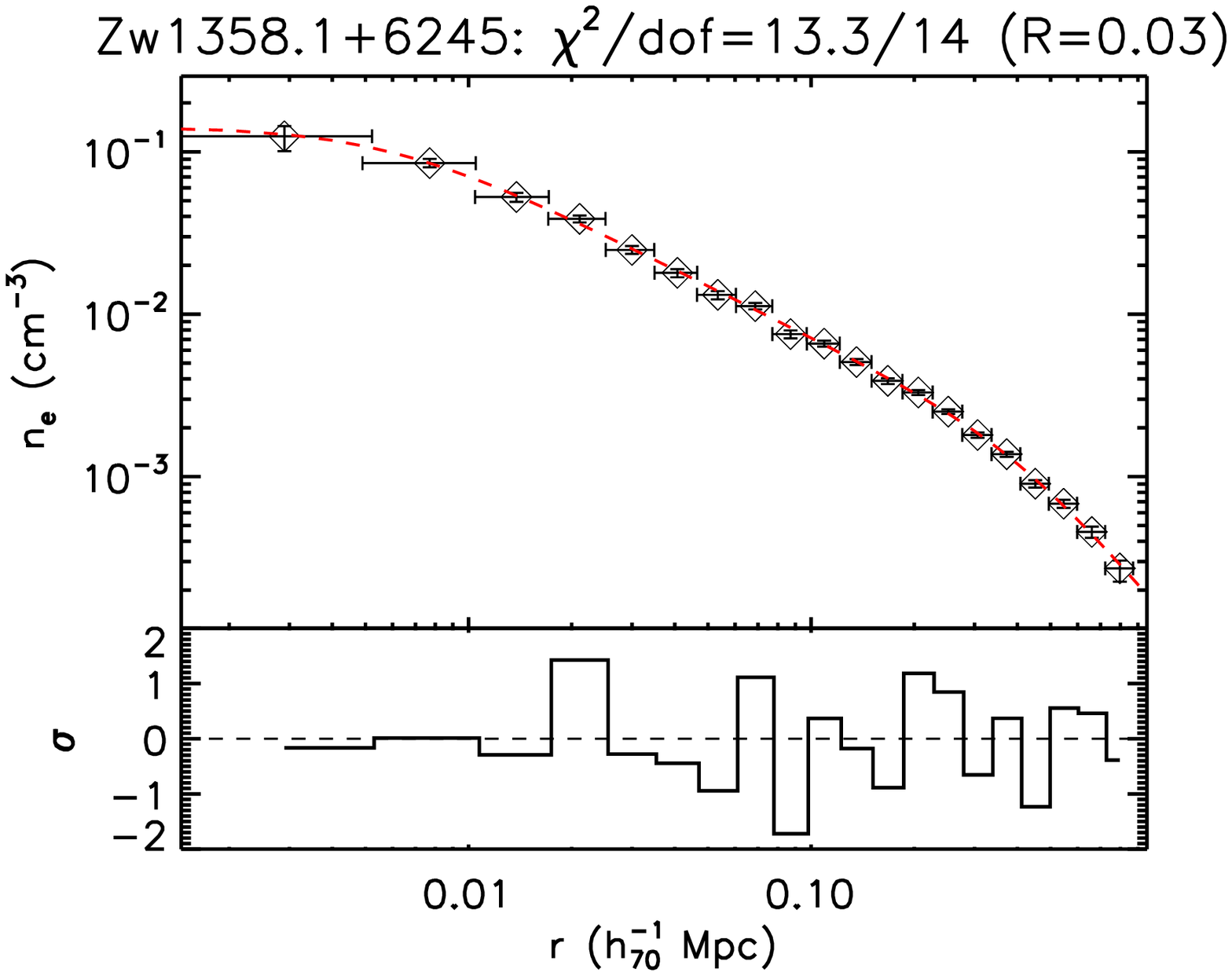,width=0.4\textwidth}
 \epsfig{figure=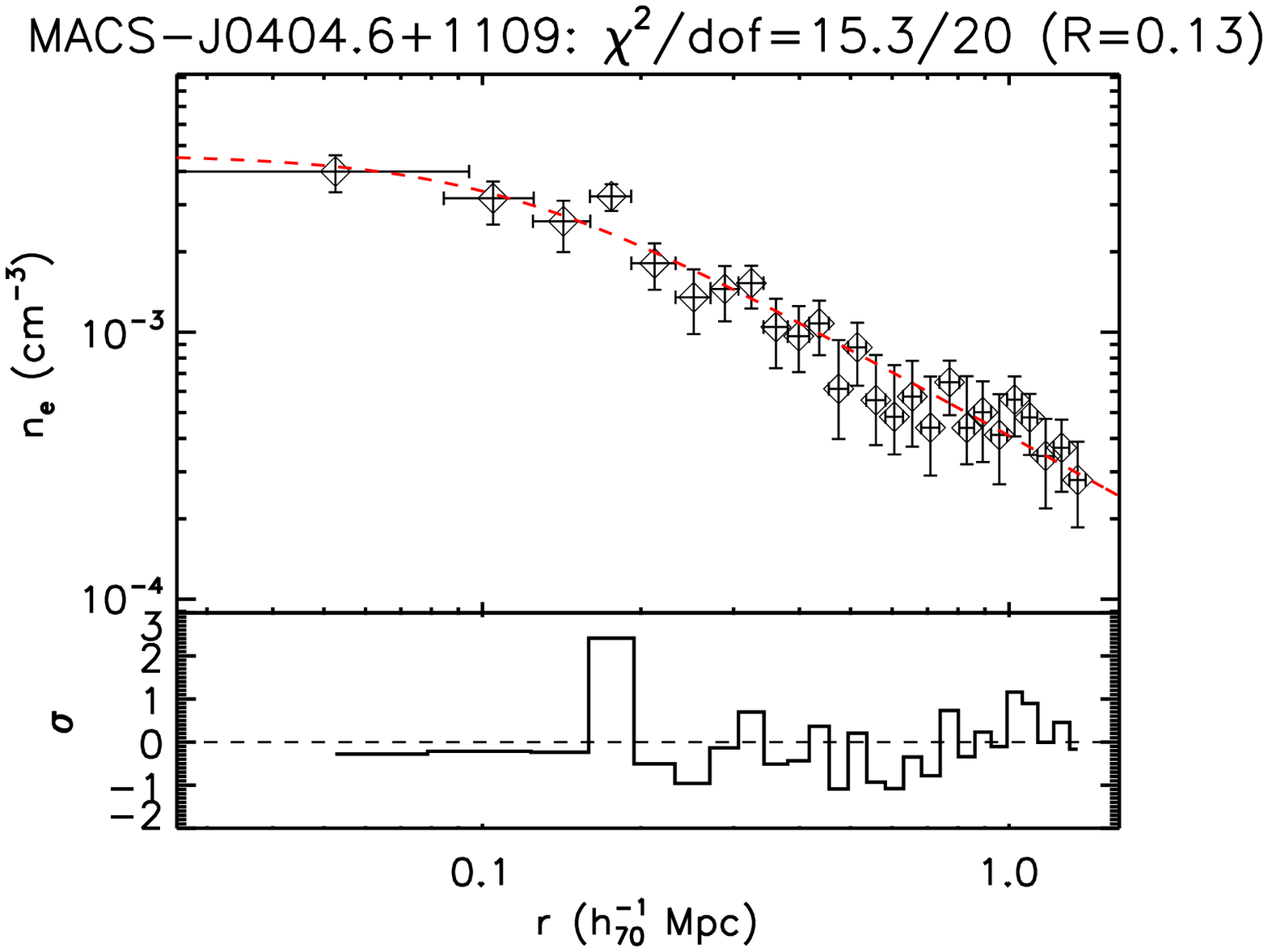,width=0.4\textwidth}
}\vspace*{-0.5cm} \hbox{
 \epsfig{figure=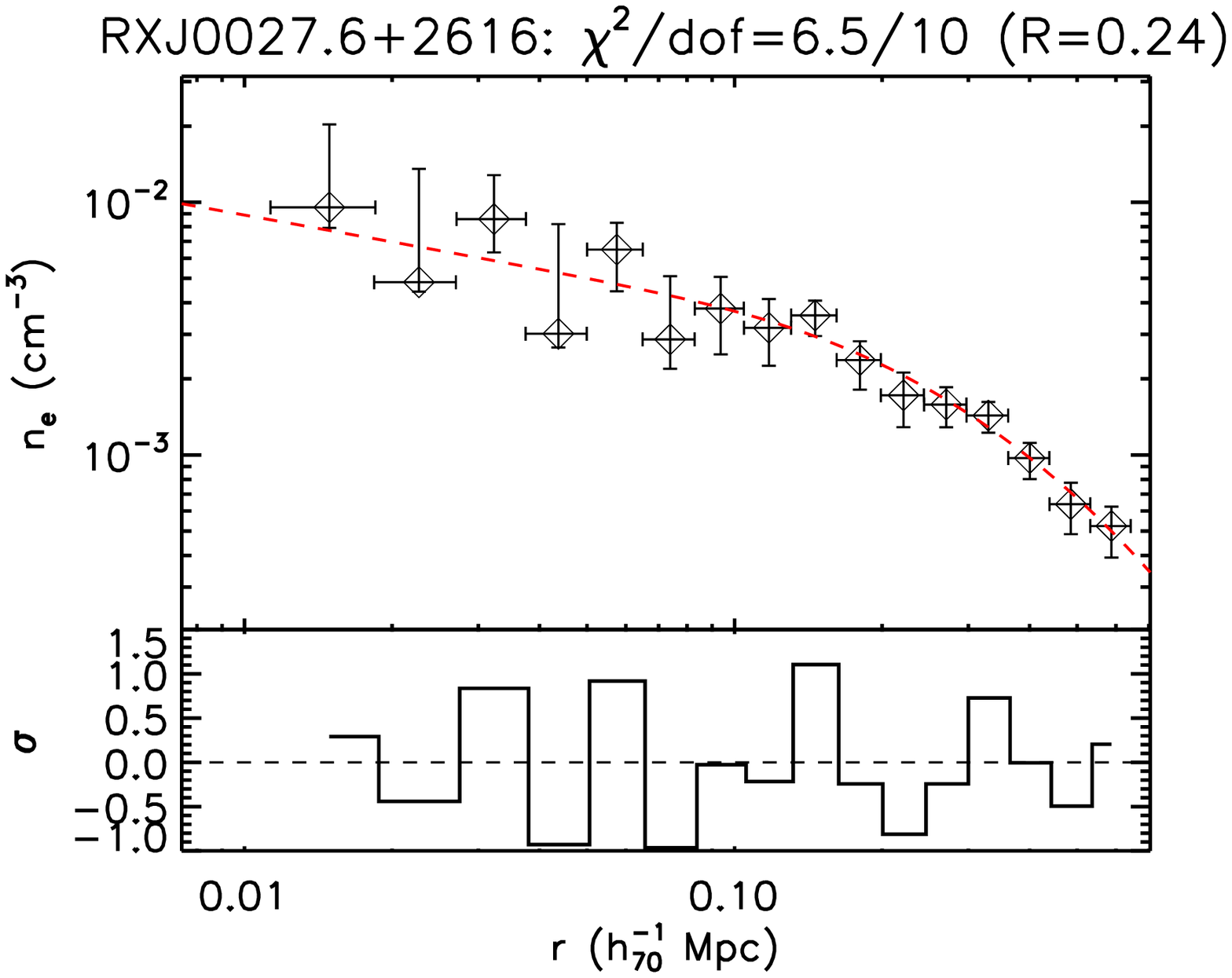,width=0.4\textwidth}
 \epsfig{figure=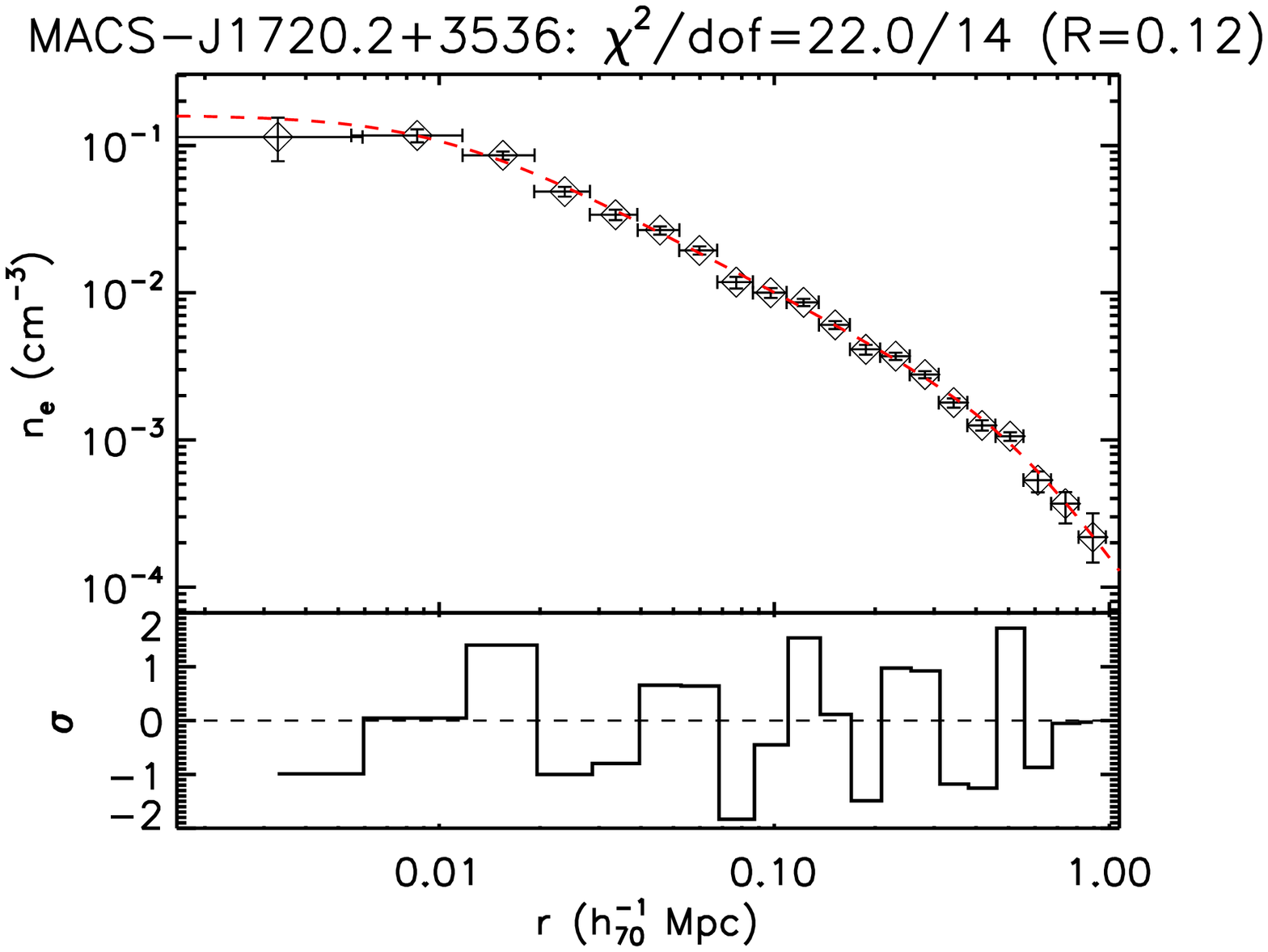,width=0.4\textwidth}
}
\caption{Plots of the deprojected electron density profiles with the
best-fit model in equation~\ref{eq:ngas} overplotted and the
residuals $\sigma_i = (d_i-m_i)/\epsilon_i$ shown in the bottom
panel, where $d_i$, $m_i$ and $\epsilon_i$ are the values of the data,
model and error on the data at a given radius, respectively.
Together with the $\chi^2 = \sum (d_i-m_i)^2/\epsilon_i^2 =
\sum \sigma_i^2$ value, the $R$ factor represents
a measure of the overall residuals and is given by
$R=\sum abs(d_i-m_i) / \sum d_i$.
In 81 per cent of the cases, $R<0.2$.
}
\label{fig:profiles}
\end{figure*}

\begin{figure*}
\vspace*{-0.0cm} \hbox{ 
 \epsfig{figure=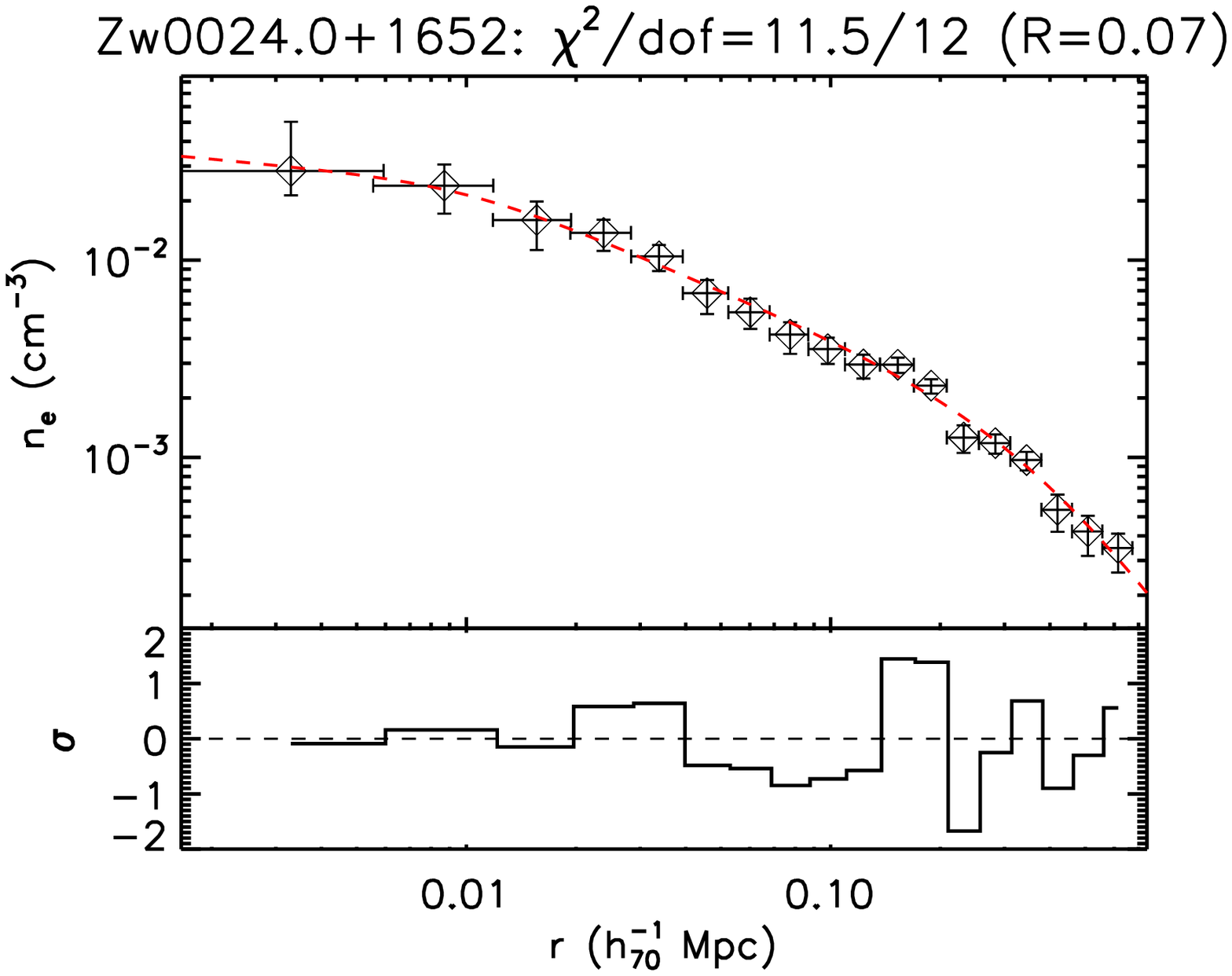,width=0.4\textwidth}
 \epsfig{figure=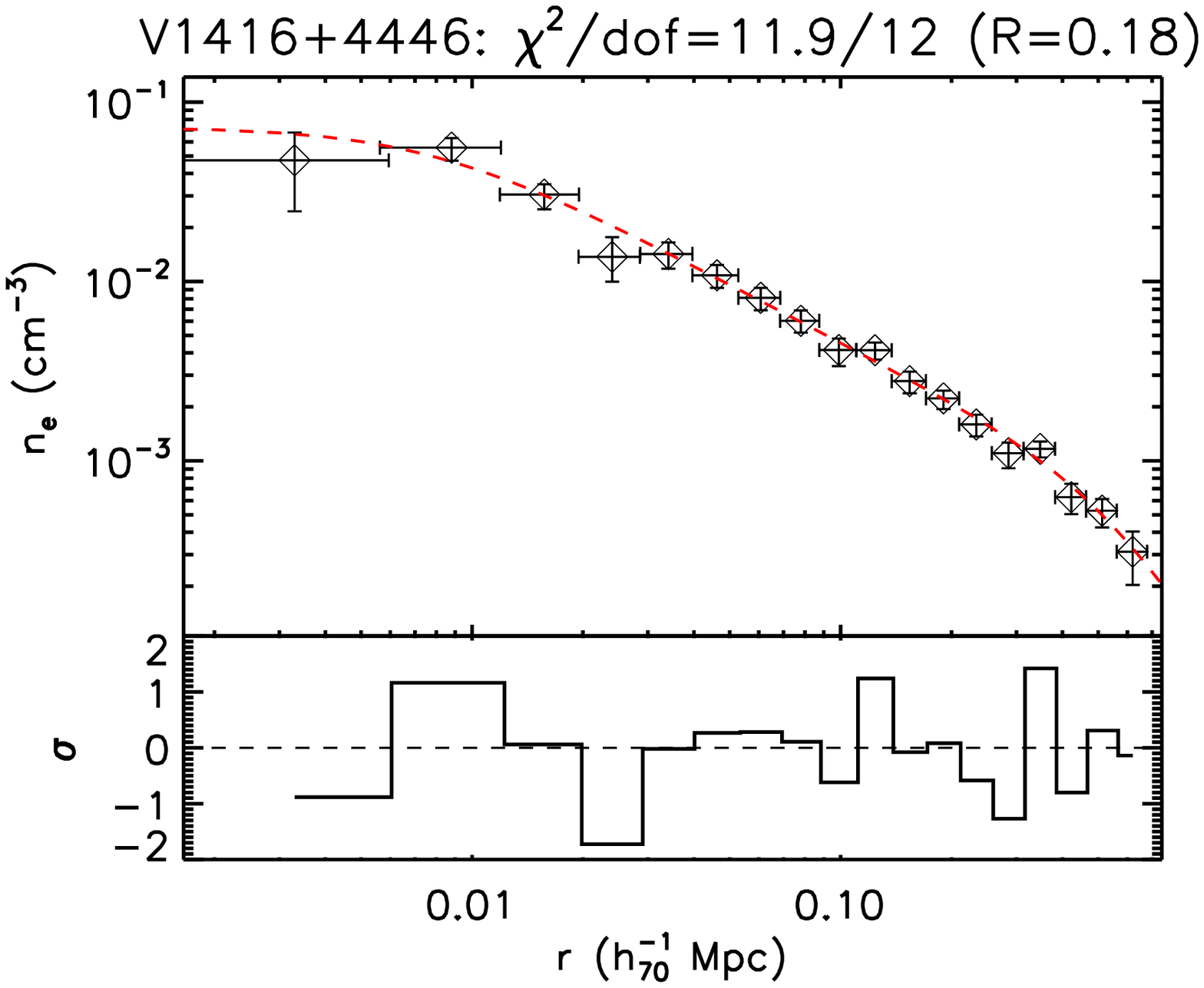,width=0.4\textwidth}
}\vspace*{-0.5cm} \hbox{ 
 \epsfig{figure=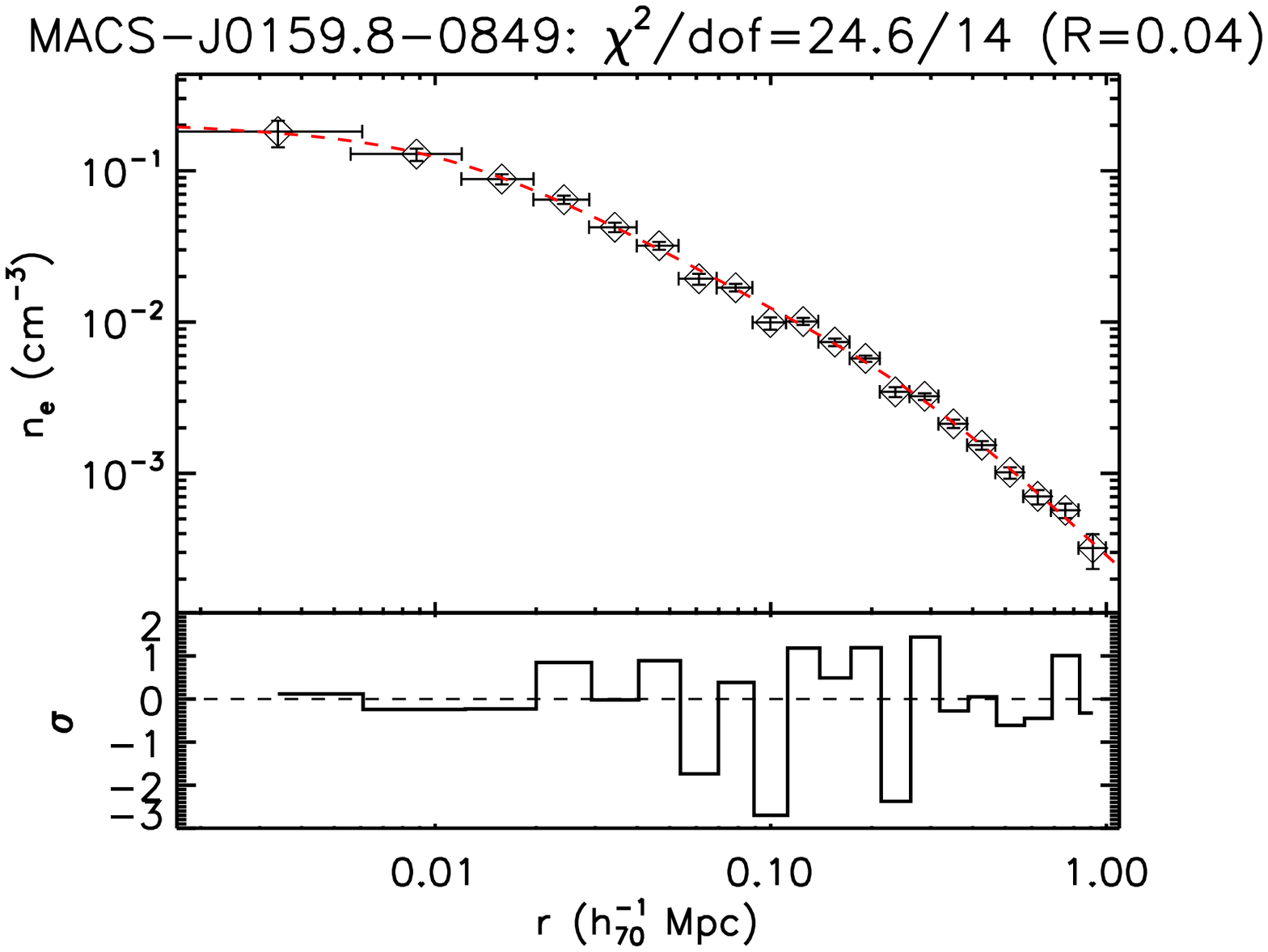,width=0.4\textwidth}
 \epsfig{figure=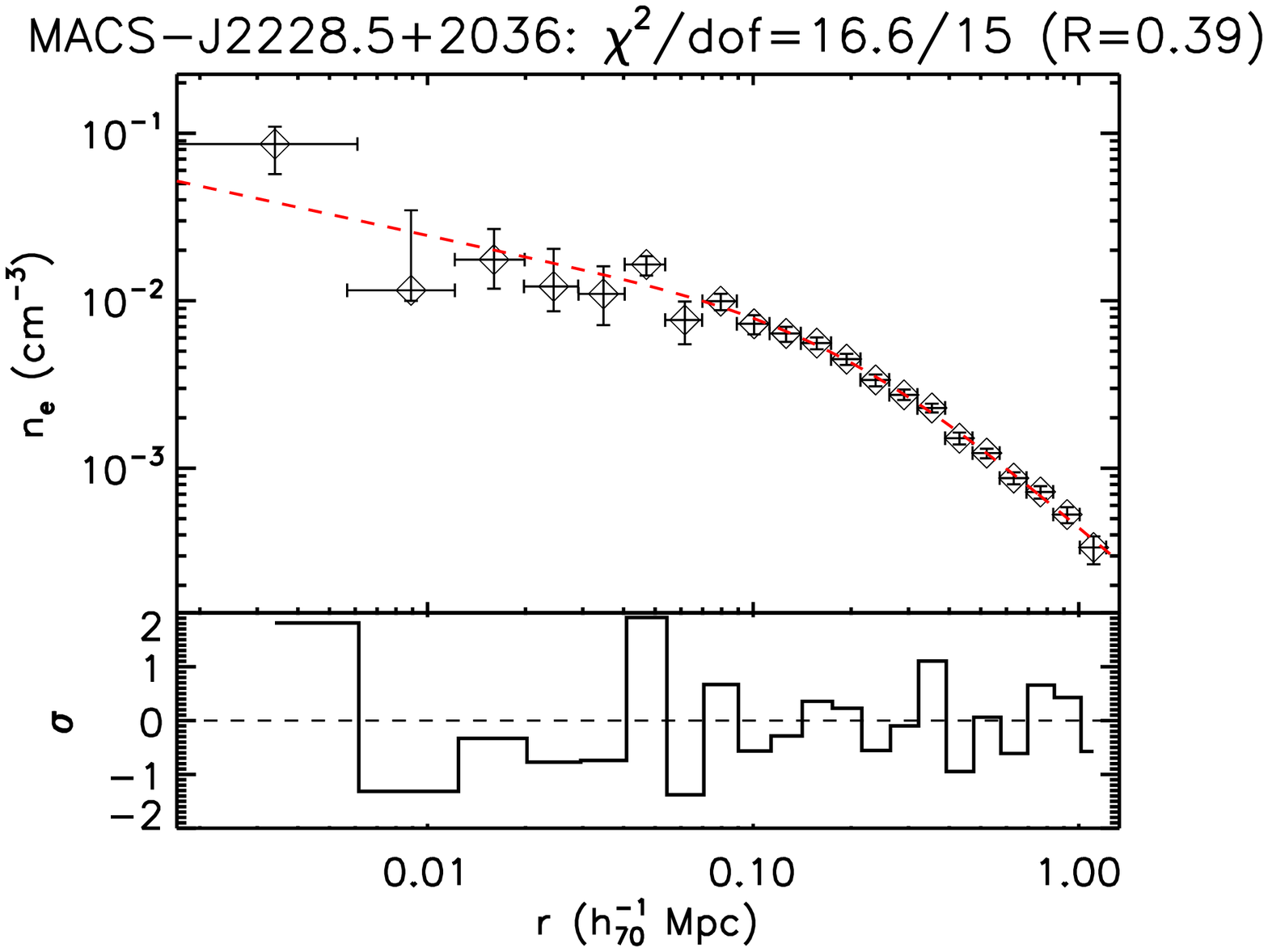,width=0.4\textwidth}
}\vspace*{-0.5cm} \hbox{ 
 \epsfig{figure=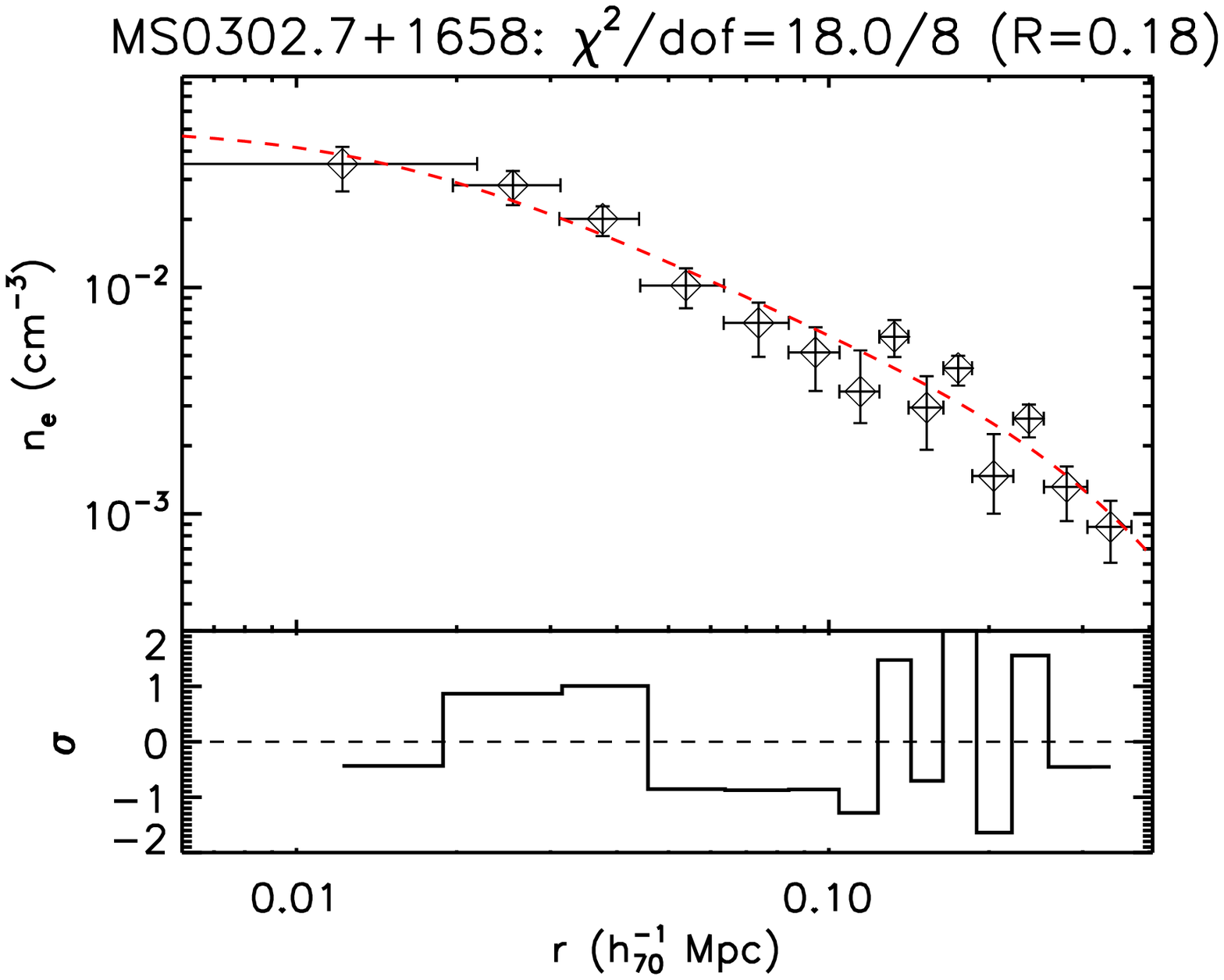,width=0.4\textwidth}
 \epsfig{figure=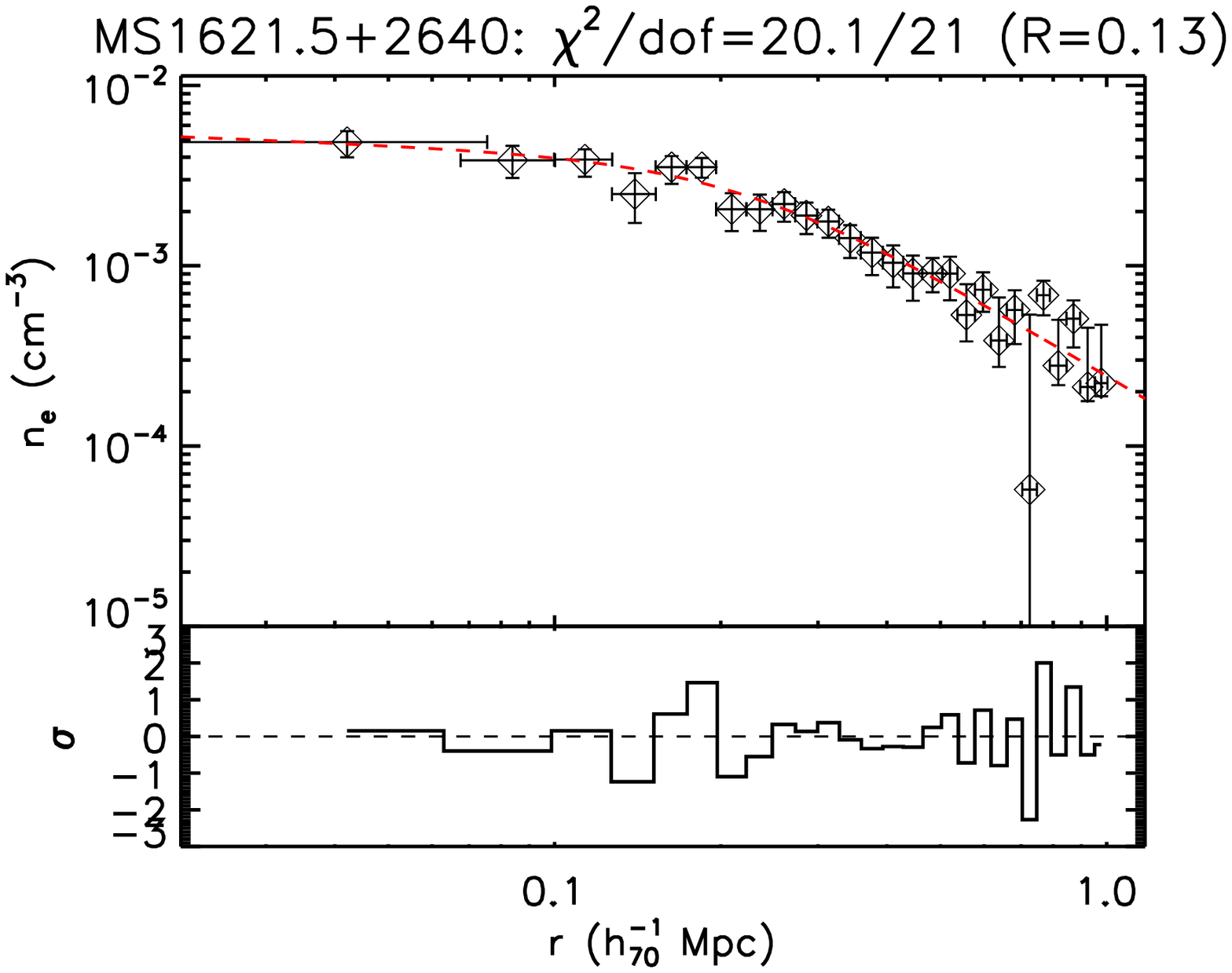,width=0.4\textwidth}
}\vspace*{-0.5cm} \hbox{ 
 \epsfig{figure=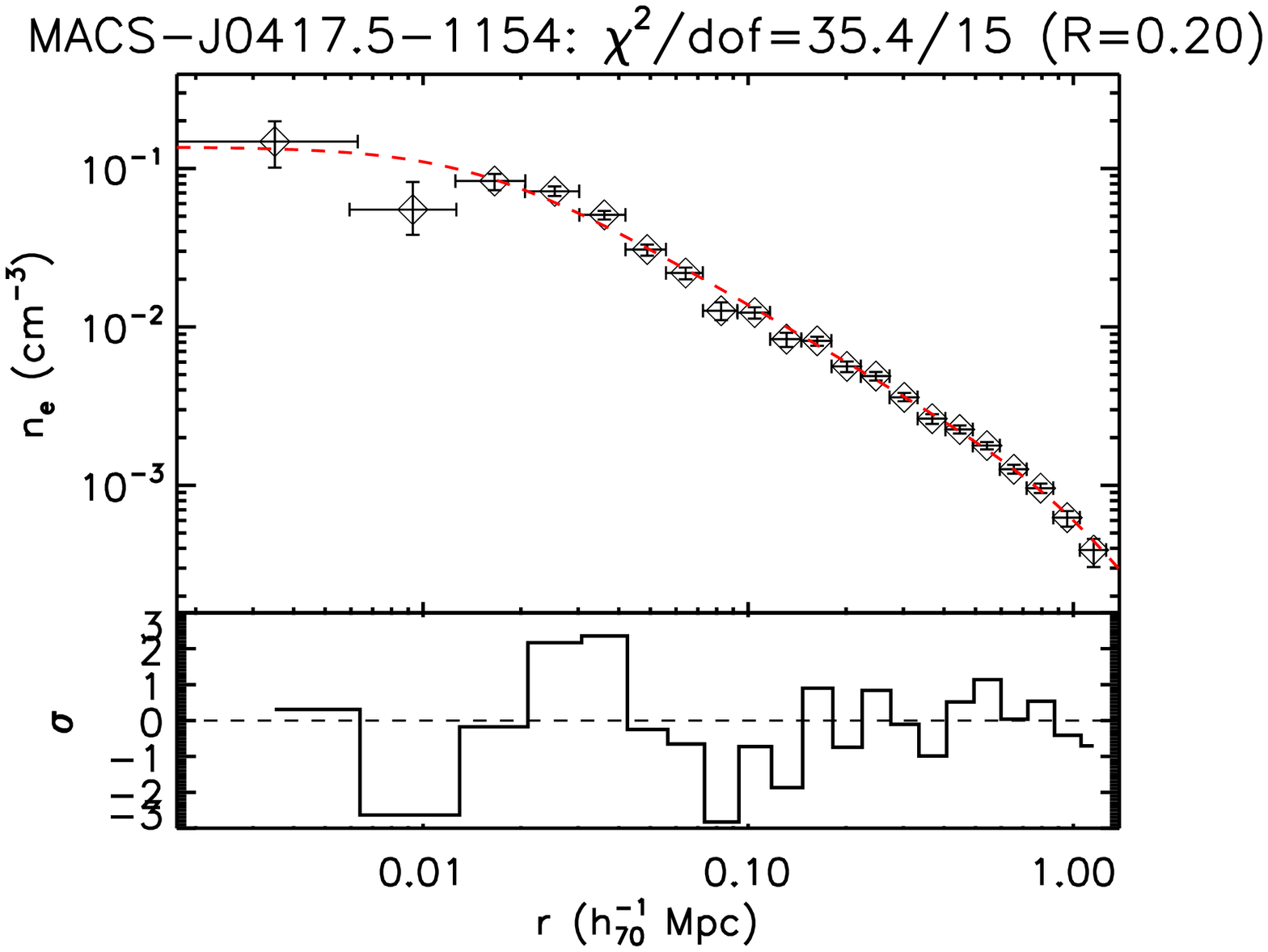,width=0.4\textwidth}
 \epsfig{figure=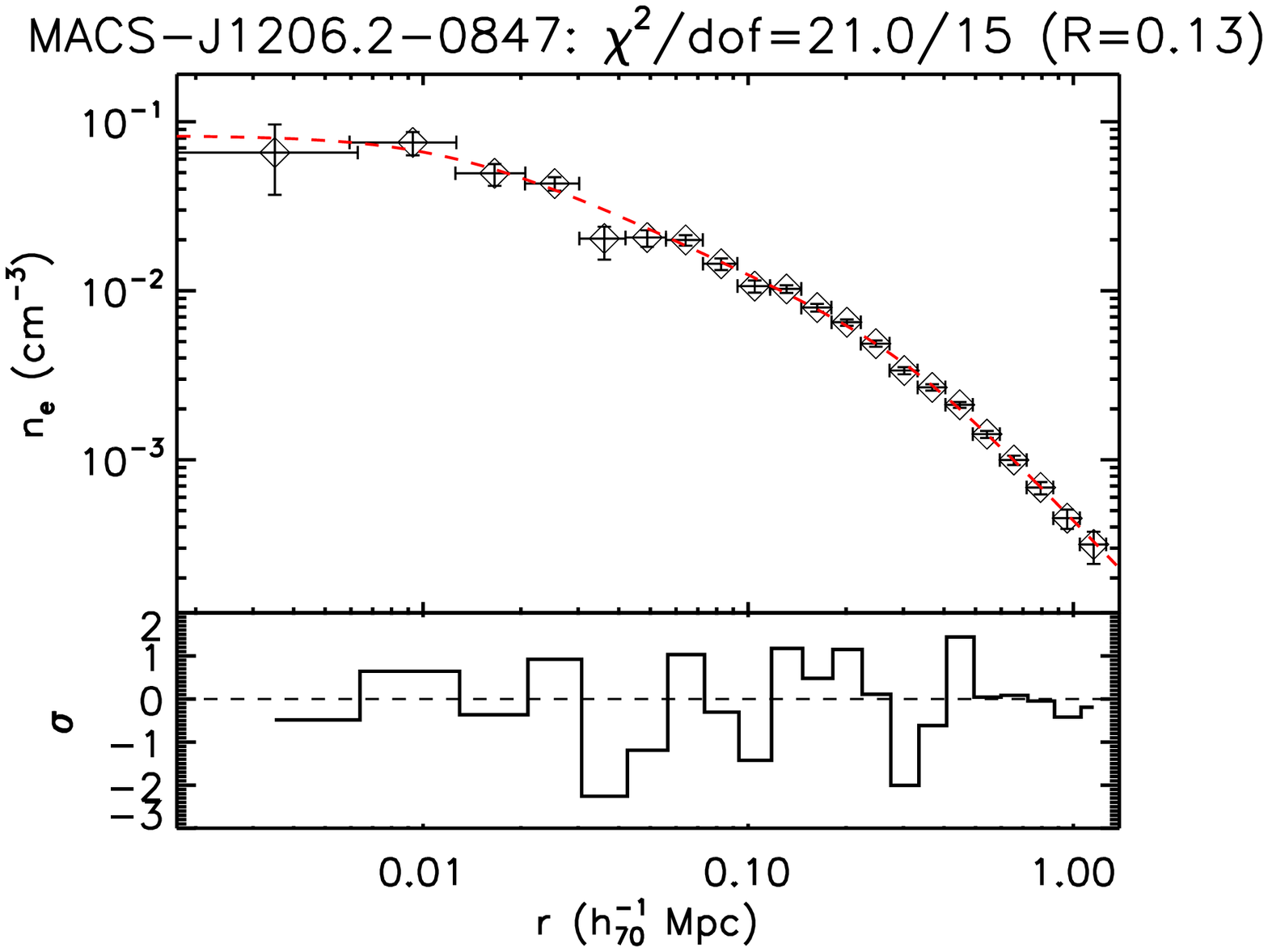,width=0.4\textwidth}
}
\end{figure*}

\begin{figure*}
\vspace*{-0.0cm} \hbox{ 
 \epsfig{figure=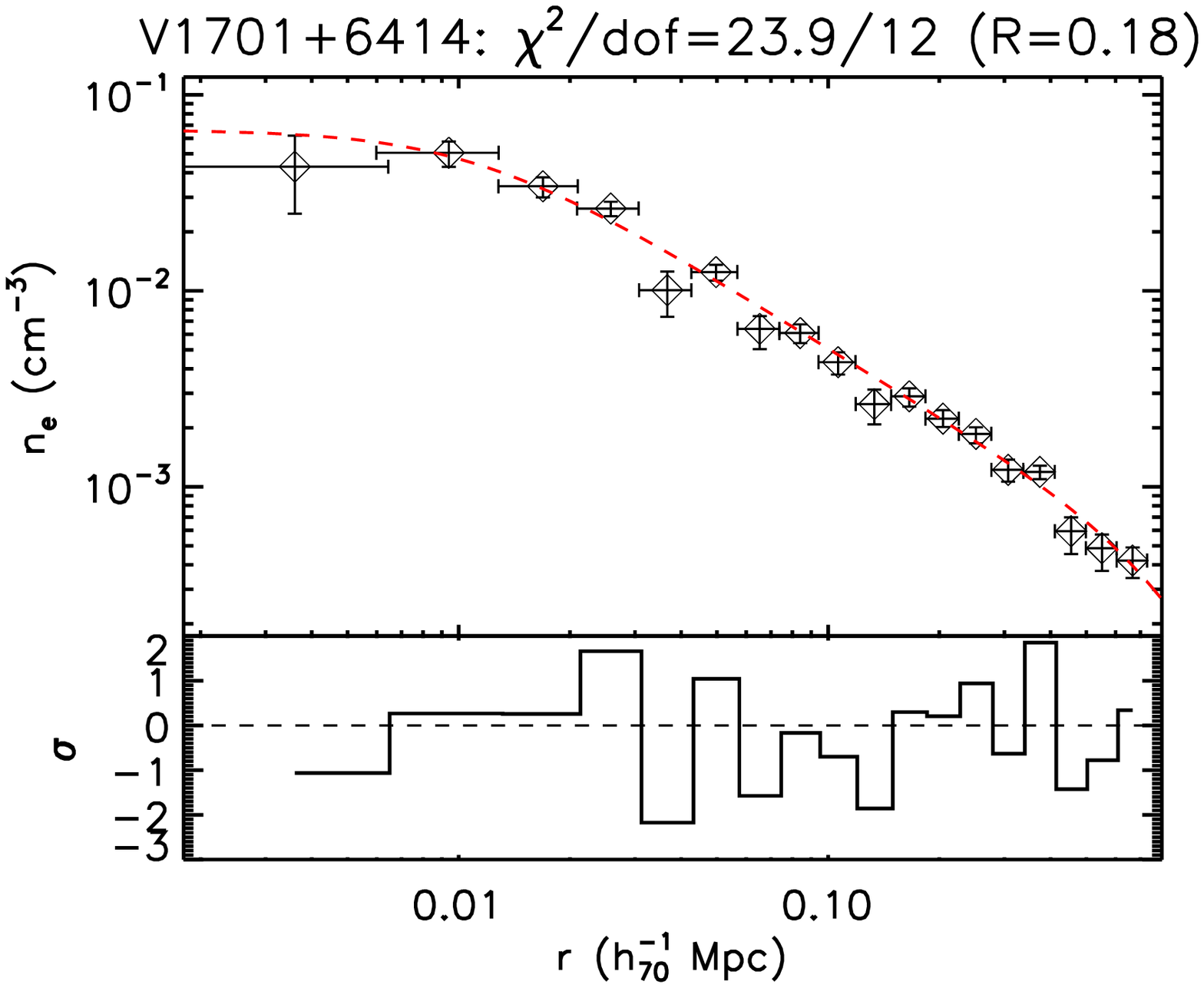,width=0.4\textwidth}
 \epsfig{figure=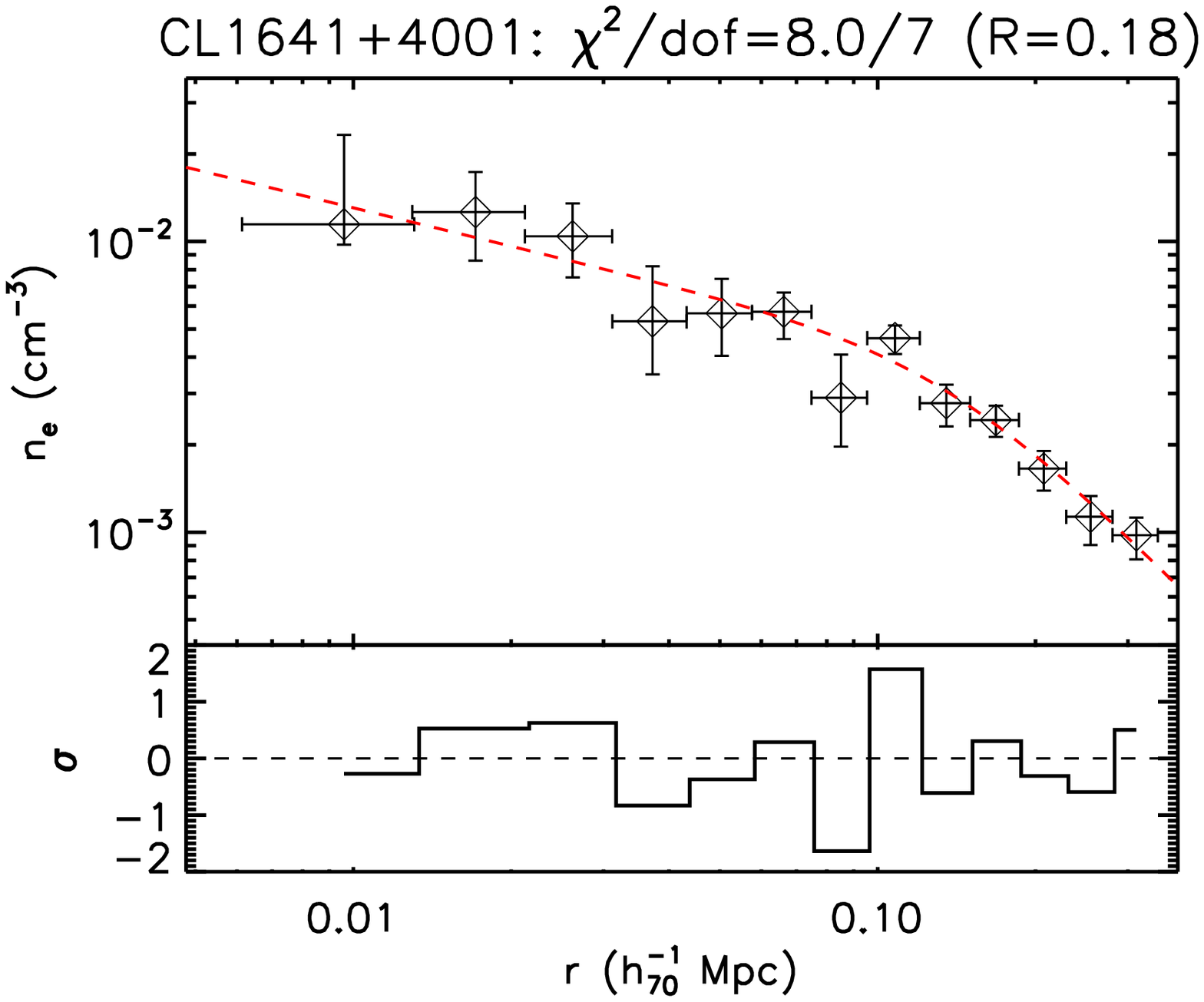,width=0.4\textwidth}
} \vspace*{-0.5cm} \hbox{ 
 \epsfig{figure=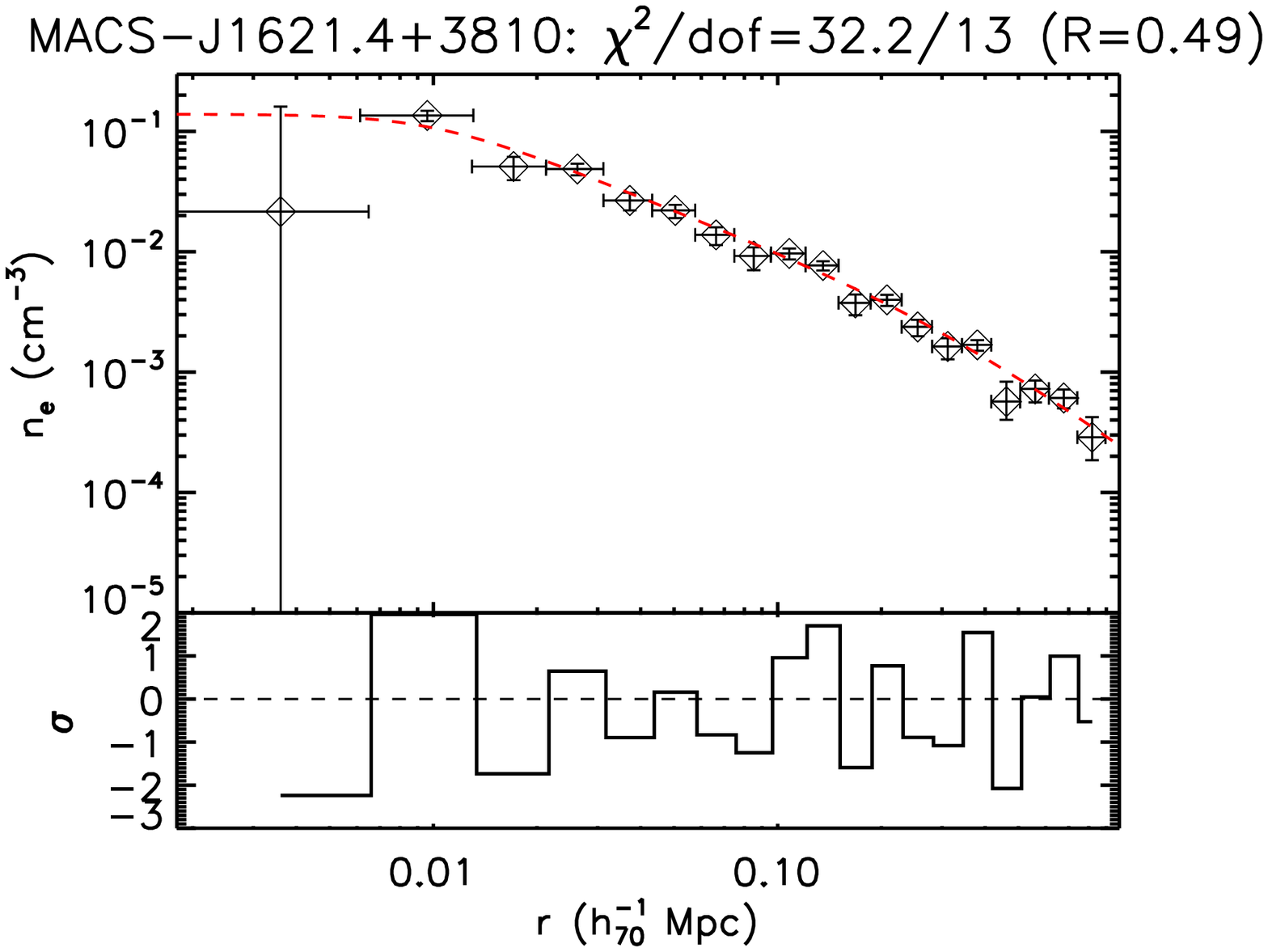,width=0.4\textwidth}
 \epsfig{figure=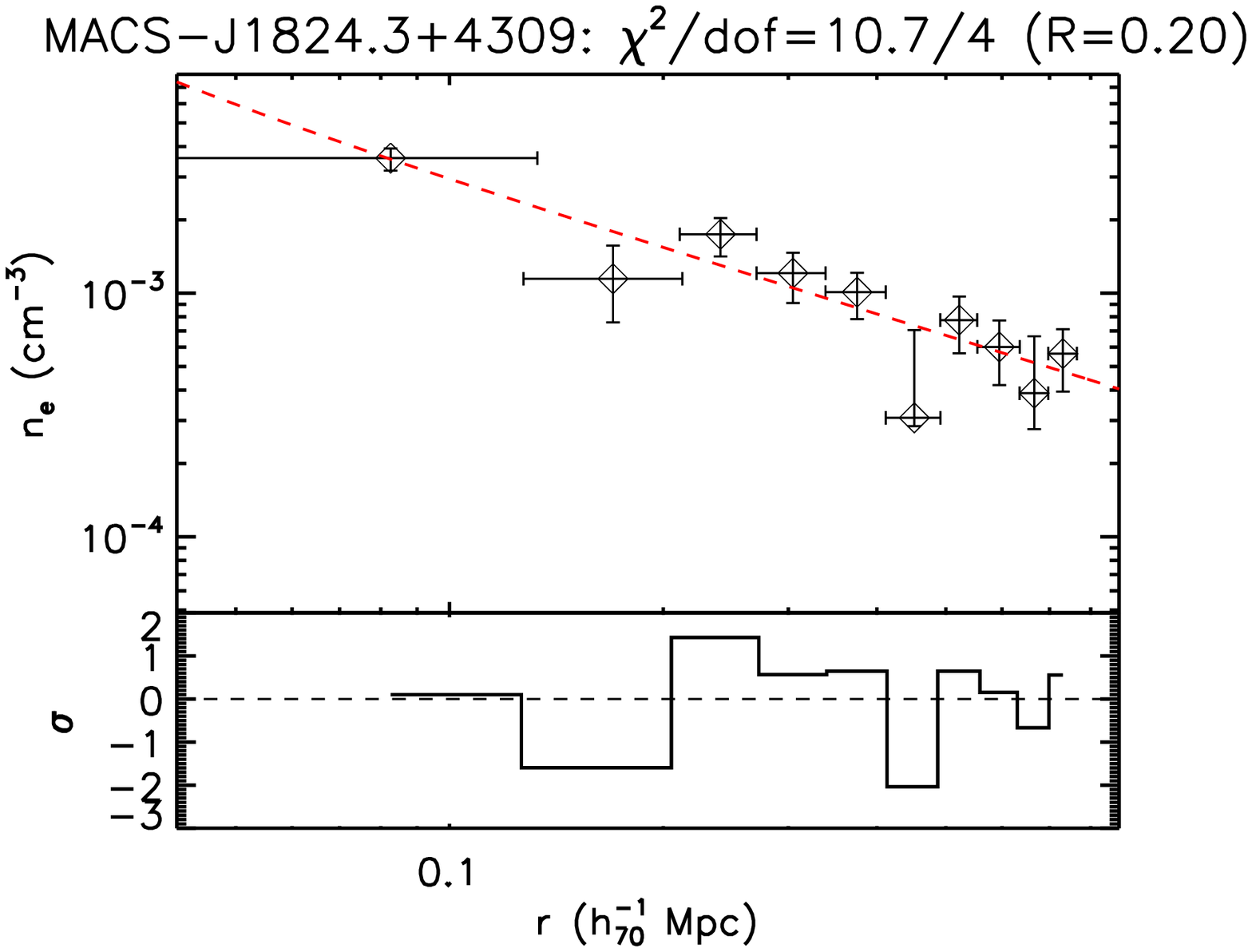,width=0.4\textwidth}
} \vspace*{-0.5cm} \hbox{ 
 \epsfig{figure=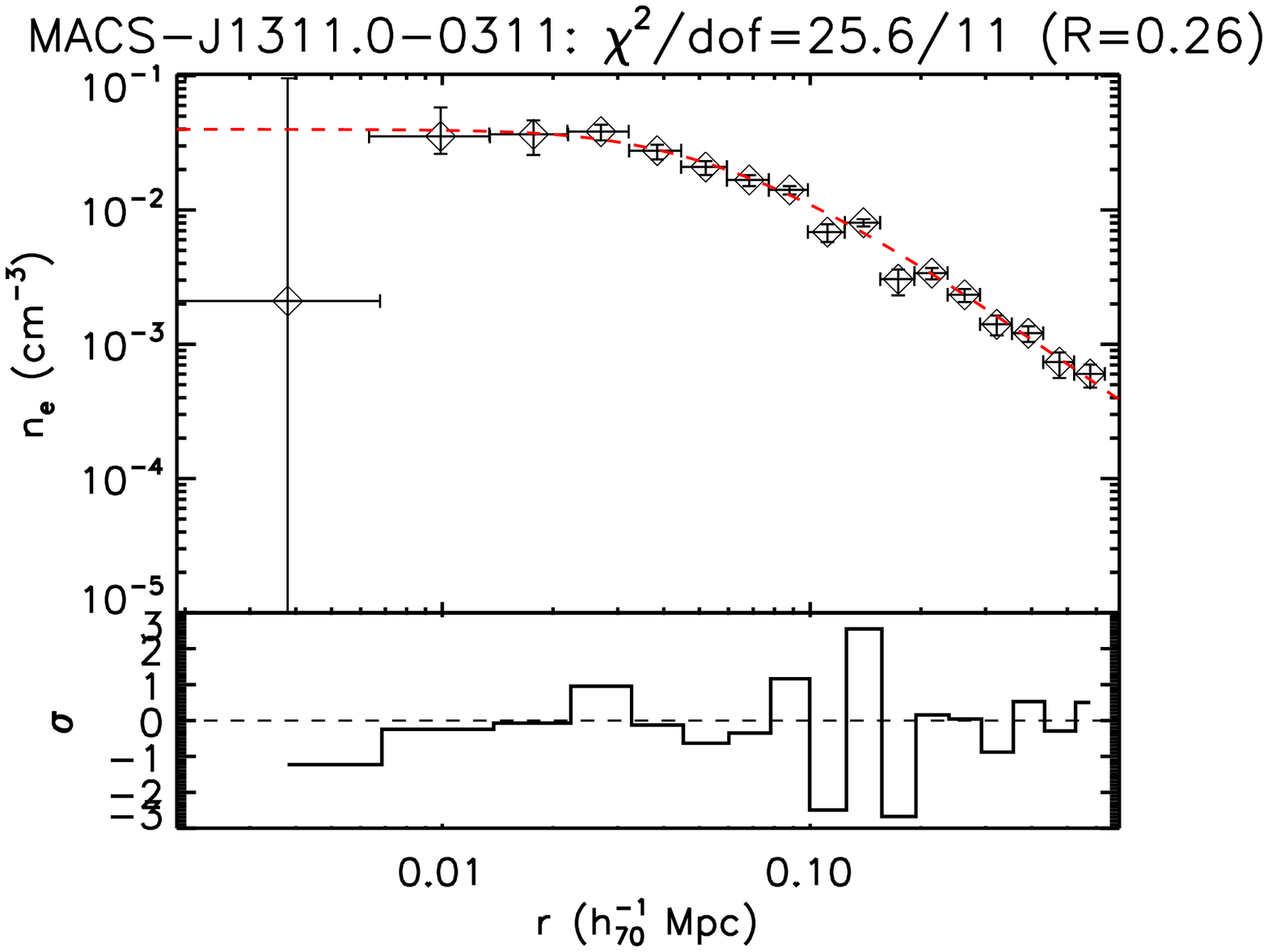,width=0.4\textwidth}
 \epsfig{figure=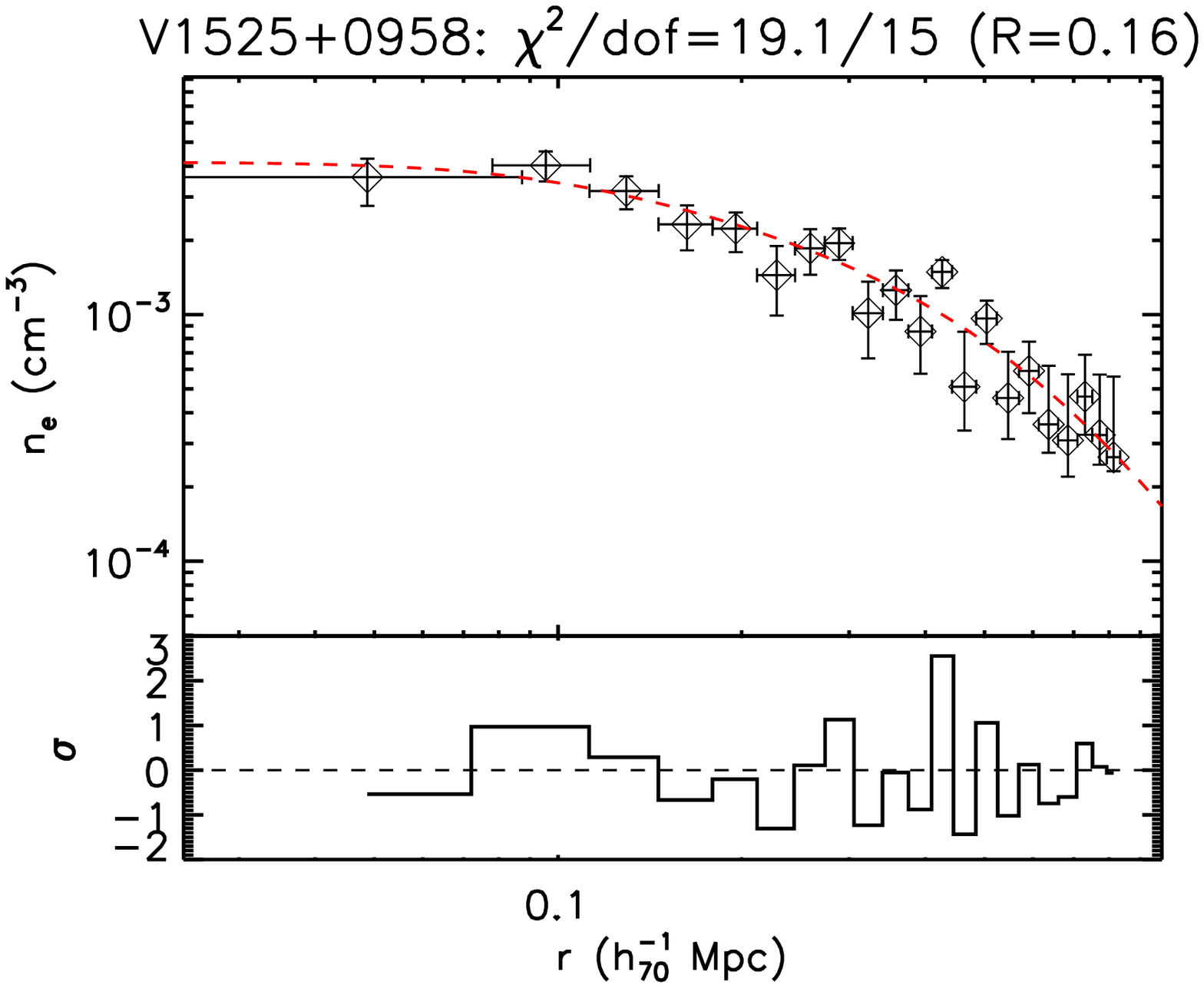,width=0.4\textwidth}
} \vspace*{-0.5cm} \hbox{ 
 \epsfig{figure=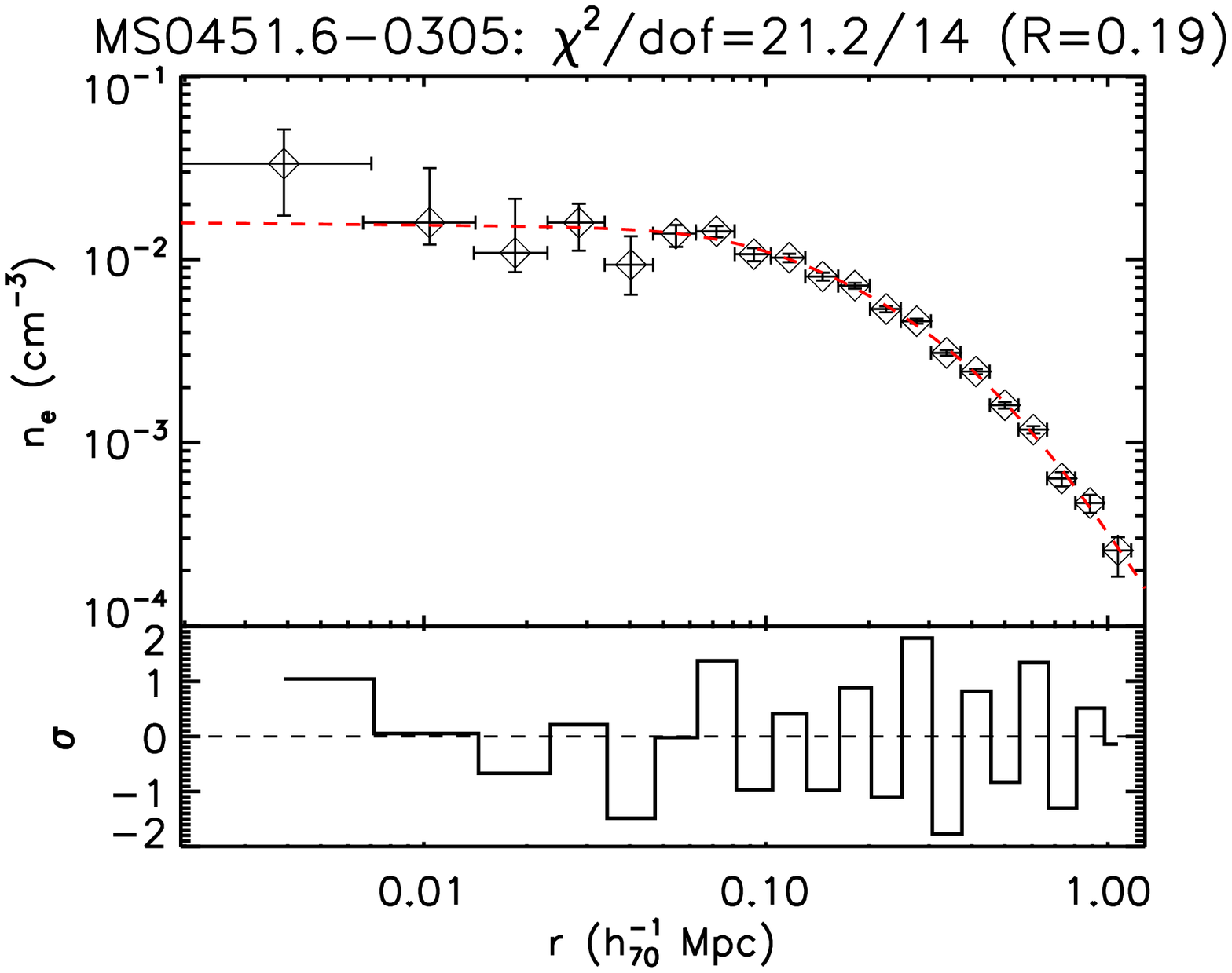,width=0.4\textwidth}
 \epsfig{figure=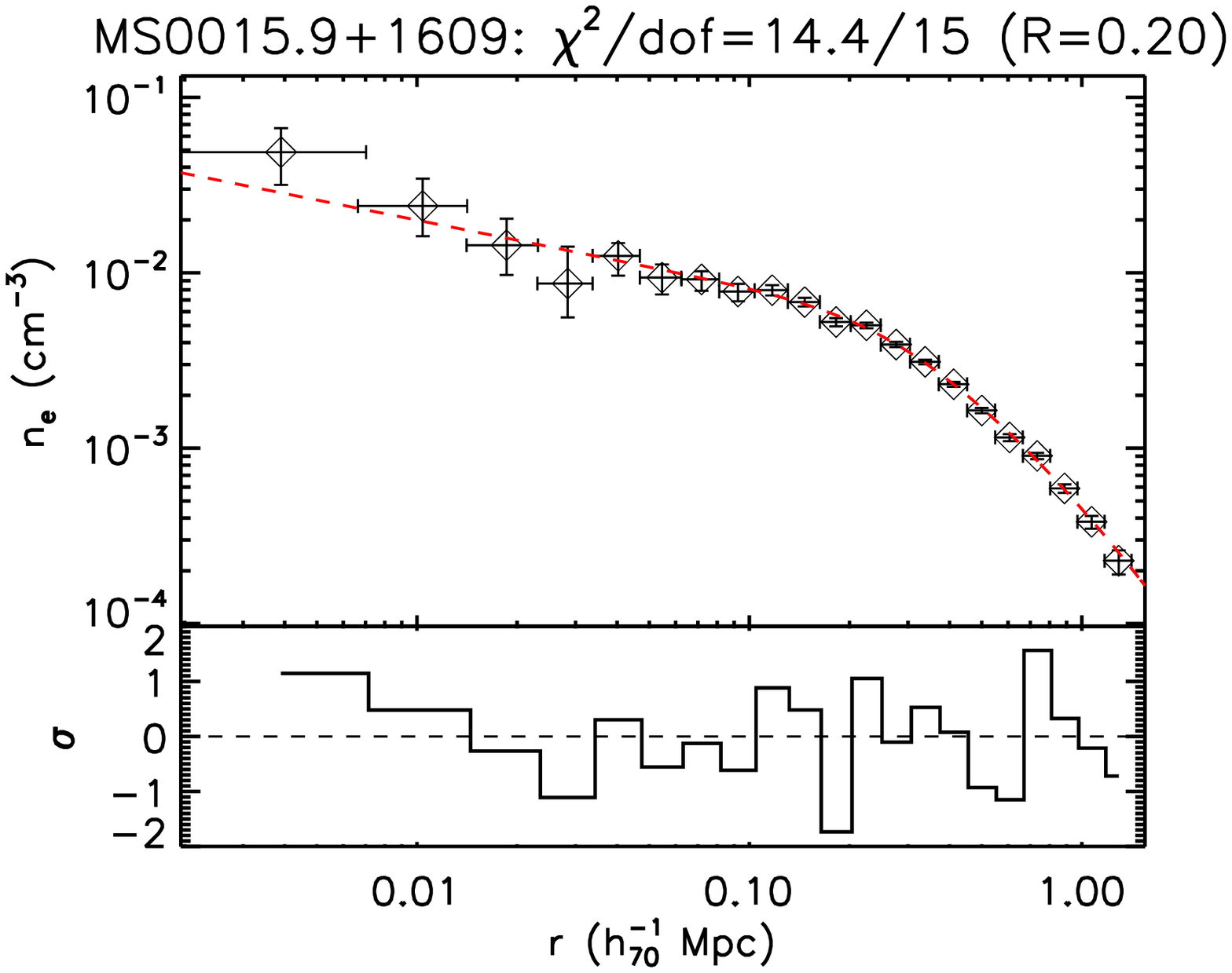,width=0.4\textwidth}
}
\end{figure*}

\begin{figure*}
\vspace*{-0.0cm} \hbox{
 \epsfig{figure=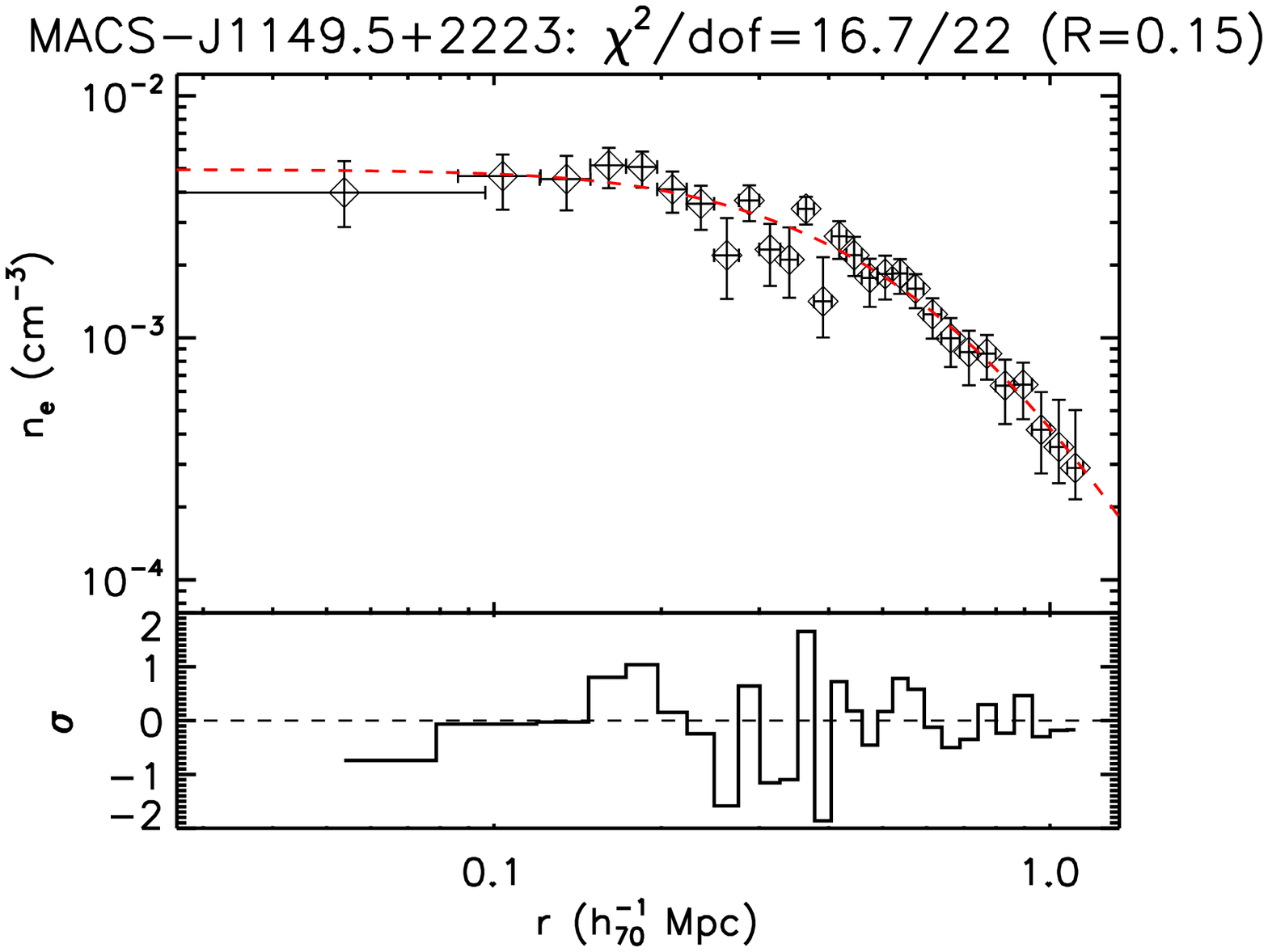,width=0.4\textwidth}
 \epsfig{figure=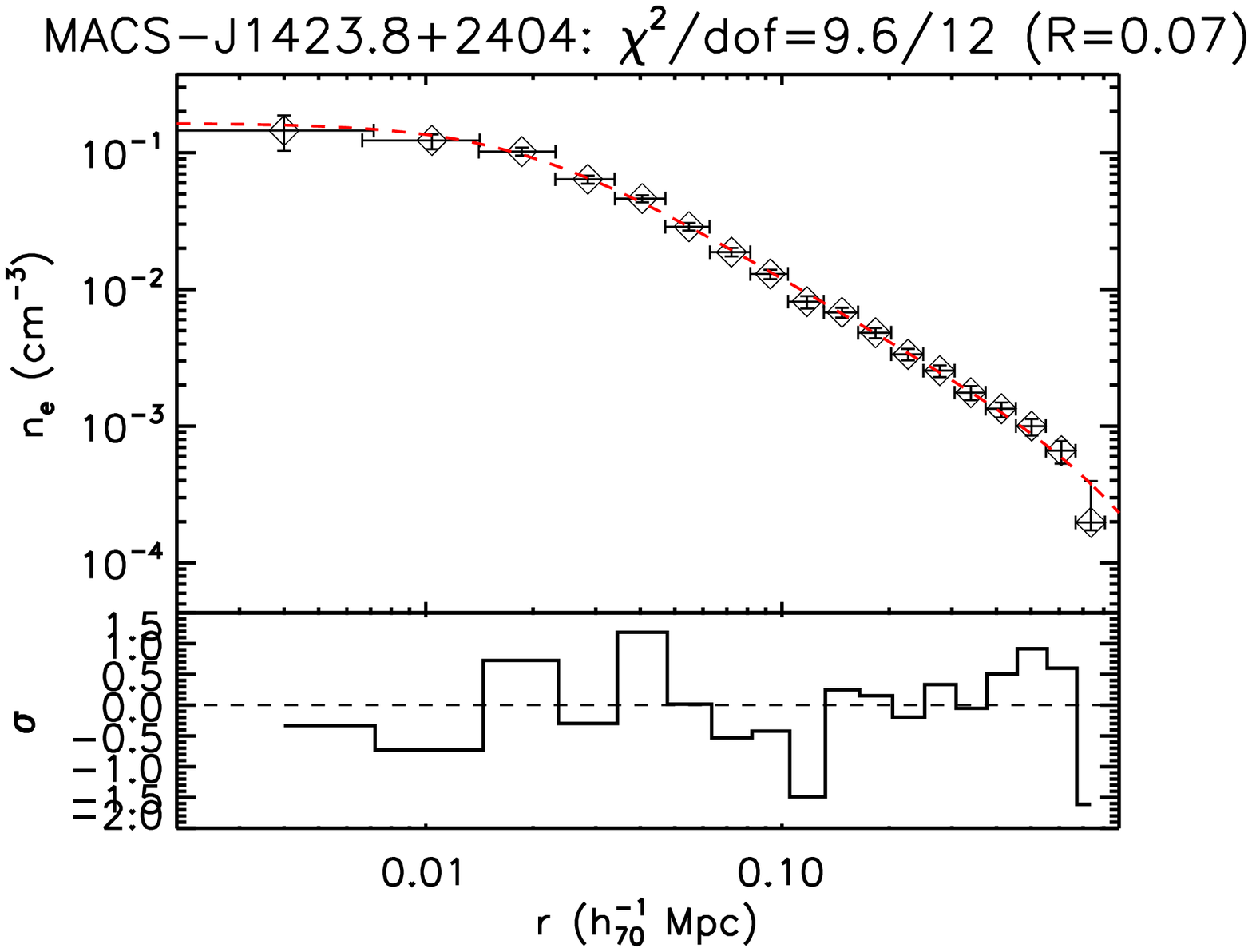,width=0.4\textwidth}
} \vspace*{-0.5cm} \hbox{
 \epsfig{figure=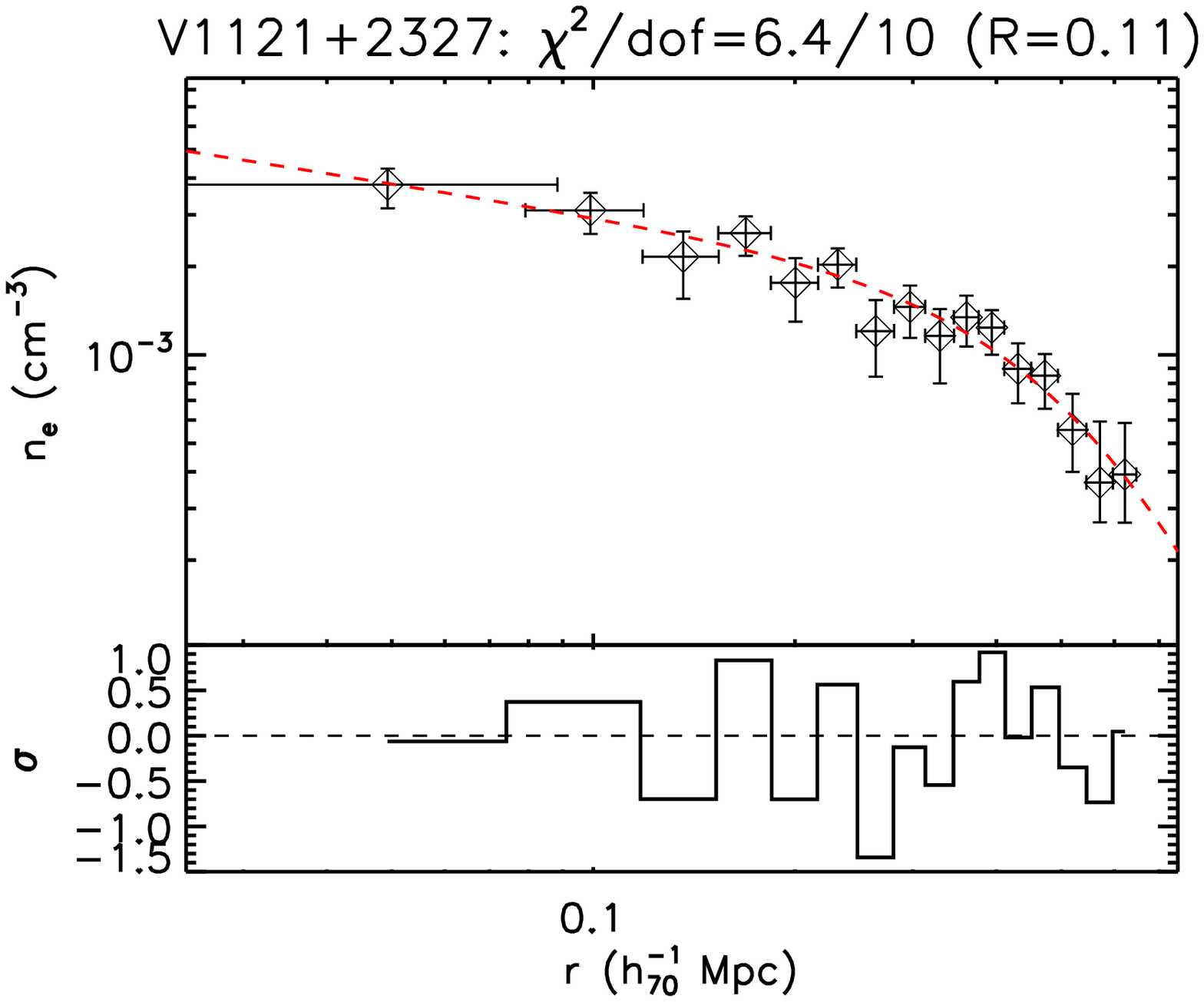,width=0.4\textwidth}
 \epsfig{figure=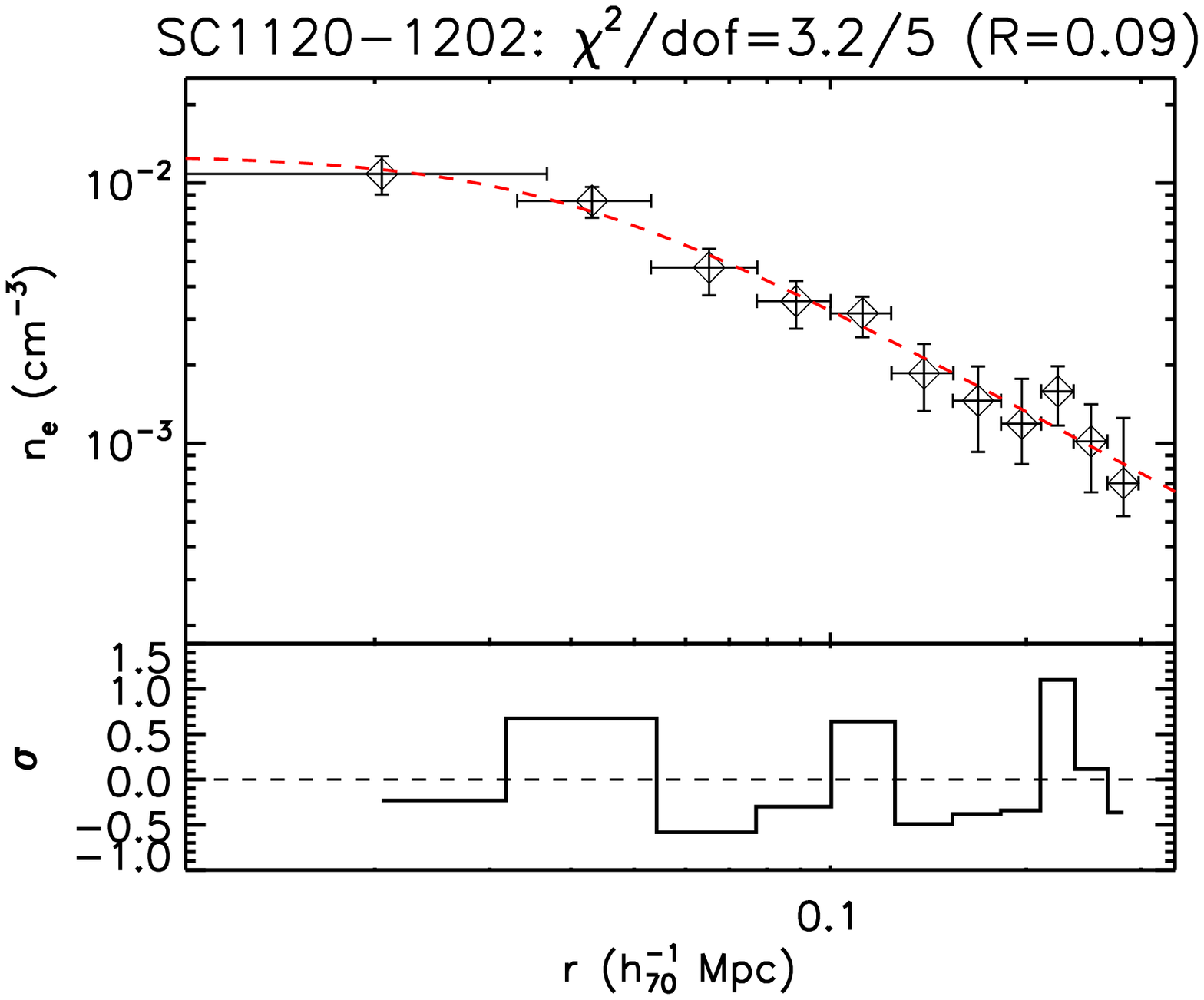,width=0.4\textwidth}
} \vspace*{-0.5cm} \hbox{
 \epsfig{figure=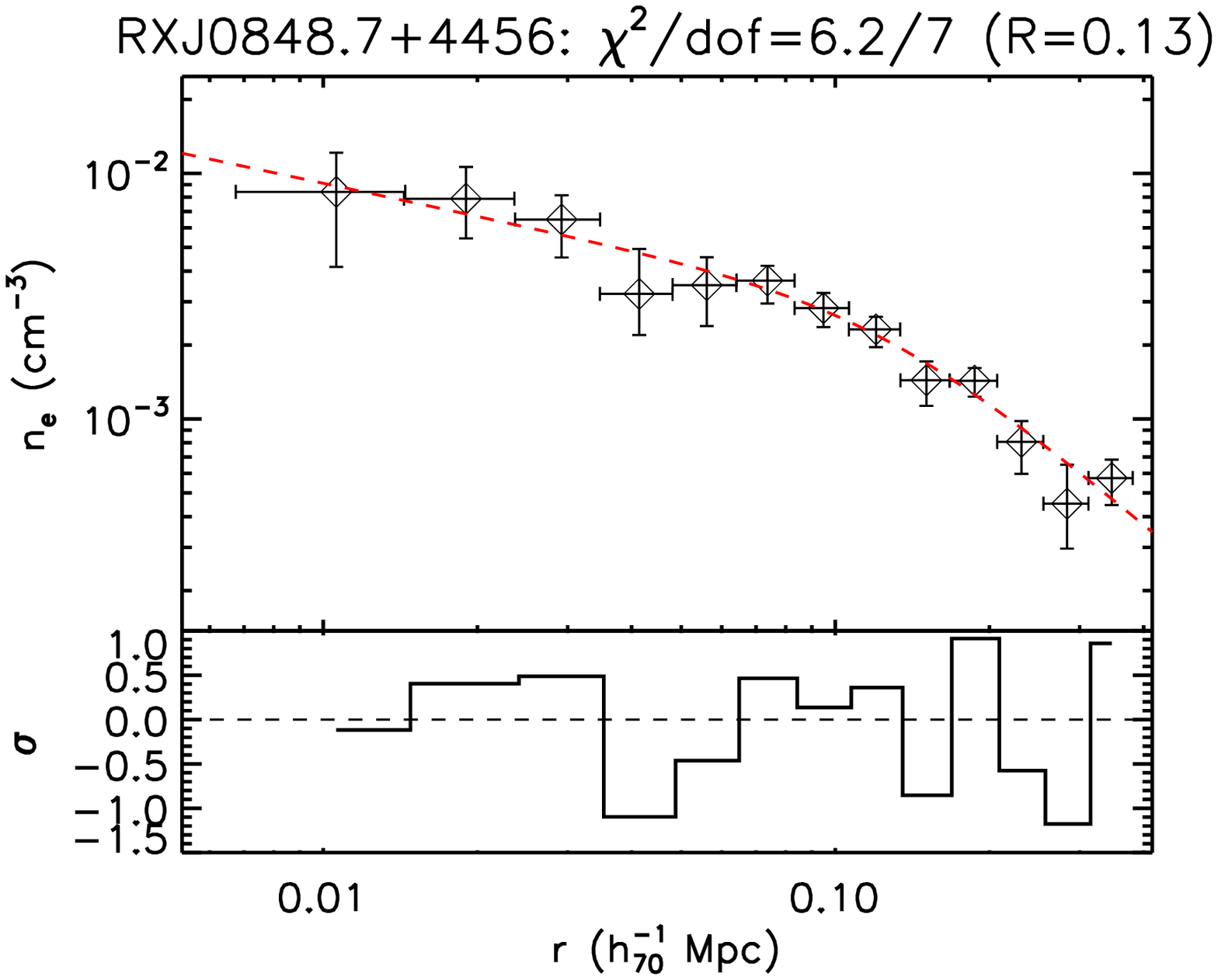,width=0.4\textwidth}
 \epsfig{figure=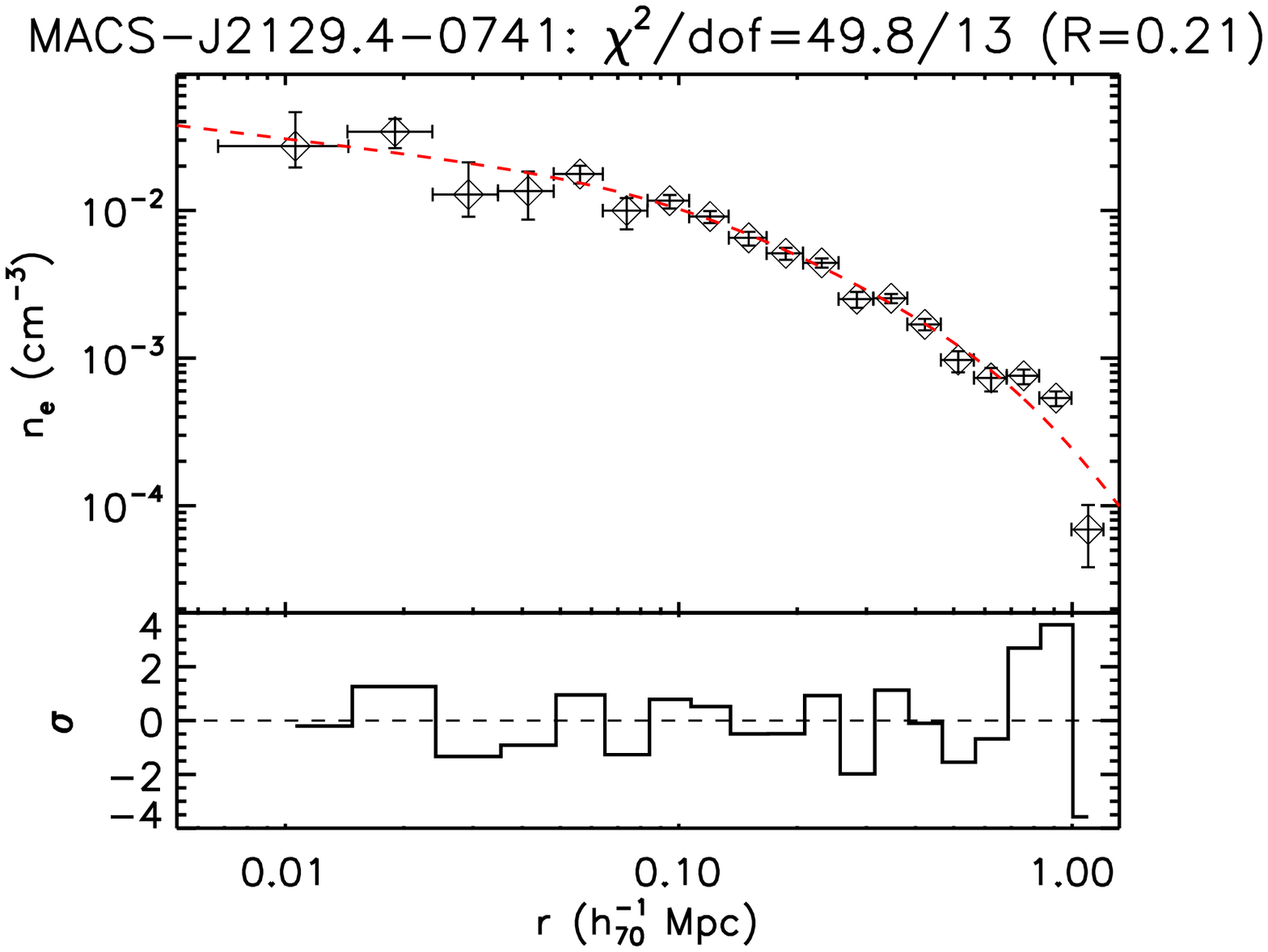,width=0.4\textwidth}
} \vspace*{-0.5cm} \hbox{
 \epsfig{figure=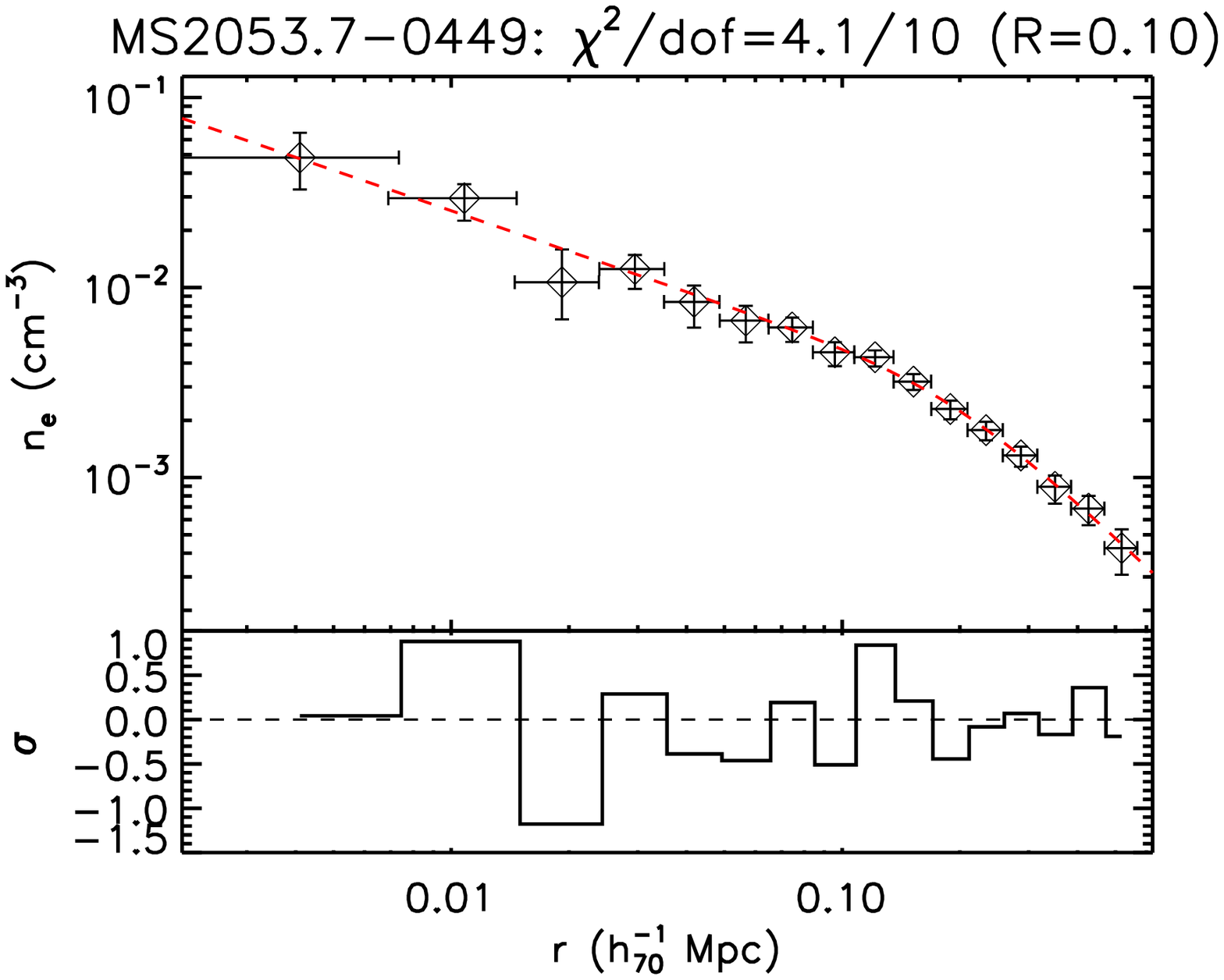,width=0.4\textwidth}
 \epsfig{figure=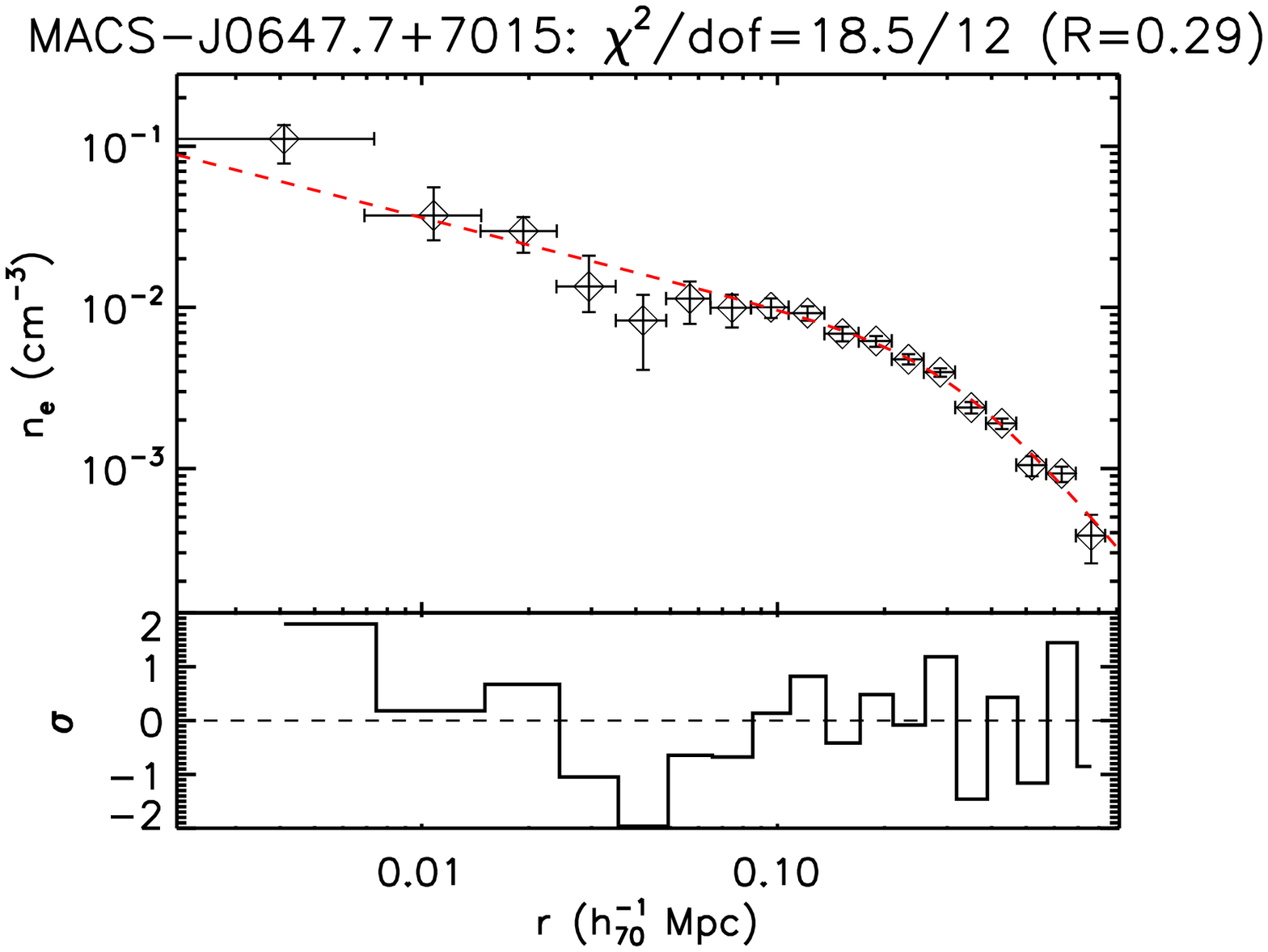,width=0.4\textwidth}
}
\end{figure*}

\begin{figure*}
\vspace*{-0.0cm} \hbox{
 \epsfig{figure=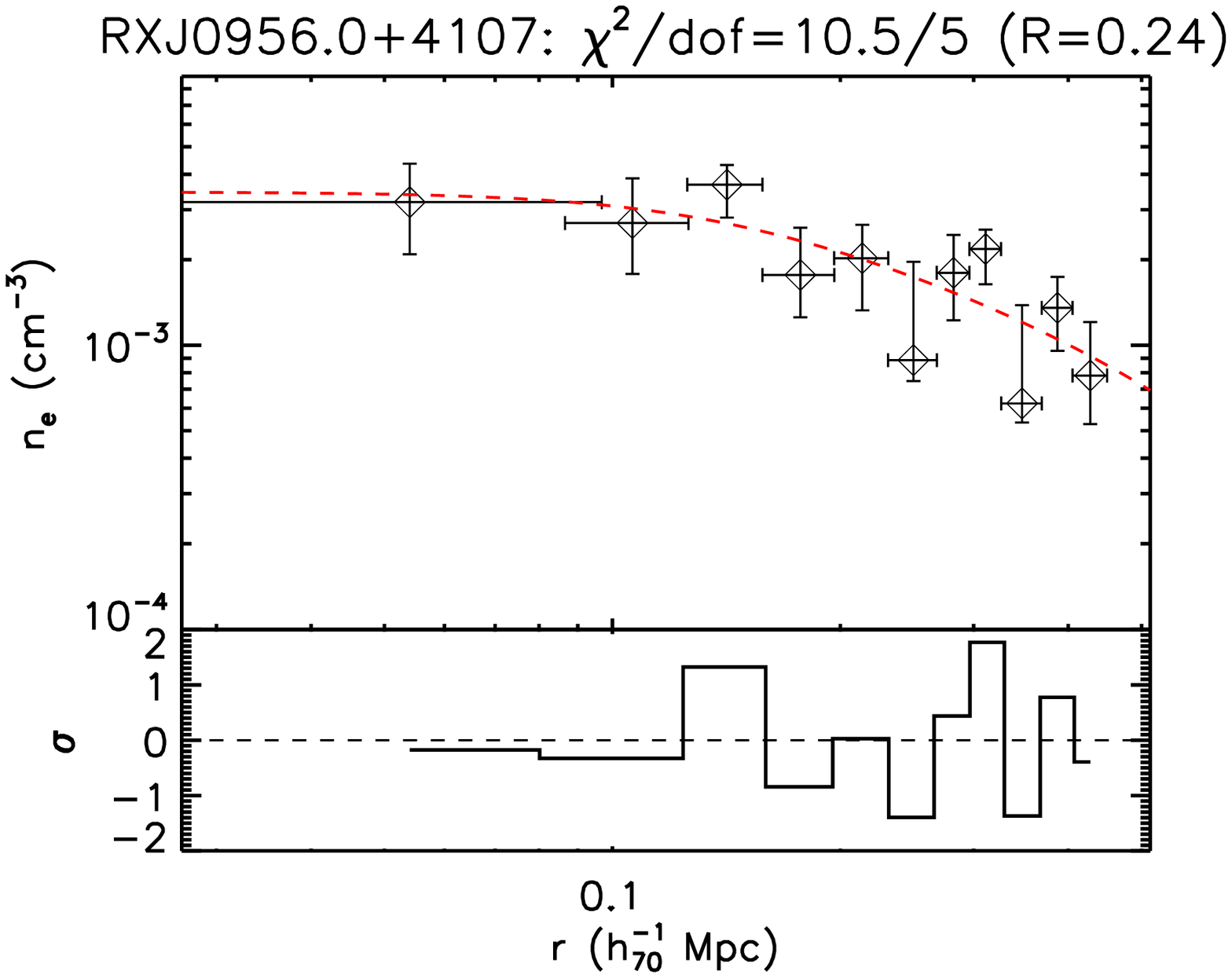,width=0.4\textwidth}
 \epsfig{figure=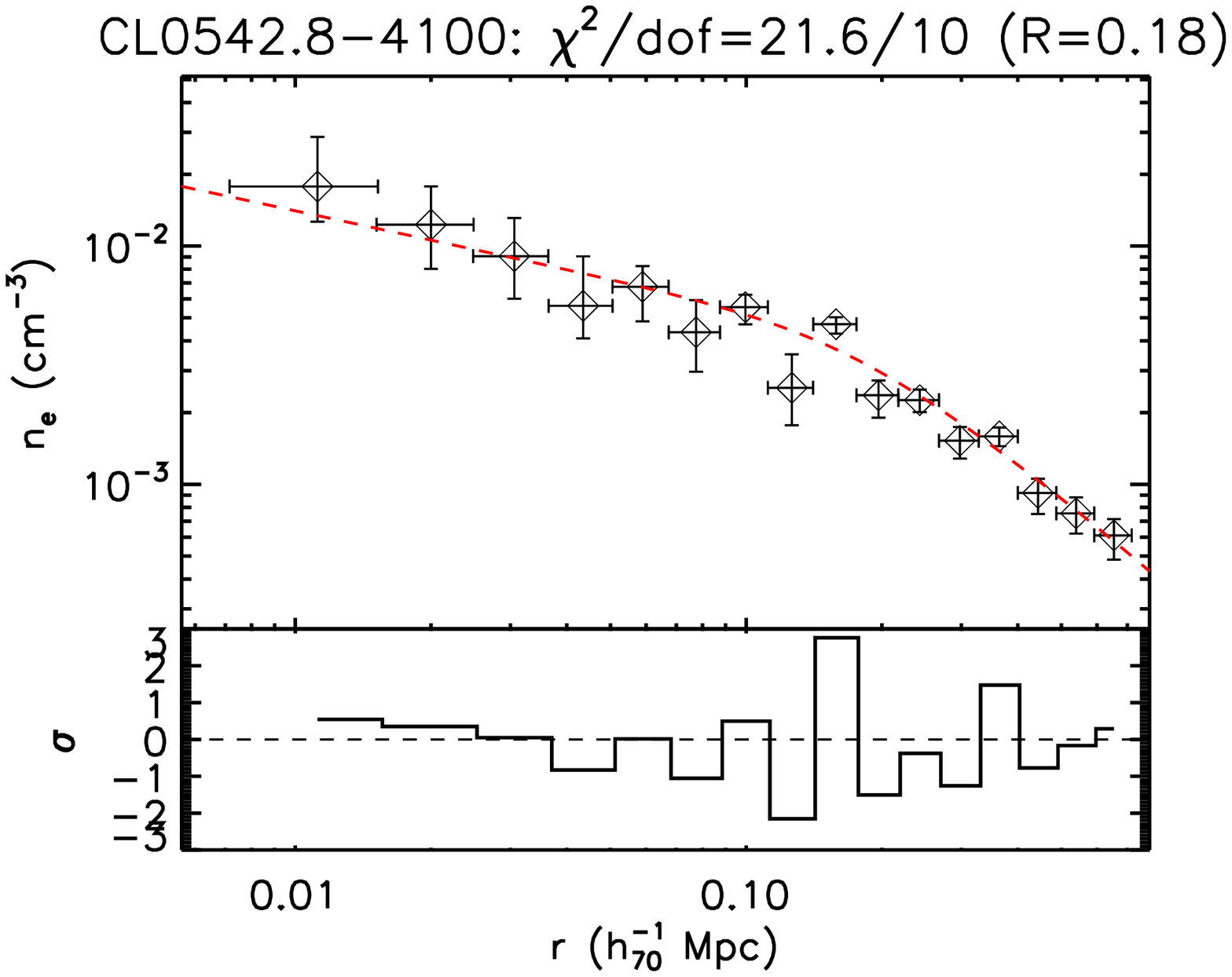,width=0.4\textwidth}
} \vspace*{-0.5cm} \hbox{
 \epsfig{figure=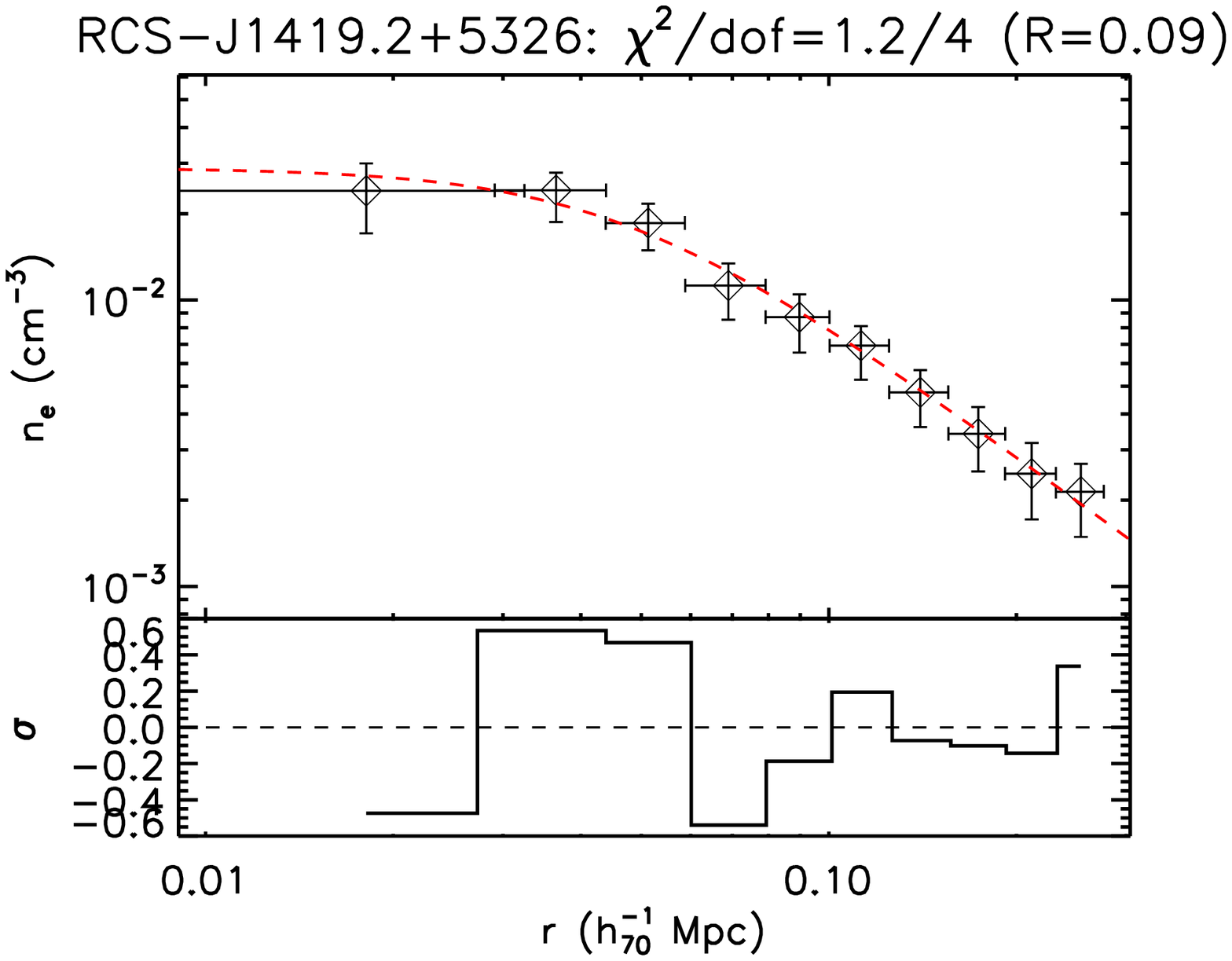,width=0.4\textwidth}
 \epsfig{figure=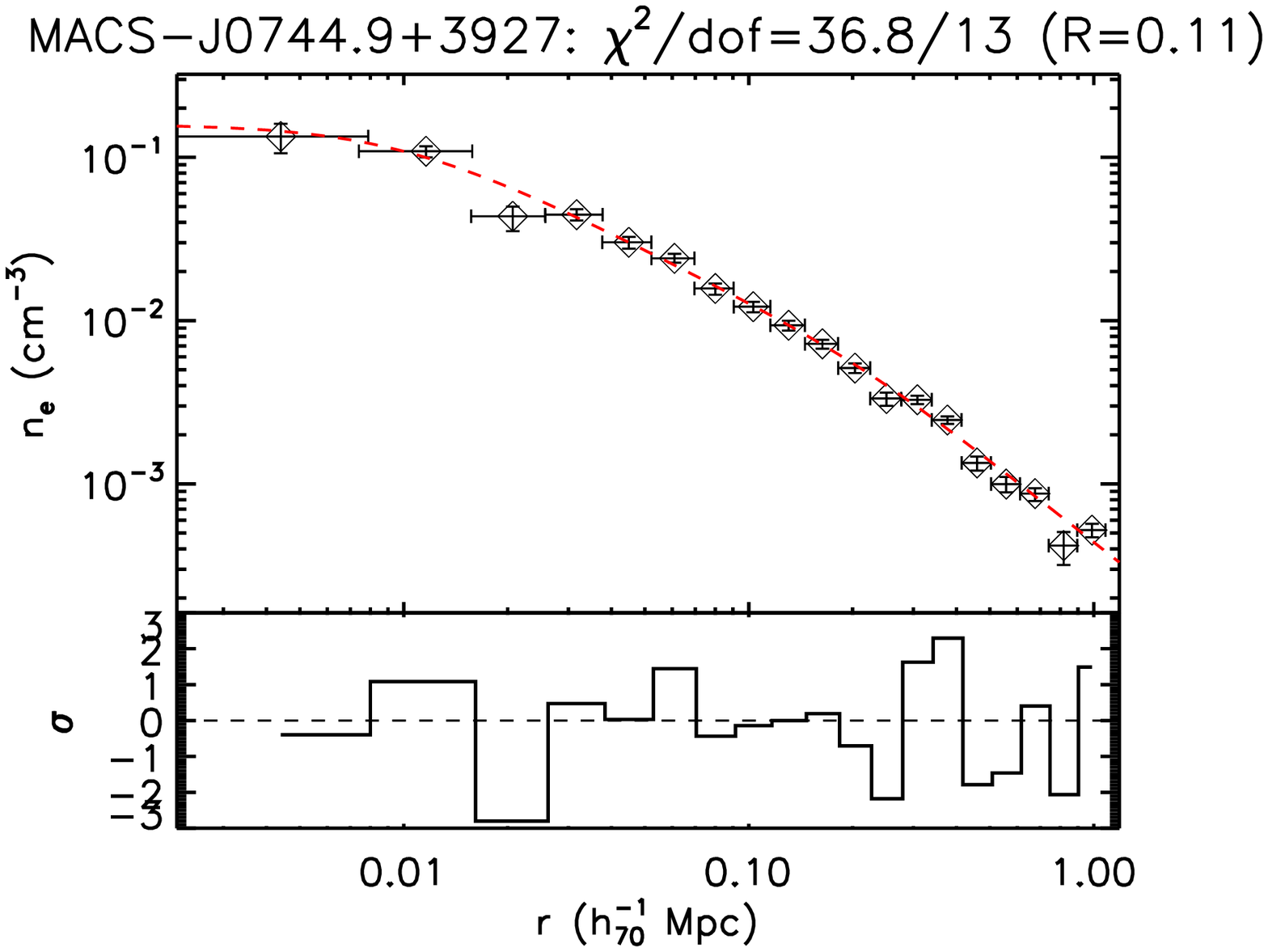,width=0.4\textwidth}
} \vspace*{-0.5cm} \hbox{
 \epsfig{figure=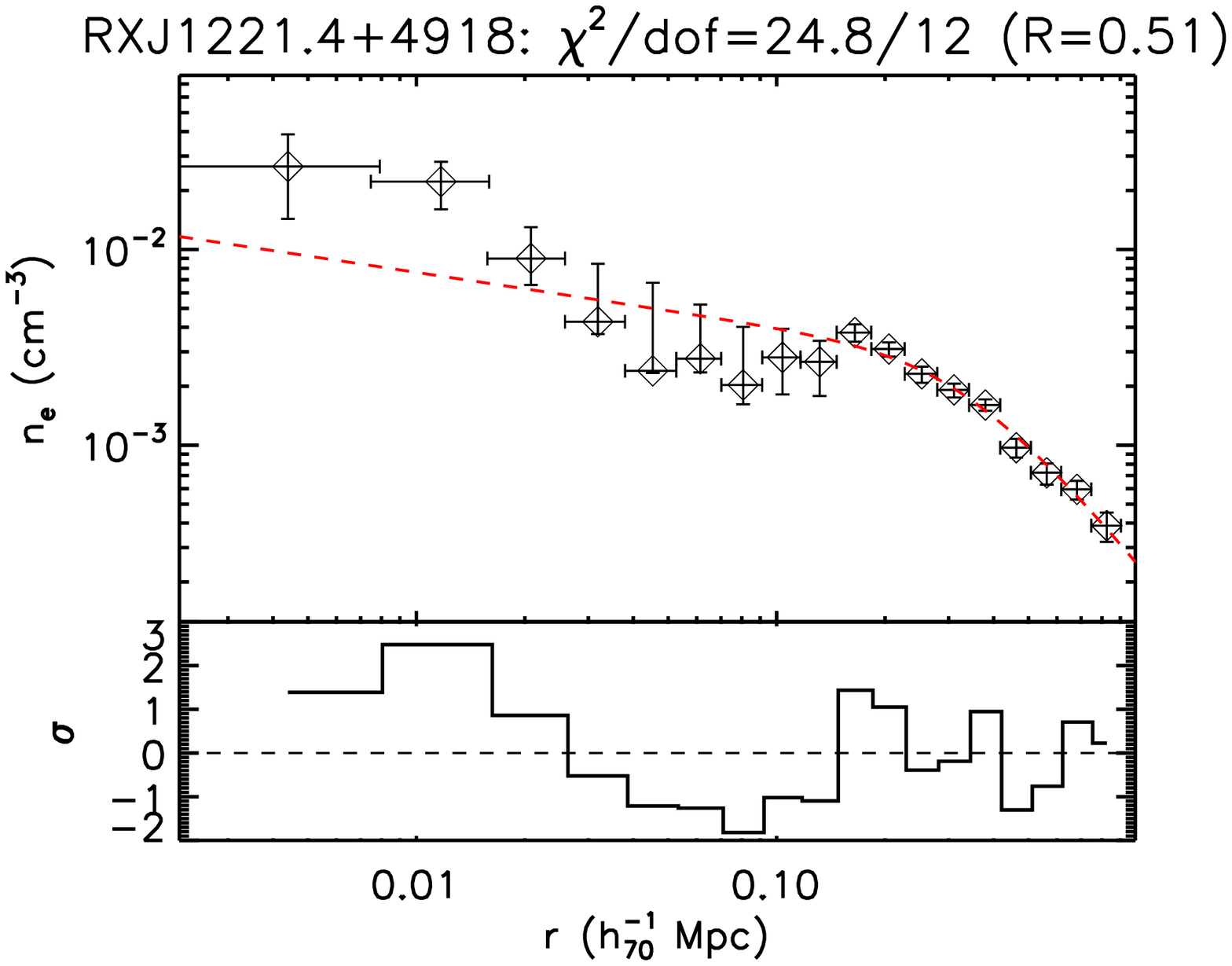,width=0.4\textwidth}
 \epsfig{figure=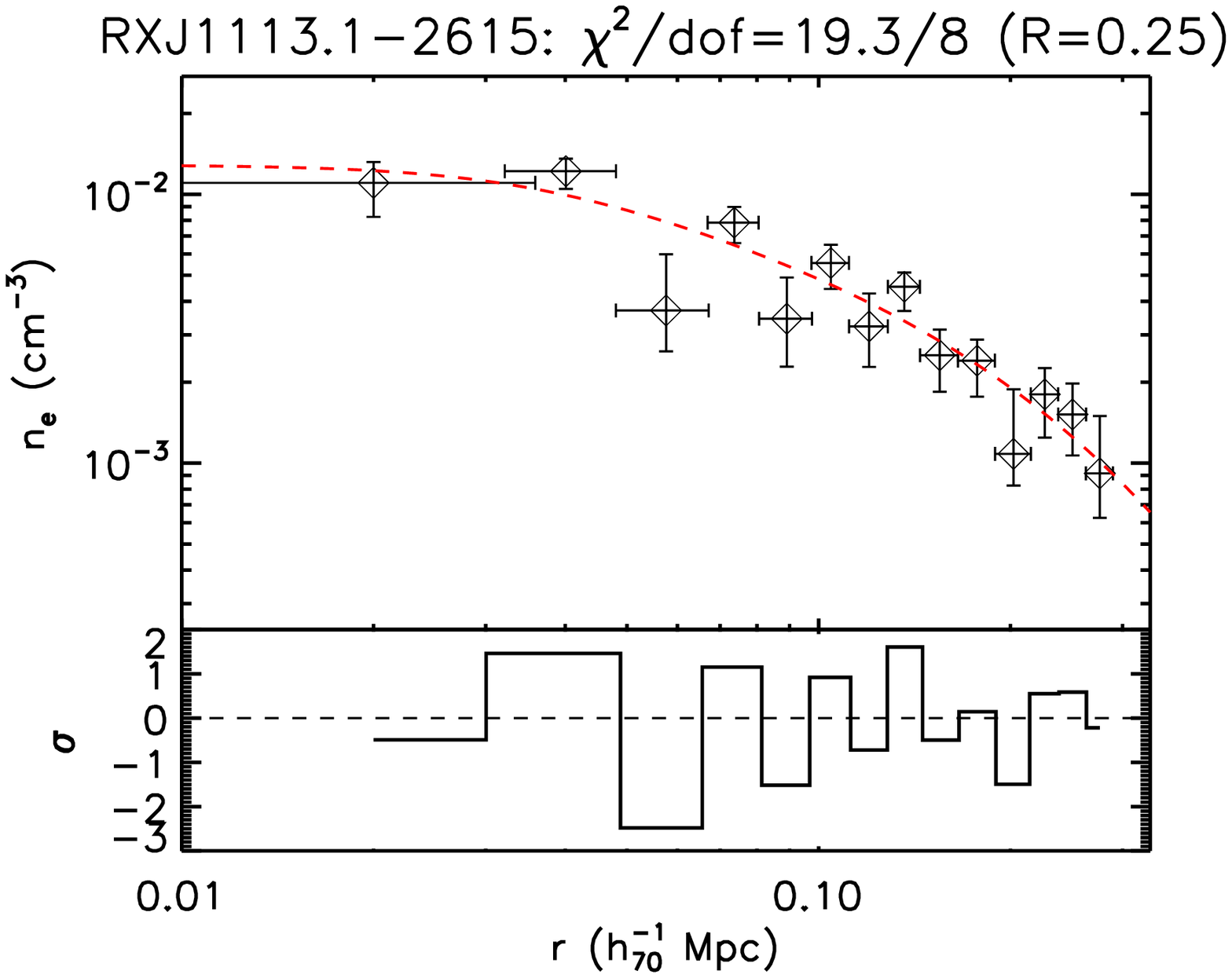,width=0.4\textwidth}
} \vspace*{-0.5cm} \hbox{
 \epsfig{figure=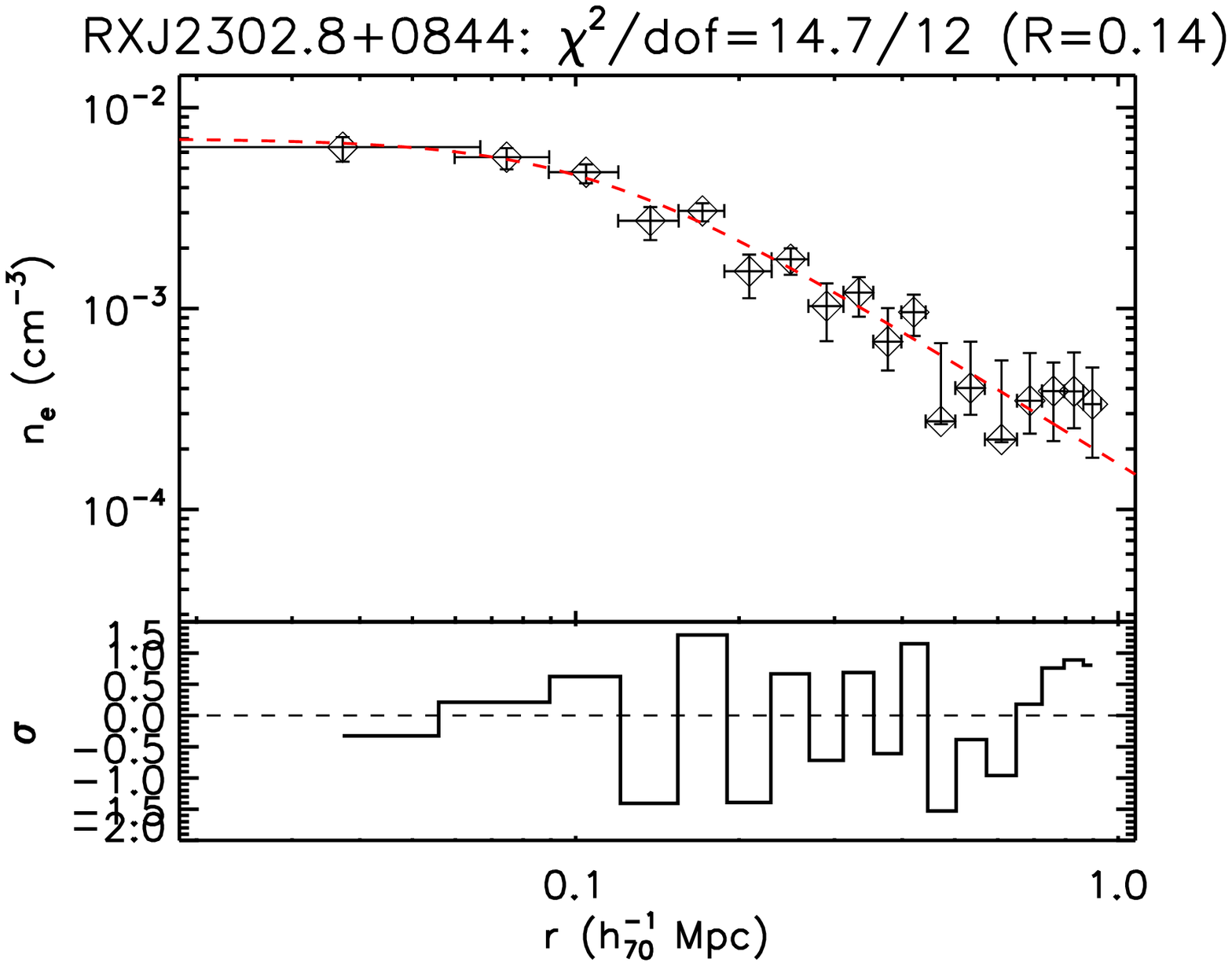,width=0.4\textwidth}
 \epsfig{figure=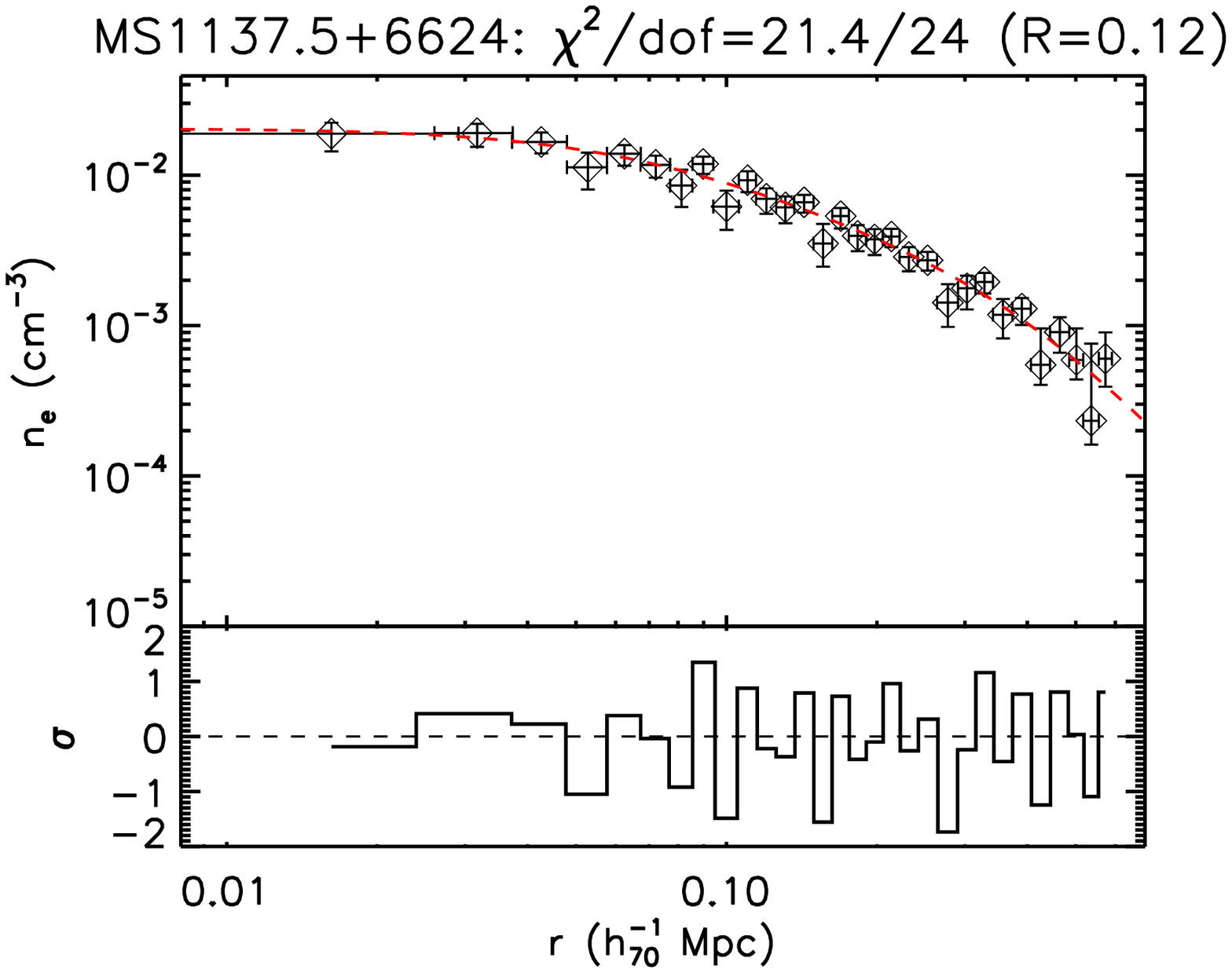,width=0.4\textwidth}
}
\end{figure*}

\begin{figure*}
\vspace*{-0.0cm} \hbox{
 \epsfig{figure=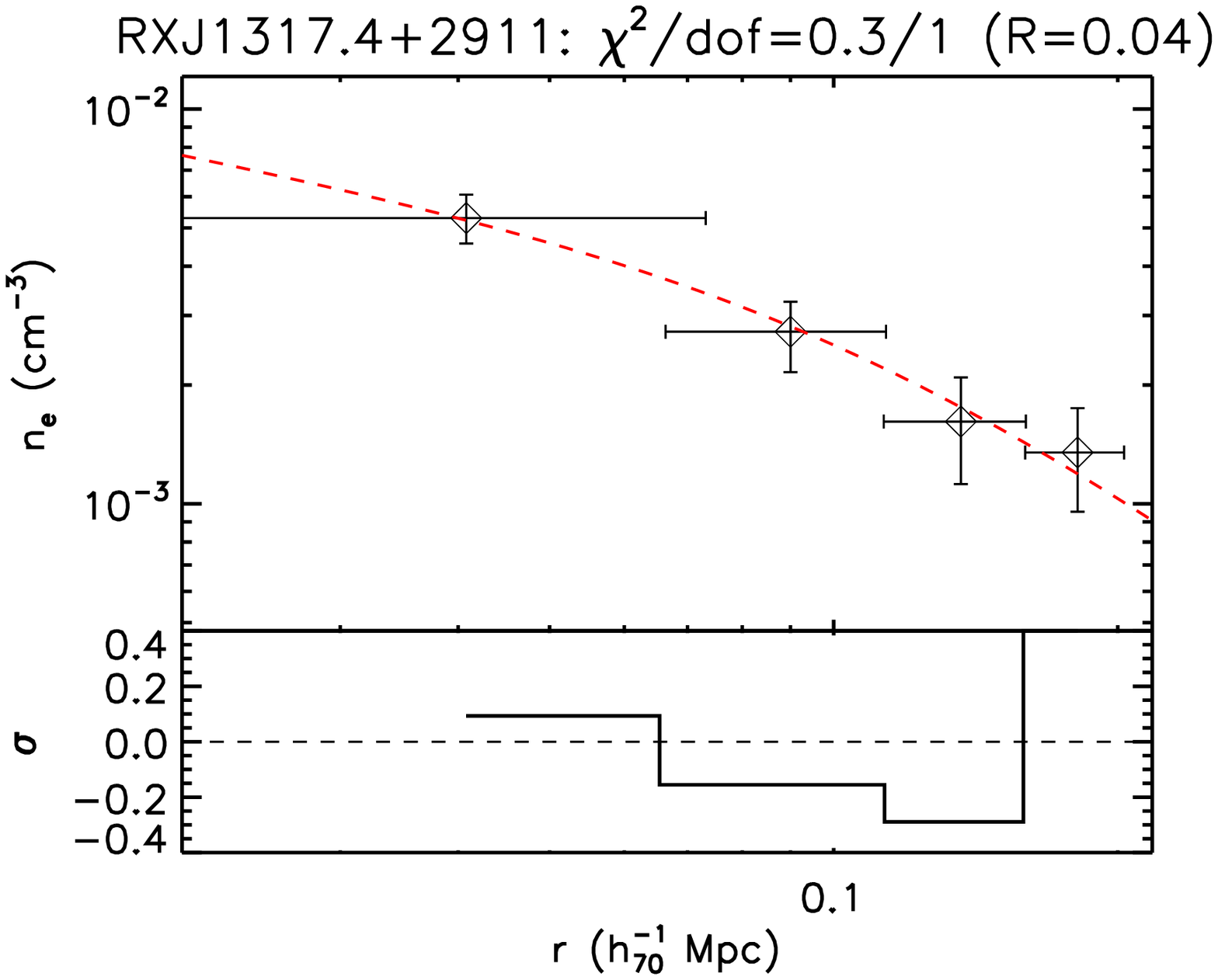,width=0.4\textwidth}
 \epsfig{figure=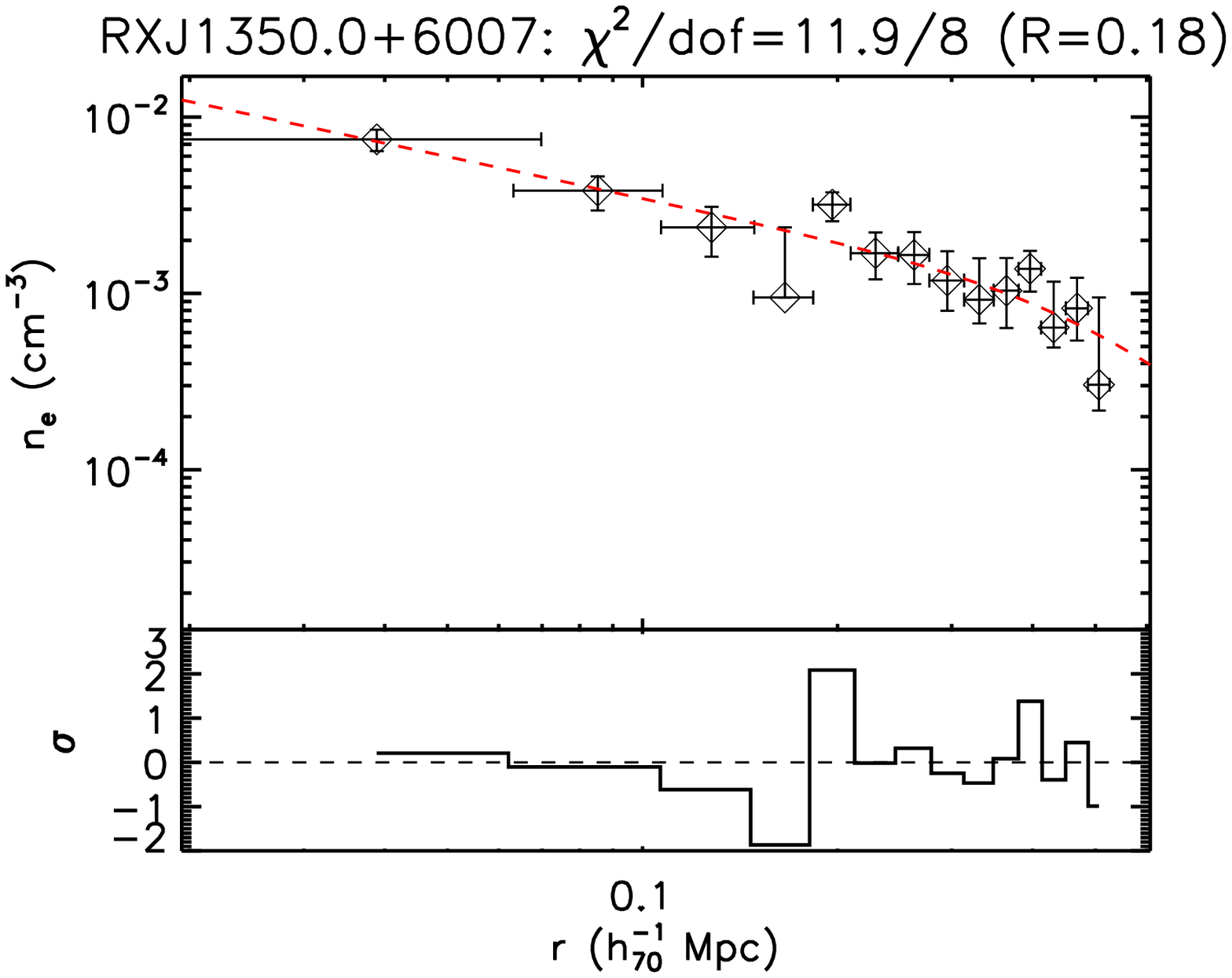,width=0.4\textwidth}
} \vspace*{-0.5cm} \hbox{
 \epsfig{figure=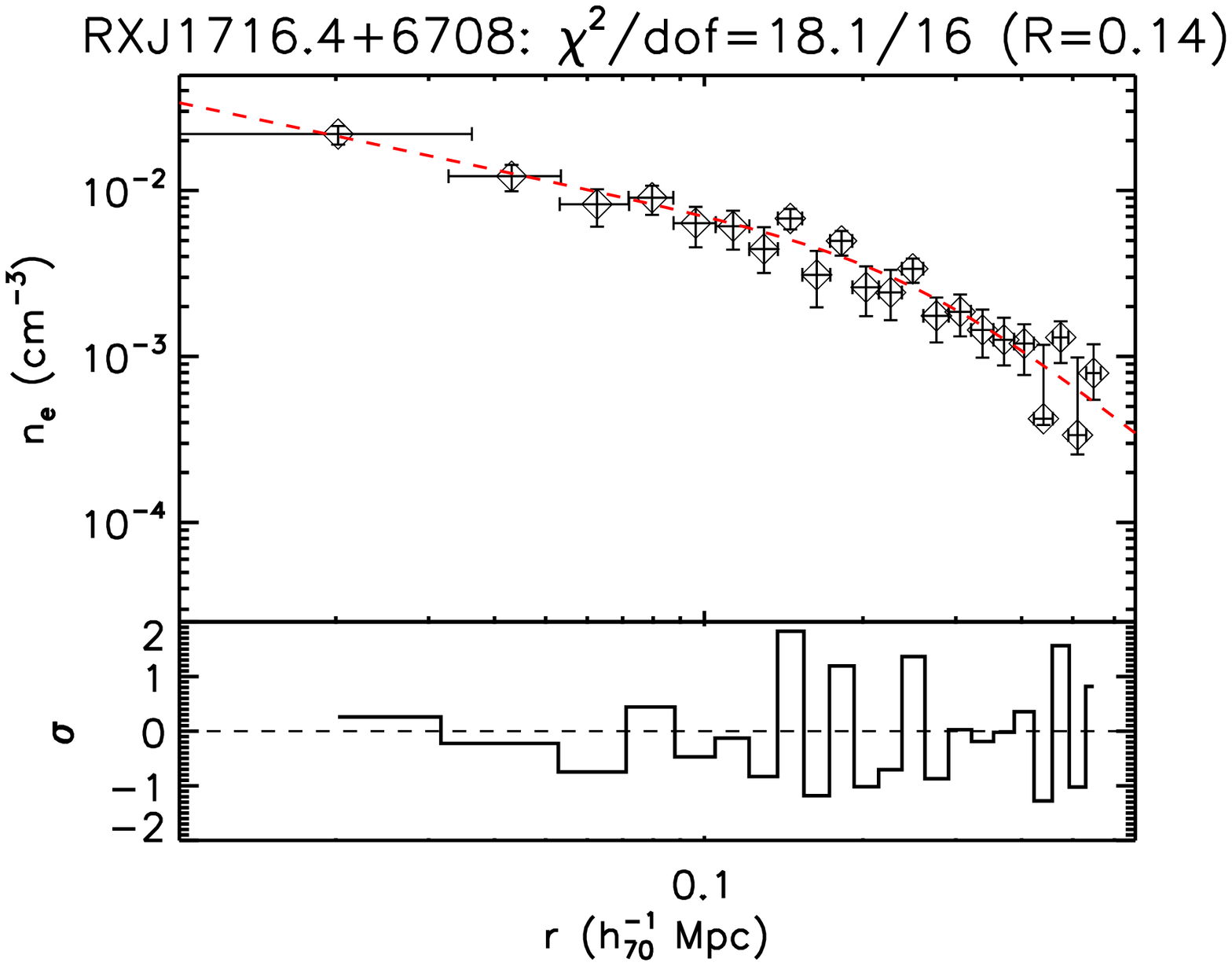,width=0.4\textwidth}
 \epsfig{figure=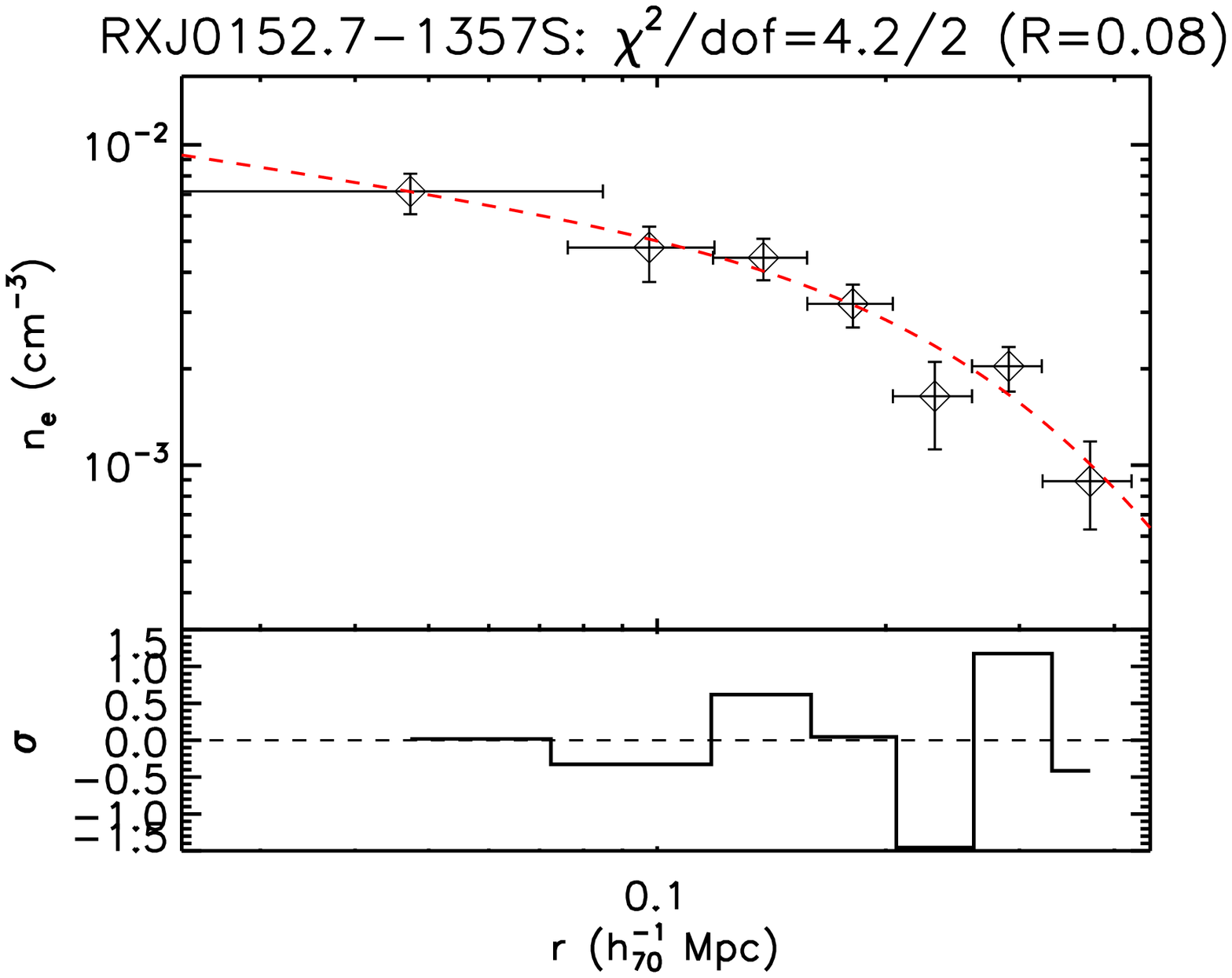,width=0.4\textwidth}
} \vspace*{-0.5cm} \hbox{
 \epsfig{figure=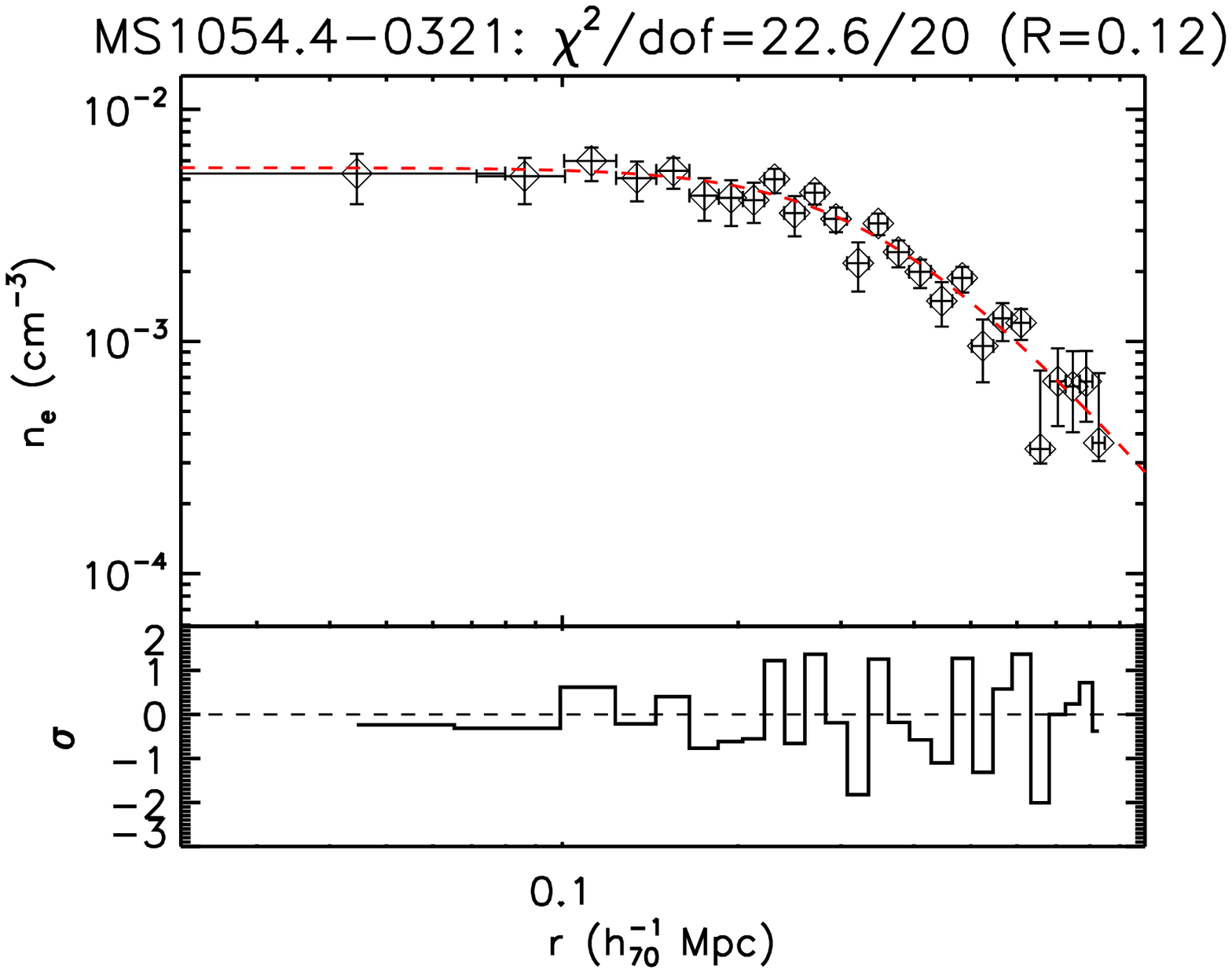,width=0.4\textwidth}
 \epsfig{figure=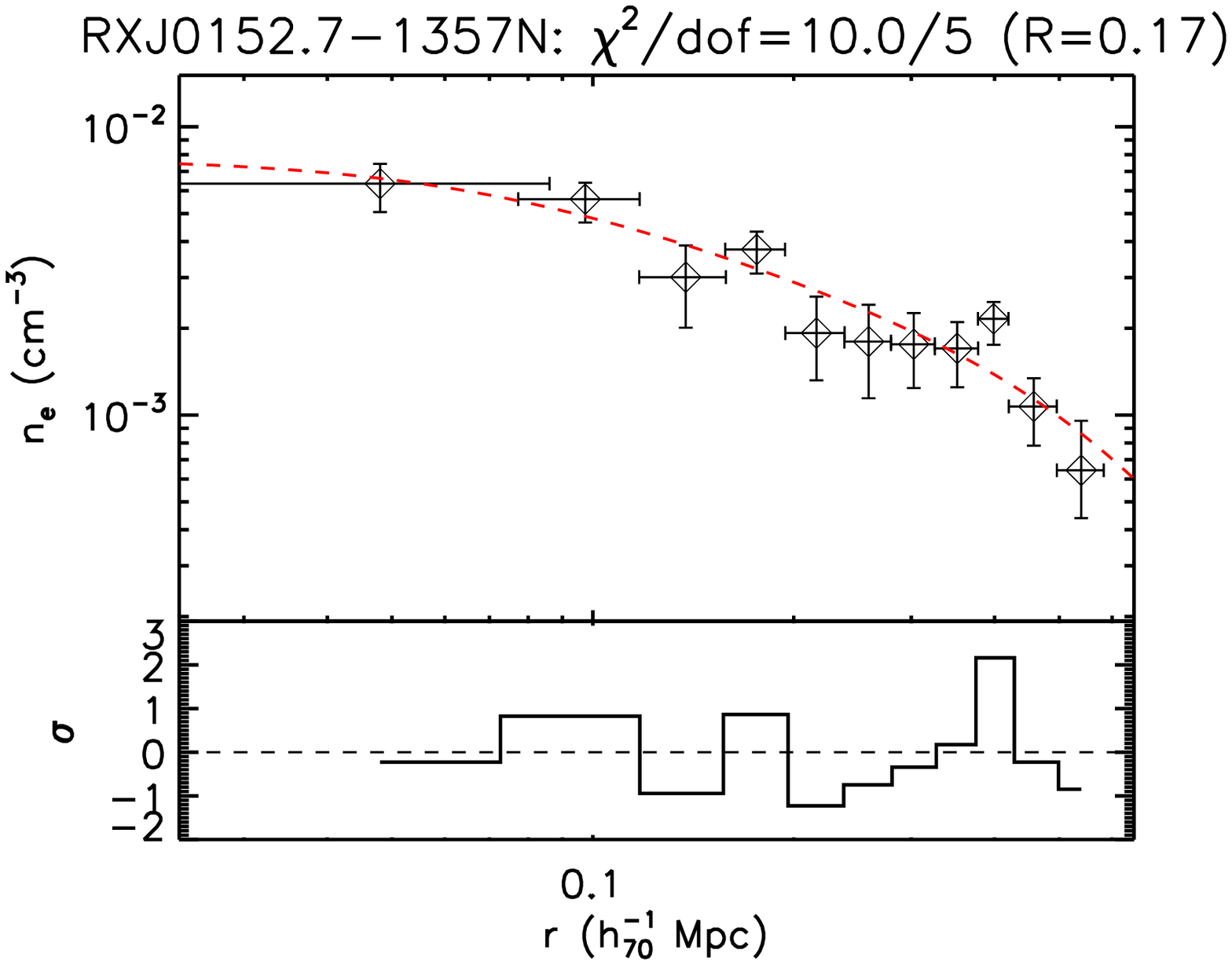,width=0.4\textwidth}
} \vspace*{-0.5cm} \hbox{
 \epsfig{figure=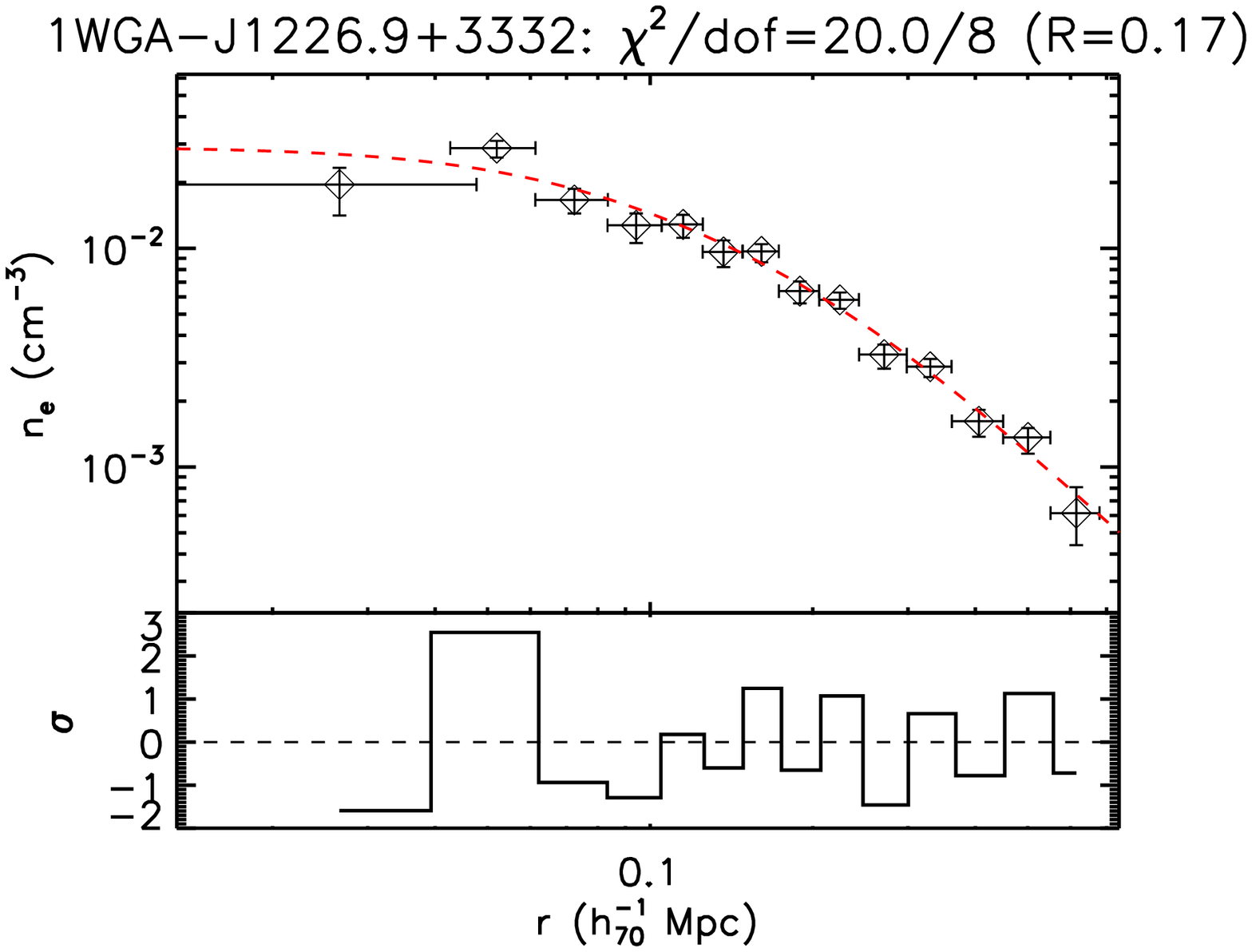,width=0.4\textwidth}
 \epsfig{figure=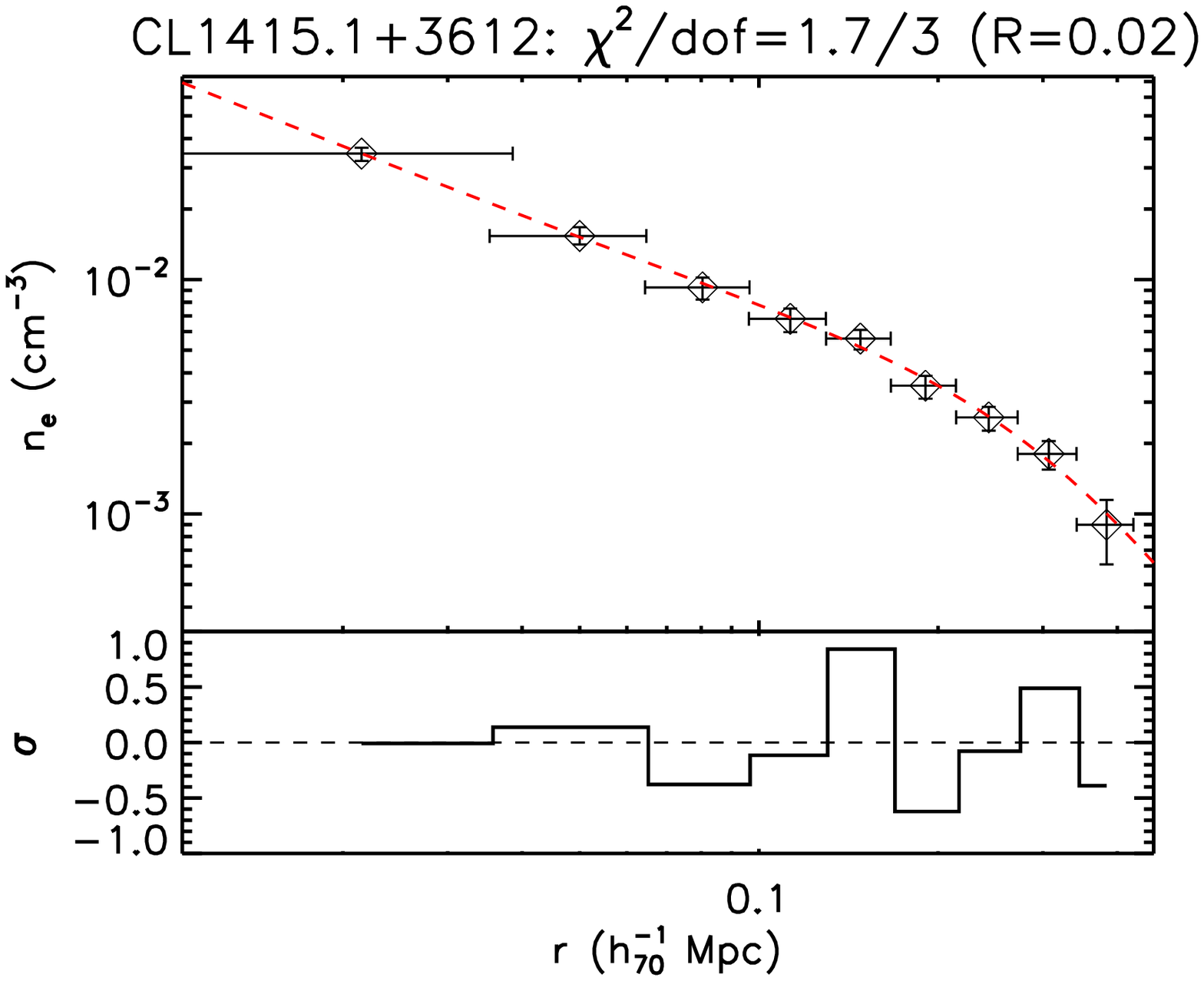,width=0.4\textwidth}
}
\end{figure*}

\begin{figure*}
\vspace*{-0.0cm} \hbox{
 \epsfig{figure=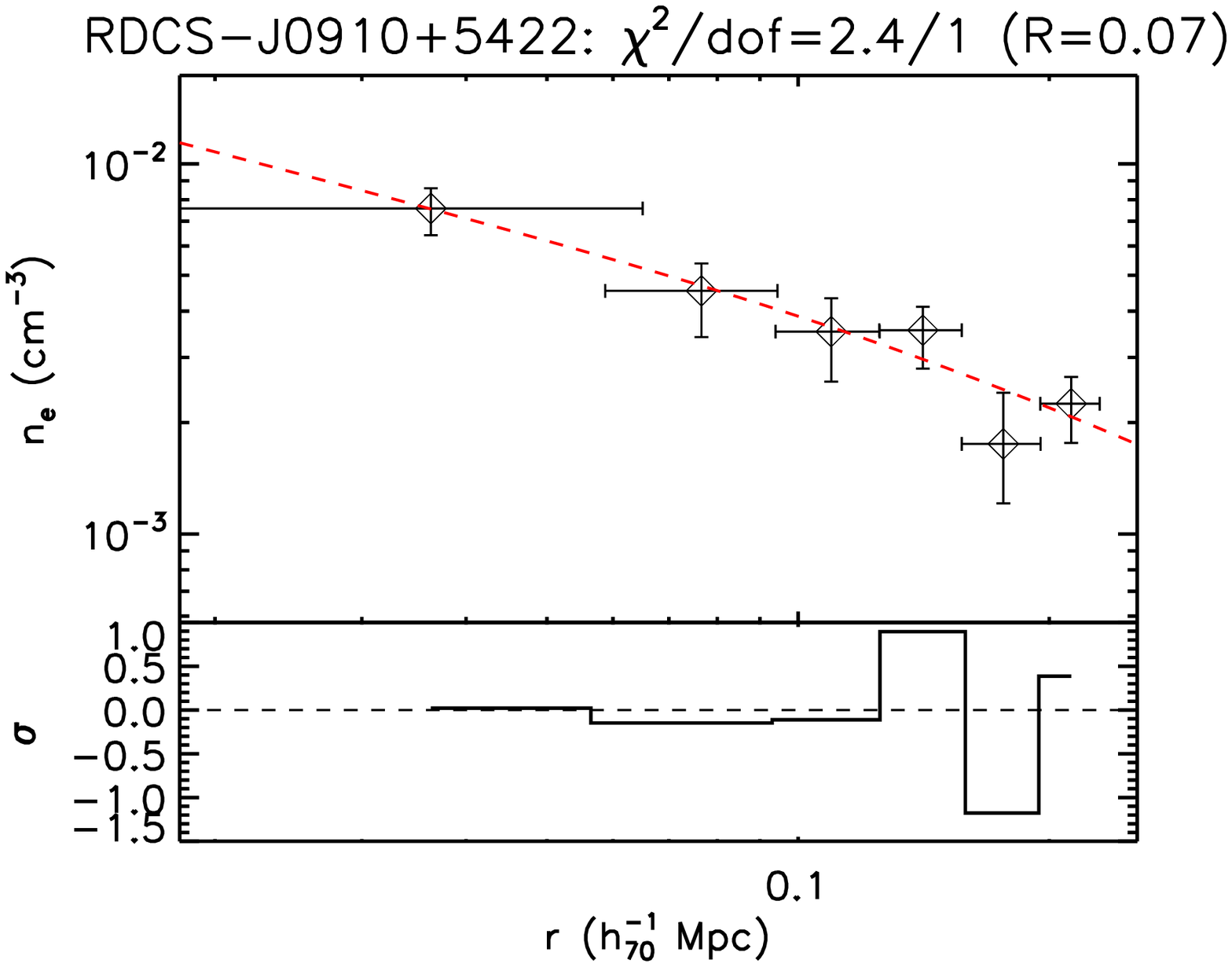,width=0.4\textwidth}
 \epsfig{figure=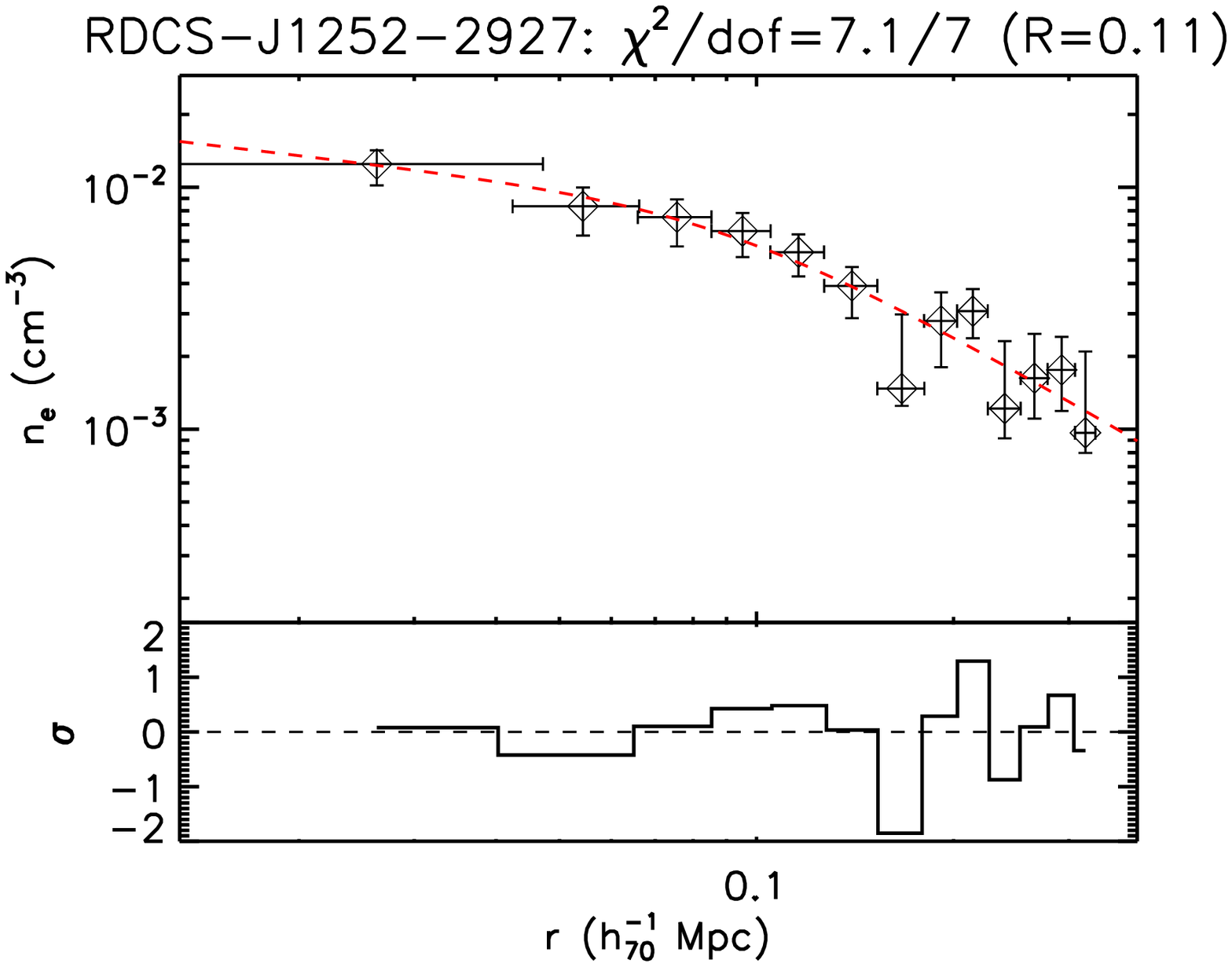,width=0.4\textwidth}
} \vspace*{-0.5cm} \hbox{
 \epsfig{figure=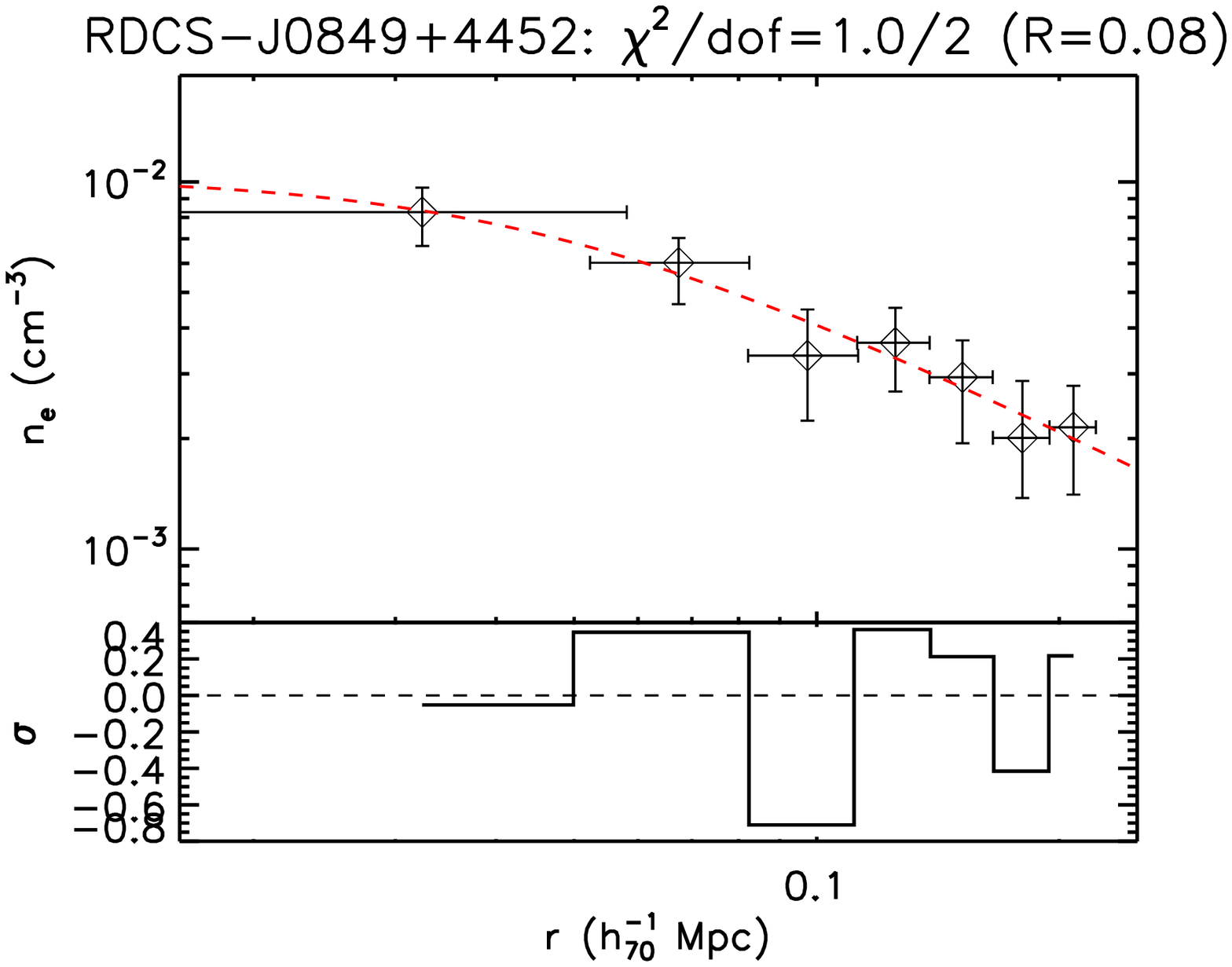,width=0.4\textwidth}
 \epsfig{figure=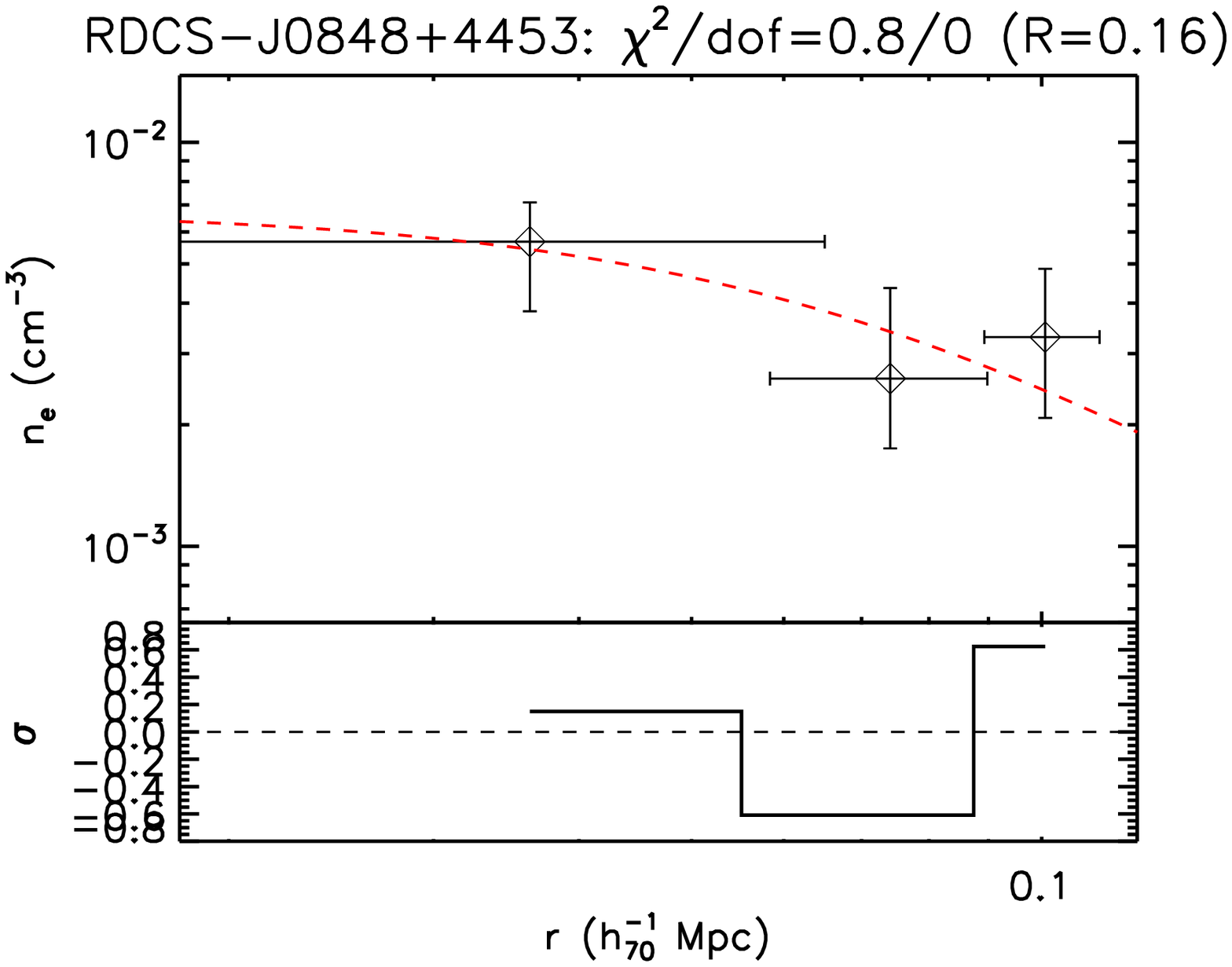,width=0.4\textwidth}
}
\end{figure*}

\end{appendix}

\end{document}